# CAGI, the Critical Assessment of Genome Interpretation, establishes progress and prospects for computational genetic variant interpretation methods

The Critical Assessment of Genome Interpretation Consortium

## ABSTRACT

The Critical Assessment of Genome Interpretation (CAGI) aims to advance the state of the art for computational prediction of genetic variant impact, particularly those relevant to disease. The five complete editions of the CAGI community experiment comprised 50 challenges, in which participants made blind predictions of phenotypes from genetic data, and these were evaluated by independent assessors. Overall, results show that while current methods are imperfect, they have major utility for research and clinical applications. Missense variant interpretation methods are able to estimate biochemical effects with increasing accuracy. Performance is particularly strong for clinical pathogenic variants, including some difficult-to-diagnose cases, and extends to interpretation of cancer-related variants. Assessment of methods for regulatory variants and complex trait disease risk is less definitive, and indicates performance potentially suitable for auxiliary use in the clinic. Emerging methods and increasingly large, robust datasets for training and assessment promise further progress ahead.

## INTRODUCTION

Rapidly accumulating data on individual human genomes hold the promise of revolutionizing our understanding and treatment of human disease.[1, 2] Effectively leveraging these data requires reliable methods for interpreting the impact of genetic variation. The DNA of unrelated individuals differs at millions of positions,[3] most of which make negligible contribution to disease risk and phenotypes. Therefore, interpretation approaches must be able to identify the small number of variants with phenotypic significance, including those causing rare disease such as cystic fibrosis,[4] those contributing to increased risk of cancer[5] or acting as cancer drivers,[6] those contributing to complex traits such as type II diabetes,[7] and those affecting the response of individuals to drugs such as warfarin.[8] Identifying the relationship between variants and phenotype can also lead to new biological insights and new therapeutic strategies. Until recently, interpretation of the role of specific variants has either been acquired by empirical observations in the clinic, thus slowly accumulating robust knowledge,[9] or by meticulous and often indirect *in vitro* experiments whose interpretation may be challenging. Computational methods offer a third and potentially powerful approach, and over one hundred have been developed,[10] but their power, reliability, and clinical utility have not been established.

The Critical Assessment of Genome Interpretation (CAGI) is an organization that conducts community experiments to assess the state of the art in computational interpretation of genetic variants. CAGI experiments are modeled on the protocols developed in the Critical Assessment of Structure Prediction (CASP) program,[11] adapted to the genomics domain. The process is designed to assess the accuracy of computational methods, highlight methodological innovation and reveal bottlenecks, guide future research, contribute to the development of new guidelines for clinical practice, and provide a forum for the dissemination of research results. Participants are periodically provided with sets of genetic data and asked to relate these to unpublished phenotypes. Independent assessors evaluate the anonymized predictions, promoting a high level of rigor and objectivity. Assessment outcomes together with invited papers from participants have been published in special issues of the journal *Human Mutation*.[12, 13] Since CAGI has stewardship of genetic data from human research participants, an essential part of the organizational structure is its Ethics Forum composed of ethicists and researchers, together with patient advocates. Further details are available at https://genomeinterpretation.org/.

Over a period of a decade, CAGI has conducted five rounds of challenges, 50 in all, attracting 738 submissions worldwide (Figure 1, Supplementary Table 1, and Supplementary Text). Challenge datasets have come from studies of variant impact on protein stability[14, 15] and functional phenotypes such as enzyme activity,[16, 17] cell growth,[18] and whole-organism fitness,[19] with examples relevant to rare monogenic disease,[20] cancer,[21] and complex traits.[22, 23] Variants in these datasets have included those affecting protein function, gene expression, and splicing, and have comprised single base changes, short insertions or deletions (indels), as well as structural variation. Genomic scale has ranged from single nucleotides to complete genomes, with inclusion of some complementary multiomic and clinical information (Figure 1).

Results are analyzed from three perspectives to provide (i) the clinical community with an assessment of the usefulness and limitations of computational methods, (ii) the biomedical research community with information on the current state of the art for predicting variant impact on a range of biochemical and cellular phenotypes, and (iii) the developers of computational methods with data on method performance with the aim of spurring further innovation. This latter perspective is particularly important because of the recent successes of artificial intelligence approaches in related fields.[24, 25]

The results are presented under a set of themes that have emerged over the decade of CAGI challenges. For each theme, specific examples of performance are provided, based on particular ranking criteria. As always in CAGI, these should not be interpreted as identifying winners and losers—other criteria might result in different selections. Further, these selections were made by authors of this paper, some of whom have been CAGI participants, rather than by independent assessors. However, the examples shown are consistent with the assessors' earlier rankings.

## RESULTS

**Biochemical effect predictions for missense variants are strongly correlated with the experimental data, but individual predicted effect size accuracy is limited.** The pathogenicity of missense variants implicated in monogenic disease and cancer is often supported by *in vitro* experiments that measure effects on protein activity, cell growth, or various biochemical properties.[26] Thirteen CAGI challenges have assessed the ability of computational methods to estimate these functional effects using datasets from both high- and low-throughput experimental assays, and ten of these have been reanalyzed here. Figure 2 shows selected results for two

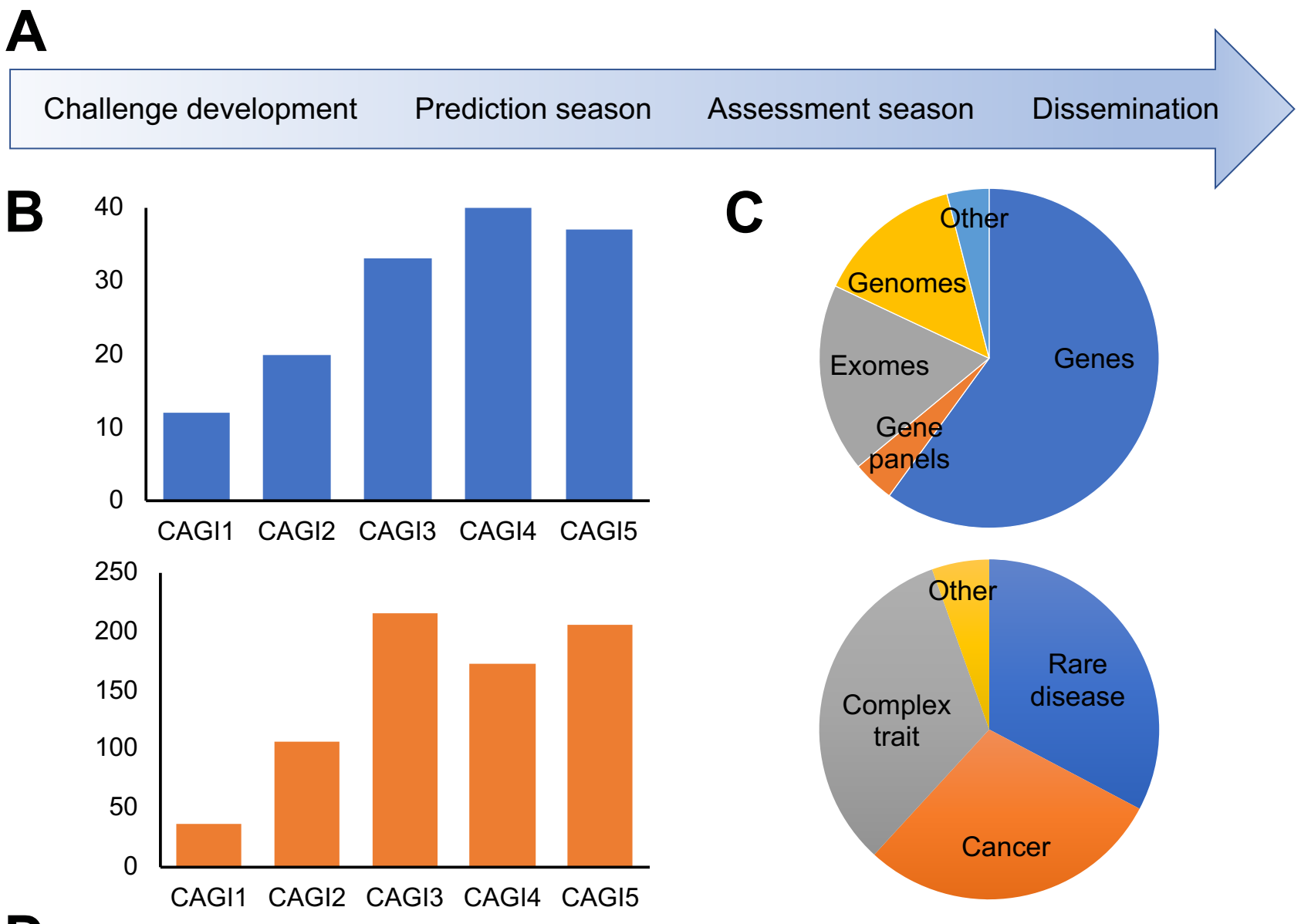

**D**

| **CAGI1** | **CAGI2** | **CAGI3** | **CAGI4** | **CAGI5** |
|---|---|---|---|---|
| CBS (RD) | Asthma twins (CT, O) | BRCA1 & BRCA2 (C) | Bipolar disorder (CT) | Annotate all missense (RD) |
| CHEK2 (C) | Breast cancer PGX (C, O) | Crohn's disease (CT) | Crohn's disease (CT) | CALM1 (RD) |
| PGP (CT, M) | CBS (RD) | FCH (CT) | eQTL causal SNPs (CT) | CHEK2 (C) |
| | Crohn's disease (CT) | HA (RD) | Hopkins panel (RD) | Clotting disease (CT) |
| | Mouse exomes (RD) | MR-1 (O) | NAGLU (RD) | ENIGMA BRCA1-2 (C) |
| | MR-1 (O) | MRE11 (C) | NPM-ALK (C) | Frataxin (C) |
| | p53 (C) | NBS1 (C) | PGP (CT, M) | GAA (RD) |
| | PGP (CT, M) | P16 (C) | Pyruvate kinase (RD) | ID panel (RD) |
| | RAD50 (C) | PGP (CT, M) | SickKids (RD) | MaPSy (CT, RD) |
| | riskSNPs (CT) | riskSNPs (CT) | SUMO ligase (C, RD) | PCM1 (CT) |
| | SCN5A (RD) | TP53 splicing (C) | Warfarin exomes (CT) | PTEN (C) |
| | | | | Regulation saturation (CT) |
| | | | | SickKids (RD) |
| | | | | TPMT (C) |
| | | | | Vex-Seq (CT, RD) |

***Figure 1:*** *CAGI timeline, participation, and range of challenges.* (**A**) Stages in a round of CAGI, typically extending over two years. Each round includes a set of challenges with similar timelines. (**B**) Number of participating unique groups (in blue) and submissions (in orange) across CAGI rounds. (**C**) Genetic scale (top) and phenotypic characterization (bottom) of CAGI challenges. Some challenges belong to more than one category and are included more than once. (**D**) CAGI challenges, listed by round. C: cancer, CT: complex trait, M: Mendelian, O: other (including multiomics), RD: rare disease. See Supplementary Table 1 for more details.

challenges, each with a different type of biochemical effect. In the NAGLU challenge,[16] participants were asked to estimate the relative enzyme activity of 163 rare missense variants in N-acetyl-glucosaminidase found in the ExAC database.[27] In the PTEN challenge,[14] participants were asked to estimate the effects of a set of 3,716 variants in the phosphatase and tensin homolog on the protein's stability as measured by relative intracellular protein abundance in a high-throughput assay.[28] Detailed descriptions of these and other challenges are provided in Supplementary Materials.

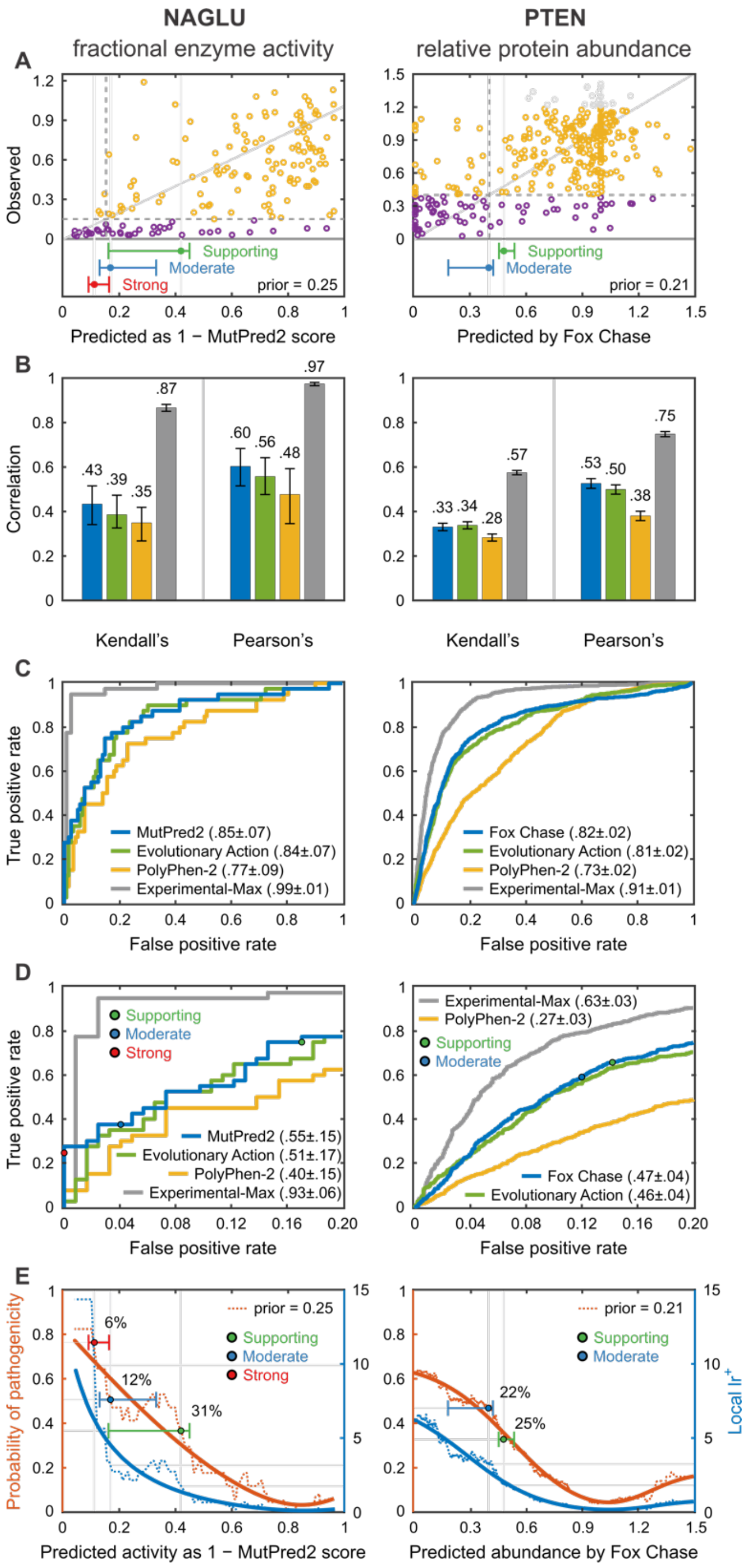

***Figure 2:*** *Predicting the effect of missense variants on protein properties: Results for two representative CAGI challenges.* Each required estimation of continuous phenotype values, enzyme activity in a cellular extract for NAGLU and intracellular protein abundance for PTEN, for a set of missense variants. Selection of methods is based on the average ranking over four metrics for each participating method: Pearson's correlation, Kendall's tau, ROC AUC and truncated ROC AUC; see Methods for definitions. (**A**) Relationship between observed and predicted values for the selected method in each challenge. "Benign" variants are yellow and "pathogenic" are purple (see text). The diagonal line represents exact agreement between predicted and observed values. Dashed lines show the thresholds for pathogenicity for observed (horizontal) and predicted biochemical values (vertical). For NAGLU, below the pathogenicity threshold, there are 12 true positives (lower left quadrant) and three false positives (upper left quadrant), suggesting a clinically useful performance. Bars below each plot show the boundaries for accuracy meeting the threshold for Supporting (green), Moderate (blue) and Strong (red) clinical evidence, with 95%

confidence intervals. (**B**) Two measures of overall agreement between computational and experimental results, for the two selected performing methods and positive and negative controls, with 95% confidence intervals. An older method, PolyPhen-2, provides a negative control against which to measure progress over the course of the CAGI experiments. Estimated best possible performance is based on experimental uncertainty and provides an empirical upper limit positive control. The color code for the selected methods is shown in panel C. (**C**) ROC curves for the selected methods with positive and negative controls, using estimated pathogenicity thresholds. (**D**) Truncated ROC curves showing performance in the high true positive region, most relevant for identifying clinically diagnostic variants. The true positive rate and false positive rate thresholds for the Supporting, Moderate, and Strong evidential support is shown for one selected method. (**E**) Estimated probability of pathogenicity (left y-axis) and positive local likelihood ratio (right y-axis) as a function of one selected method's score. Predictions with probabilities over the red, blue, and green thresholds provide Strong, Moderate, and Supporting clinical evidence, respectively. Solid lines show smoothed trends. Prior probabilities of pathogenicity are the estimated probability that any missense variant in these genes will be pathogenic. For NAGLU, the probabilities of pathogenicity reach that needed for a clinical diagnosis of "likely pathogenic". For predicted enzyme activity less than 0.11, the probability provides Strong evidence, below 0.17 Moderate evidence, and below 0.42, Supporting evidence. The percent of variants encountered in the clinic expected to meet each threshold are also shown. Performance for PTEN shows that the results are consistent with providing Moderate and Supporting evidence levels for some variants.

For both challenges, the relationship between estimated and observed phenotype values shows high scatter (Figure 2A). There is modest improvement with respect to a well-established older method, PolyPhen-2,[29] which we consider a baseline. This is a trend consistently seen in other missense challenges (Supplementary Table 2). How much of this improvement is due to the availability of larger and more reliable training sets rather than methodological improvements is unknown. Consistent with the scatter plots, there is moderate agreement between predicted and experimental values as measured by Pearson's correlation and Kendall's tau (Figure 2B).

Over all ten analyzed missense functional challenges (Supplementary Table 3, Supplementary Figures 1-6), Pearson's correlation for the selected methods ranges between 0.24 and 0.84 (average correlation $\bar{r}$ = 0.55) and Kendall's tau ranges between 0.17 and 0.63 ($\bar{\tau}$ = 0.40), both showing strong statistical significance over the random model ($\bar{r}$ = 0, $\bar{\tau}$ = 0). The PolyPhen-2 baseline achieves $\bar{r}$ = 0.36 and $\bar{\tau}$ = 0.23. Direct agreement between observed and predicted values is measured by $R^2$ (Supplementary Table 3); see Methods. $R^2$ is 1 for a perfect method and zero for a control method that assigns the mean of the experimental data for every variant. For NAGLU, the highest $R^2$ achieved is 0.16, but for PTEN it is only –0.09. Over the ten biochemical challenges, the highest $R^2$ value ranges between –0.94 and 0.40, with an average of –0.19. The relatively poorer showing by this criterion compared with Pearson's and Kendall's correlation metrics suggests that the methods are often not well calibrated to the experimental value, reflecting the fact that they are rarely designed for predictions of continuous values and scales of this kind. Overall, performance is far above random but modest in terms of absolute accuracy.

***Diversity of methods***. A diverse set of methods was used to address the biochemical effect challenges, varying in the type of training data, input features, and statistical framework. Most were trained on pathogenic versus presumed benign variants.[10, 30] At first glance, a binary classification approach appears ill-suited to challenges which require prediction across a full range of phenotype values. In practice, function and pathogenicity are related,[31] and so these methods performed as well as the few trained specifically to identify alteration of function.[32]

Many methods are based on measures that reflect the evolutionary fitness of substitutions and population dynamics, rather than pathogenicity or functional properties. The relationship between fitness, pathogenicity and function is complex, perhaps limiting performance. To partly address this, some methods also exploit functional roles of specific sequence positions, particularly by utilizing UniProtKB annotations and predicted structural and functional properties.[32-35]

Despite the broad range of algorithms, training data, features and the learning setting there is a strong correlation between results of the leading methods (Pearson's correlation ranges from 0.6 to 0.9), almost always stronger than the correlation between specific methods and experiment (Supplementary Figure 8). The level of inter-method correlation is largely unrelated to the level of correlation with experiment, which varies widely from about 0.24 (CALM1) to 0.6 (NAGLU). Why correlation between methods is stronger than with experiment is unclear, though it may be affected by the relatedness of functional disruption, evolutionary conservation, and pathogenicity as well as common training data and experimental bias. The assessor for the NAGLU challenge identified 10 variants where experiment disagrees strongly with predicted values for all methods.[16] When these are removed, the correlation between the leading methods' results and experiment increases from 0.6 to 0.73 (Supplementary Figure 8), although it is still lower than the correlation between the two leading methods (0.82), of which, surprisingly, one is supervised[34] and the other is not.[36] It could be that these 10 variants are cases where the computational methods systematically fail, or it could be that most are some form of experimental artifact. In situations like this, follow-up experiments are needed.

***Structure-informed approaches.*** Some methods use only biophysical input features, and in some cases are trained on the effect of amino acid substitutions on protein stability, rather than pathogenicity or functional impact. Benchmarking suggests that a large fraction of rare disease-causing and cancer driver missense mutations act through destabilization,[37, 38] so there is apparently considerable potential for these approaches. These methods have been effective on challenges directly related to stability, being selected as first and second for the PTEN and TMPT protein abundance challenges and first for the Frataxin change of free energy of folding challenge. They have been amongst top performers in a few other challenges, sometimes in combination with sequence feature methods, for example, cancer drivers in CDKN2A and rescue mutations in TP53.[15] Generally, however, these methods, along with the structure-based machine learning methods, have not been as successful as expected compared to the methods that are primarily sequence-based. Three factors may improve their performance in future. First, better combination with the sequence methods will likely mitigate the problem of false negatives; that is, pathogenic variants that are not stability related. Second, until recently, use of structure has been restricted by low experimental structural coverage of the human proteome (only about 20% of residues). Because of recent dramatic improvements in the accuracy of computed protein structures,[39] variants in essentially all residues in ordered structural regions are now amenable to this approach. Third, better estimation of the effect of missense mutations on stability[40] should improve accuracy. An advantage of biophysical and related methods is that they can sometime provide greater insight into underlying molecular mechanisms (Supplementary Figure 13).

Domain-level information had the potential to deliver improved performance in other instances, such as CBS, where the heme-binding domain present in humans was absent from the yeast ortholog, and RAD50, for which assessment showed that restricting predictions of deleteriousness to the specific domain involved in DNA repair would have substantially improved the accuracy of several methods.

**Computational methods can substantially enhance clinical interpretation of missense variants.** The most direct test of the clinical usefulness of computational methods is to assess their ability to correctly assign pathogenic or benign status for clinically relevant variants. CAGI challenges have addressed this for rare disease variant annotations and for germline variants related to cancer risk.

***Results for predicting the biochemical effects of missense mutations inform clinical applications.*** For some biochemical challenges, it is possible to relate the results to clinical utility of the methods. For NAGLU, some rare variants in the gene cause recessive Sanfilippo B disease. Disease strongly correlates with variants conferring less than 15% enzyme activity,[16] allowing variants in the study to be classified as pathogenic or benign on that basis (purple and yellow circles in Figure 2A). Figure 2A shows that 12 out of the 15 variants with less than 15% predicted activity using the selected method also have less than 15% experimental activity, suggesting high positive predictive value and clinical usefulness for assigning pathogenicity. On the other hand, 28 of the 40 variants with measured activity below 15% are predicted to have higher activity so there are also false negatives. For PTEN, information on the relationship to disease is less well established, but data fall into low and high abundance distributions,[14] and the assessor suggested a pathogenicity threshold at the distribution intersection.

Performance in correctly classifying variants as pathogenic is often represented by ROC curves (Figure 2C), showing the tradeoff between true positive (y-axis) and false positive (x-axis) rates as a function of the threshold used to discretize the phenotype value returned by a prediction method, and summarized by the area under that curve (AUC). The selected methods return AUCs greater than 0.8 for both challenges. Over all reanalyzed biochemical effect challenges, the top AUC ranges from 0.68 to 1.0, with an average of $\overline{\text{AUC}} = 0.83$, and with high statistical significance over a random model (AUC = 0.5). The PolyPhen-2 baseline has $\overline{\text{AUC}} = 0.74$; see Supplementary Table 3. However, all models fell well short of the empirical limit ($\overline{\text{AUC}} = 0.98$) estimated from variability in experimental outputs. Because the experimental uncertainties are based on technical replicates, the experimental AUCs are likely overestimated, so it is difficult to judge how much further improvement might be possible. The full ROC curve areas provide a useful metric to measure the ability of the methods to separate pathogenic from other variants. In a clinical setting though, the left portion of the curve is often the most relevant; that is, the fraction of pathogenic variants identified without incurring too high a level of false positives, where the level of tolerated false positives is application dependent. Figure 2D uses truncated ROC curves to show the performance in this region, with the selected methods' AUCs reaching 0.55 for NAGLU and 0.47 for PTEN. The smaller value for the PTEN truncated ROC curve AUC reflects the higher fraction of false positives at the left of the PTEN scatter plot, particularly those variants predicted to have near-zero protein abundance but with high observed values.

For use in the clinic, the quantity of most interest is the probability that a variant is diagnostic of disease (i.e., can be considered pathogenic), given the available evidence. In addition to the information provided by a computational method, initial evidence is also provided by knowledge of how likely any variant in a particular gene is to be diagnostic of the disease of interest.[41] For example, for NAGLU, about 25% of the rare missense variants in ExAC were found to have less than 15% enzyme activity,[42] suggesting that there is an approximately 25% prior probability that any rare missense variant found in the gene will be pathogenic (a prior odds of pathogenicity of 1:3). To obtain the desired posterior probability of pathogenicity, which is also the local positive predictive value at the score $s$ returned by a method, one can use a standard Bayesian odds formulation[43]

$$\text{posterior odds of pathogenicity} = \text{lr}^{+} \times \text{prior odds of pathogenicity},$$

where the local positive likelihood ratio, $\text{lr}^{+}$, is the slope of the ROC curve at the score value $s$; see Methods for a formal discussion.

Figure 2E shows $\text{lr}^{+}$ and the posterior probability of pathogenicity for NAGLU and PTEN. For NAGLU, at low predicted enzyme activities the positive likelihood ratio rises sharply to about 15. The corresponding posterior probability is 0.8. For PTEN, $\text{lr}^{+}$ reaches a value of about 6. Using a pathogenicity prior of 0.21 wildtype protein abundance (Supplementary Materials) the corresponding posterior probability of pathogenicity is 0.6. ACMG/AMP sequence variant interpretation guidelines recommend a probability of pathogenicity of ≥0.99 to label a variant "pathogenic" and ≥0.90 to label one "likely pathogenic", the thresholds for clinical action.[26, 44, 45] So for these and other biochemical challenges, the computational evidence alone is not sufficiently strong to classify the variants other than as Variant of Uncertain Significance.

However, the clinical guidelines integrate multiple lines of evidence to contribute to meeting an overall probability of pathogenicity threshold, so that it is not necessary (or indeed possible) for computational methods alone to provide a pathogenicity assignment. The guidelines provide rules that classify each type of evidence as Very Strong, Strong, Moderate, and Supporting.[26] For example, a null mutation in a gene where other such loss-of-function mutations are known to cause disease is considered Very Strong evidence, while at the other extreme, a computational assignment of pathogenicity for a missense mutation is currently considered only Supporting. Although these guidelines were originally defined in terms of evidence types, Tavtigian et al.[45] have shown that the rules can be approximated using a Bayesian framework, with each threshold corresponding to reaching a specific positive likelihood ratio; e.g., $\text{lr}^{+} = 2.08$ for Supporting evidence when the prior probability is 0.1 (Methods). The resulting thresholds for each level of evidence are shown below the scatter plots in Figure 2A and in the posterior probability of pathogenicity plots in Figure 2E. For NAGLU, for the selected method, predicted enzyme activities lower than 0.11 correspond to Strong evidence, below 0.17 to Moderate, and below 0.42 to Supporting. These thresholds correspond to approximately 31% of rare variants in this gene providing Supporting evidence, 12% Moderate, and 6% Strong. The top performing methods for the ten biochemical missense challenges all reach Supporting, and sometimes Moderate and Strong evidential support (Supplementary Figures 1-6, Supplementary Tables 2 and 3). These results are encouraging in that they suggest this framework can supply a means of quantitatively evaluating the clinical relevance of computational predictions and that under appropriate circumstances, computational evidence can be given more weight than at present. The next section explores these properties further.

***Identifying rare disease variants.*** The ClinVar[9] and HGMD[46] databases provide an extensive source of rare disease-associated variants against which to test computational methods. A limitation is that most methods have used some or all of these data in training, making it difficult to perform unbiased assessments. The prospective "Annotate All Missense" challenge assessed the accuracy of those predictions on all missense variants that were annotated as pathogenic or benign in ClinVar and HGMD after May 2018 when predictions were recorded, through December 2020, so avoiding training contamination. All predictions directly submitted for the challenge as well as all precomputed predictions deposited in the dbNSFP database[47] before May 2018 (dbNSFP v3.5) were evaluated, predictions from a total of 26 groups.

All selected methods, including PolyPhen-2, achieved high AUCs, ranging from 0.85 to 0.92 for separating "pathogenic" from "benign" variants, and only slightly lower values (maximum AUC 0.88) when "likely pathogenic" and "likely benign" are included (Figure 3A). The two metapredictors, REVEL[48] and Meta-LR,[49] tools that incorporate predictions from multiple other methods, perform slightly better than primary methods, although VEST3 and VEST4[33] outperformed Meta-LR. There is a substantial improvement over the performance of PolyPhen-2, especially in the left part of the ROC curve (Figure 3B), though as with the biochemical effect challenges, some of that may be due to the availability of larger and more reliable training sets. Supplementary Table 4 shows slightly higher performance on ClinVar pathogenic variants than HGMD; however, these resources use different criteria for assigning pathogenicity.

The lower panels in Figure 3 show positive likelihood ratios and posterior probability of pathogenicity for a selected metapredictor, REVEL.[48] With a prior probability of pathogenicity of 0.1 (approximating the prior when examining possible pathogenic variants in one or a few genes in a diagnostic setting; see Methods) 14, 10, and 4% of variants reach the Supporting, Moderate, and Strong evidence thresholds, respectively. With the much smaller prior of 0.01, representative of screening for possible secondary variants, about 6% of variants will provide Supporting evidence and 2% reach Moderate. These estimates are not exact since there may be significant differences between the distribution and properties of variants in these databases and those encountered in the clinic. For example, some genes have only benign variant assignments in the databases, and these might be excluded from consideration in the clinic. Performing the analysis on only genes with both pathogenic and benign assignments slightly reduced performance—highest AUC on the confident set of variants is 0.89 instead of 0.92. The selected methods also change slightly; see Supplementary Figure 9. In spite of this and other possible caveats, the overall performance of the computational methods is encouraging and, as with biochemical effect challenges, suggests that the computational methods can provide greater benefit in the clinic than recognized by the current standards.

***Identifying germline cancer risk variants.*** About a quarter of CAGI experiments have involved genes implicated in cancer (Figure 1), and have included variants in BRCA1, BRAC2, PTEN, TPMT, NSMCE2 (coding for SUMO-ligase), CHEK2, the MRN complex (RAD50, MRE11, and NBS1), FXN, NPM-ALK, CDKN2A and TP53. An additional challenge addressed breast cancer pharmacogenomics. From a cancer perspective, the most informative of these is a challenge provided and assessed by members of the ENIGMA consortium,[21] using a total of 321 germline BRCA1/BRCA2 missense and in-frame indel variants. Performance on this challenge was impressively high, with four groups providing submissions that gave AUCs greater than 0.9 and evidence (red threshold). The estimated % of variants encountered in a clinical setting expected to meet each threshold are also shown. For example, about 14% of variants provide Supporting evidence. Dotted lines show results obtained with a prior probability of 0.01.

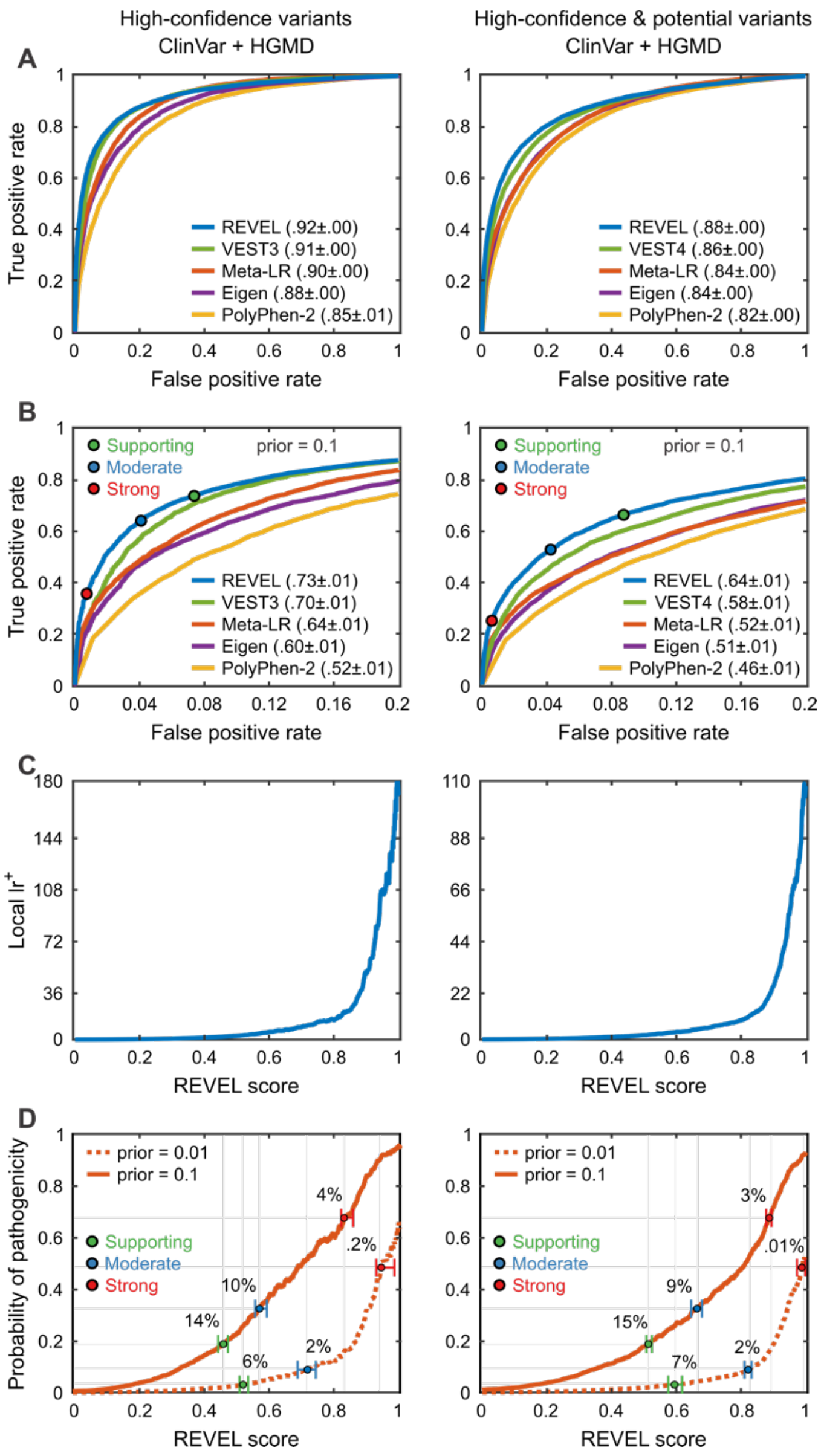

***Figure 3:*** *Performance of computational methods in correctly identifying pathogenic variants in the two principal rare disease variant databases, HGMD and ClinVar.* The left panels show data for variants labeled as "pathogenetic" in ClinVar and "DM" in HGMD together with "benign" in ClinVar. The right panels add variants labelled as "likely pathogenic" and "likely benign" in ClinVar as well as "DM?" in HGMD. Meta and single method examples were selected on the basis of the average ranking of each method for the ROC and truncated ROC AUCs. See Supplementary Materials for more details and selection criteria. (**A**) ROC curves for the selected metapredictors and single methods, together with a baseline provided by PolyPhen-2. Particularly for pathogenic variants alone, impressively high ROC areas are obtained, above 0.9, and there is a substantial improvement over the older method's performance. (**B**) Blowup of the left-hand portion of the ROC curves, most relevant to high confident identification of pathogenic variants. Clinical thresholds for Supporting, Moderate and Strong clinical evidence are shown. (**C**) Local positive likelihood ratio as a function of the confidence score returned by REVEL. Very high values (>100) are obtained for the most confident pathogenic assignments. (**D**) Local posterior probability of pathogenicity; that is, probability that a variant is pathogenic as a function of the REVEL score for the two prior probability scenarios. For a prior probability of 0.1, typical of a single candidate gene situation (solid line) and database pathogenic and benign variants (left panel) the highest scoring variants reach posterior probability above 0.9, strong enough evidence for a clinical assignment of "likely pathogenic". In both panels, variants with a score greater than 0.45 provide Supporting clinical evidence (green threshold), and scores greater than 0.8 provide Strong

two with AUCs exceeding 0.97. In the other BRCA1/BRCA2 variant challenge, the highest AUC is 0.88 on a total of 10 missense variants. The strong results may reflect the fact these are highly studied genes. More and larger scale challenges with a variety of genes are required in order to draw firm conclusions. Further details of cancer challenges are provided in Supplementary Figure 10 and Supplementary Table 5.

**Assessing methods that estimate the effect of variants on expression and splicing is difficult, but results show these can contribute to variant interpretation.** Variants that regulate the abundance and isoforms of mRNA either through altered splicing or through altered rates of transcription play a significant role in disease, particularly complex traits. CAGI has included four challenges using data from high-throughput assays of artificial gene constructs, two for splicing and two for expression. For all four, evaluation of the results is limited by a combination of small effect sizes (changes larger than two-fold in splicing or expression are rare in these challenges) and experimental uncertainty, but some interesting properties can be identified.

***Splicing.*** The CAGI splicing challenges used data from high-throughput minigene reporter assays.[50]

The MaPSy challenge asked participants to identify which of a set of 797 exonic single-nucleotide HGMD disease variants affect splicing and by how much. Two experimental assays were available, one *in vitro* on a cell extract and the other by transfection into a cell line. Only variants that produced a statistically robust change of at least 1.5-fold were considered splicing changes. Figure 4 summarizes the results. The top-performing groups achieved moderately high AUCs of 0.84 and 0.79 and the highest $lr^+$ is about 6. Notably, very few variants qualify as significant splicing changes, and there are inconsistences between the two assays, with a number of variants appearing to have a fold change substantially greater than 1.5 in one assay but not the other. Additionally, the experimental noise significantly overlaps with many splicing differences. For these reasons, it is unclear what maximum AUC could be achieved by a perfect method.

The Vex-Seq challenge required participants to predict the extent of splicing change introduced by 1,098 variants in the vicinity of known alternatively spliced exons.[50] Supplementary Figure 11 shows that performance was rather weak for identification of variants that increase splicing (top AUC = 0.71), but that may be because many points classified as positive are experimental noise. There are more variants that show a statistically robust decrease in splicing, and prediction performance is correspondingly stronger (top AUC = 0.78). The assessor noted additional nuances in performance.[50]

The selected high-performing method for both challenges (MMSplice[51]) decomposes the sequence surrounding alternatively spliced exons into distinct regions and evaluates each region using separate neural networks.[51] More detailed splicing results are provided in Supplementary Table 6.

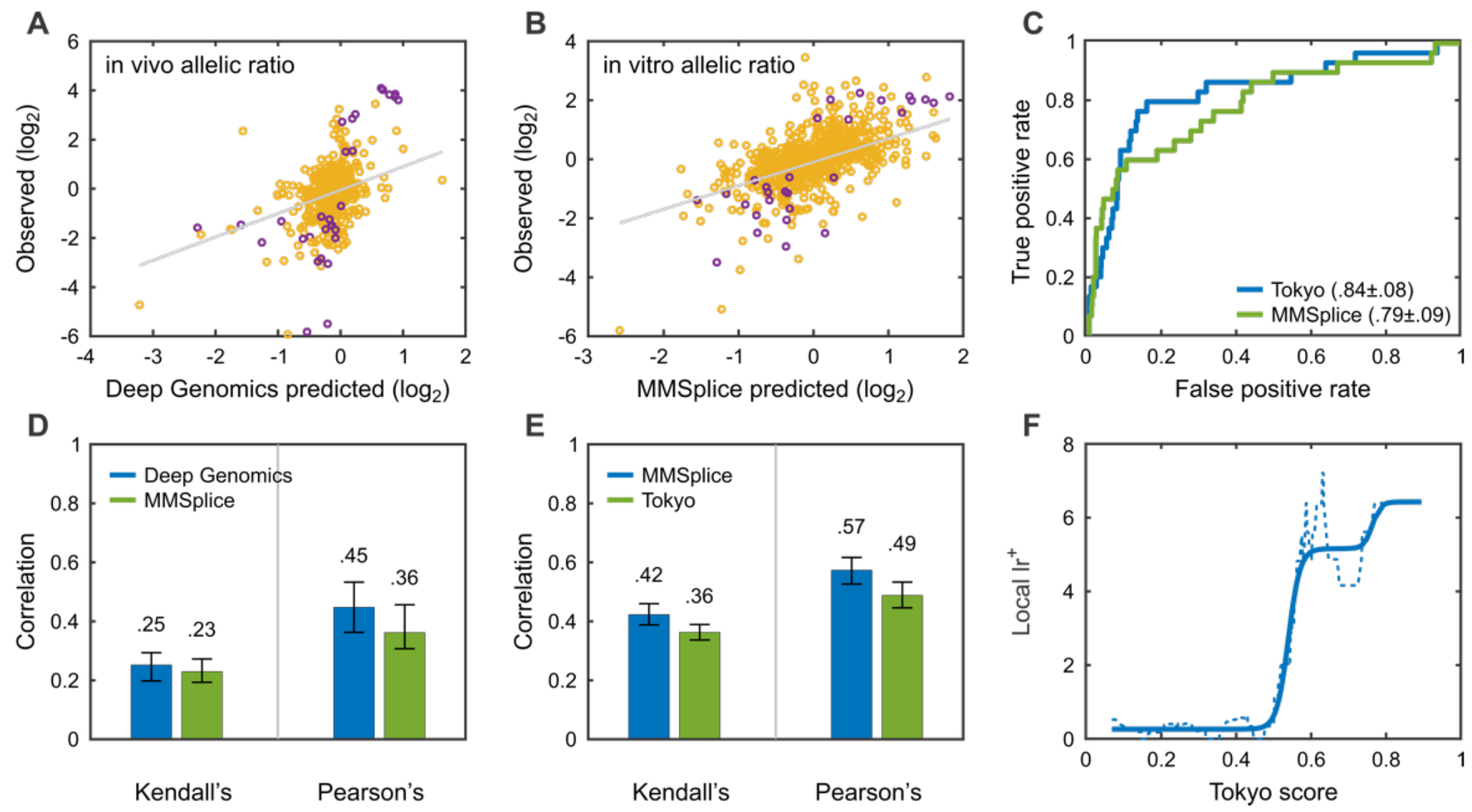

***Figure 4:*** *Performance of computational methods in identifying variants that affect splicing in the MaPSy challenge.* Methods were selected based on the average ranking over three metrics: Pearson's correlation, Kendall's tau, and ROC AUC. Scatter plots, Kendell's tau, and Pearson's correlation results are shown for *in vivo* (**A**, **D**) and *in vitro* assays (**B**, **E**) separately. The small number of purple points in the scatter plots represent splicing fold changes greater than 1.5-fold. The ROC curve (**C**) shows performance in variant classification for the two selected methods. The maximum local positive likelihood ratio ($lr^+$, **F**) may be large enough for use as auxiliary information, see Discussion (solid line is smoothed fit to the data).

***Transcription.*** Several CAGI challenges have assessed the ability of computational methods to identify single base changes that affect the expression level of specific genes. The CAGI4 eQTL challenge assessed whether methods could find causative variants in a set of eQTL loci, using a massively parallel reporter assay.[52] Because of linkage disequilibrium and sparse sampling, variants associated with an expression difference in an eQTL screen are usually not those directly causing the observed expression change. Rather, a nearby variant will be. The challenge had two parts. Participants were asked to predict whether insertion of the section of genomic DNA around each variant position into the experimental construct produced any expression. Supplementary Figure 12A shows that the top performing methods were effective at this—the largest AUC is 0.81. The second part of the challenge required participants to predict which variants affect expression levels. Here the results are much less impressive (Supplementary Figure 12B). The scatter plot shows a weak relationship between observed and predicted expression change and Pearson's or Kendall's correlation are also small. The best AUC is only 0.66 and the maximum $lr^+$ is about 5. Most of the experimental expression changes are small (less than 2-fold), and may be largely experimental noise, partly accounting for the apparent poor performance. But as the scatter plot shows, a subset of the variants with largest effects could not be identified by the top performing method. A combination of experimental and computational factors contributed to poor performance, and more challenges of this sort are needed.

The CAGI5 regulation saturation mutagenesis challenge examined the impact on expression of every possible base pair variant within five disease-associated human enhancers and nine disease-associated promoters.[53] As shown in Figure 5, performance is stronger in promoters than enhancers and stronger for decreases in expression compared with increases. Fewer variants show experimental increases and these tend to be less well distinguished from noise. Performance for small expression changes is hard to evaluate because of overlap with experimental noise. Nevertheless, the highest AUC for promoter impact prediction is 0.81 while the highest AUC for enhancer impact prediction is 0.79, relatively respectable values. In addition, the scatter plots show that large decreases in expression are well predicted, suggesting the methods are quite informative for the most significant effects.

The CAGI splicing and expression challenges are not as directly mappable to disease and clinical relevance as in some other challenge areas. Variants have been a mixture of common and rare and the use of artificial constructs in high-throughput experiments limits relevance of challenge performance in the whole-genome context. Nevertheless, the results do have potential applications. In complex trait disease genome-wide association studies (GWAS), the variants found to be associated with a phenotype are usually not those causing the effect. Identifying the functional variants is not straightforward, and current regulatory prediction methods can provide hypotheses as to possible effects on expression or splicing.

**CAGI participants identified diagnostic variants that were not found by clinical laboratories.** A major goal of CAGI is to test the performance of computational methods under as close to clinical conditions as possible. In the area of rare disease diagnosis, four challenges have addressed this by requiring participants to identify diagnostic variants in sets of clinical data. The Johns Hopkins and intellectual disability (ID) challenges employed diagnostic panels, covering a limited set of candidate genes in particular disease areas. As compared with genome-wide data, diagnostic panels inherently restrict the search to only variants belonging to a known set of relevant genes.

For a number of genetically undiagnosed cases in the Johns Hopkins panel, CAGI participants found high-confidence deleterious variants in genes associated with a different disease from that reported, suggesting physicians may have misdiagnosed the symptoms.[54] However, because of the clinical operating procedures of the diagnostic laboratory, it has not been possible to further investigate these cases. In the ID panel, some plausible calls were made on novel variants that had not been reported to the patient partly because the majority of standard computational methods returned assignments of “benign”.[55]

The two SickKids challenges (SickKids4 and SickKids5) were based on whole-genome sequence data for children with rare diseases from the Hospital for Sick Children in Toronto. These are all cases that were undiagnosed by the state-of-the-art SickKids pipeline,[20] and so were particularly challenging compared with those normally encountered in the clinic. In the SickKids4 challenge, variants proposed by challenge participants were deemed diagnostic by the referring physicians for two of the cases in part due to matching detailed phenotypes. This was the first instance of the CAGI community directly contributing in the clinic. In SickKids5, two of the highest confidence nominated diagnostic variants provided correct genome-patient matches. While not meeting ACMG/AMP criteria for pathogenicity[26] these were considered interesting candidates for further investigation, again potentially resolving previously intractable cases.

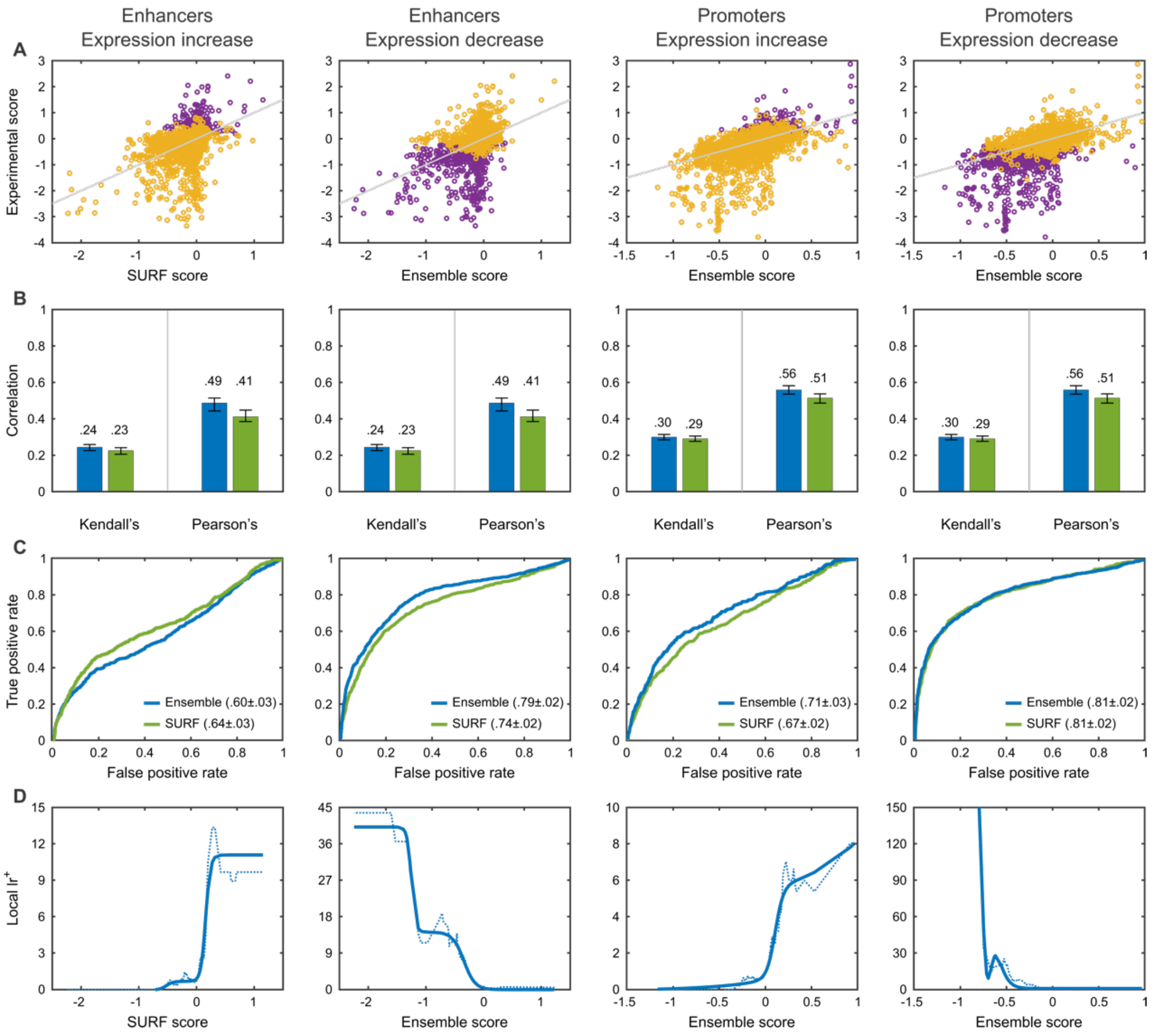

***Figure 5:*** *Performance on the Regulation Saturation expression challenge.* The two left columns show performance in predicting increased (left) and decreased (right) expression in a set of enhancers (purple points represent variants that significantly change expression). The right pair of columns show equivalent results for promoters. The scatter plots (**A**) show strong performance in identifying decreases in expression (purple points), but weaker results for expression increases. Performance on promoters is stronger than on enhancers. Overlap of changed and non-changed experimental expression points suggests that experimental uncertainty reduces the apparent performance of the computational methods. Panel (**B**) shows correlation coefficients for selected methods. Panel (**C**) shows ROC curves for predicting under and overexpression. Panel (**D**) shows local $lr^+$, where the solid lines are smoothed fits to the data.

These clinical challenges required participants to develop full analysis pipelines, including quality assessment for variant calls, proper inclusion of known pathogenetic variants from databases such as HGMD[46] and ClinVar,[9] and an evaluation scheme for weighing the evidence. The SickKids challenges also required compilation of a set of candidate genes or some equivalent approach. Varying success in addressing these factors will have influenced the results, so it is not possible to effectively compare the core computational methods. Overall, current approaches have limitations in this setting—they tend to ignore or fail to reliably evaluate synonymous and noncoding variants; if the relevant gene is not known its variants will usually not be examined; and data for epigenetic causes are not available. Nevertheless, the CAGI results for these challenges again make it clear that current state-of-the-art computational approaches can make valuable contributions in real clinical settings.

**Complex trait interpretation is often complicated by confounders in the data.** Many common human diseases, such as Alzheimer's disease, asthma, and type II diabetes, are complex traits and as with monogenic disorders, genetic information should in principle be useful for both diagnosis and prognosis. Individual response to drugs (pharmacogenomics) also often has a complex trait component. As opposed to monogenic disease, where one or two genetic variants are judged responsible for a phenotype, and cancer, where a small number (4-5 on average) of driver variants are key in each individual,[6] complex traits have relatively small contributions from many variants. Environmental factors usually also play a substantial role so that methods using genome information alone have limited maximum accuracy. Further complicating interpretation, variants have relatively small effects on gene expression, splicing, and molecular function. Also, most CAGI complex trait challenges have been based on exome data, whereas at many GWAS risk loci lie outside coding regions.[56] To some extent, the status of relevant common variants not present in the exome data can be imputed on the basis of linkage disequilibrium, but this places an unclear limit on achievable accuracy. Limited or no availability of training data also restricted method performance. Phenotypes tend to be less precise than for other types of disease; for example, phenotypes underlying a Crohn's disease diagnosis are quite varied. Altogether, these factors make this a difficult CAGI area. Nevertheless, these challenges have been informative and have drawn new investigators into the rapidly developing area of Polygenic Risk Score (PRS) estimation.[57] One challenge, CAGI4 Crohn's, has yielded apparently robust conclusions on the performance of methods in this area.

***Crohn's disease (CAGI4).*** Participants were provided with exome data for 111 individuals and asked to identify the 64 who had been diagnosed with Crohn's disease. A variety of computational approaches were used, including clustering by genotypes, analysis of variants in pathways related to the disease, and evaluation of SNPs in known disease-associated loci. The highest-scoring method (AUC  0.72; Figure 6A) used the latter approach together with conventional machine learning, and trained on data from an earlier GWAS.[58] Figure 6B shows case and control score distributions for that method. A perfect method would have no distribution overlap. These results are far from that, but there is clear signal at the extremes, and as Figure 6C shows, that translates into a positive likelihood ratio with an approximately twenty-fold range (0.3 to 6), only a little lower than that obtained for the biochemical effect and clinical missense challenges. With a prior probability of disease of 1.3%,[59] relative risk (see Methods) also has a range of about twenty-fold (Figure 6D), with the highest-risk individuals having six-fold higher risk compared to that estimated for the population average. For some complex trait diseases, for example coronary heart disease,[60] this is discriminatory enough to support clinical action, and for many diseases would provide a valuable additional factor to more standard risk measures such as age and sex. Newer PRS methods, which aim to incorporate many weak contributions from SNPs, were not evaluated in this CAGI challenge.

Other complex trait CAGI challenges (Supplementary Materials) revealed batch effects (CAGI2 Crohn's and Bipolar Disorder) and population structure effects (CAGI3 Crohn's) in the data, leaked clinical data (Warfarin and VET challenges), or discrepancies between self-reported traits and those predicted from genetic data (PGP challenges), thereby complicating assessment.

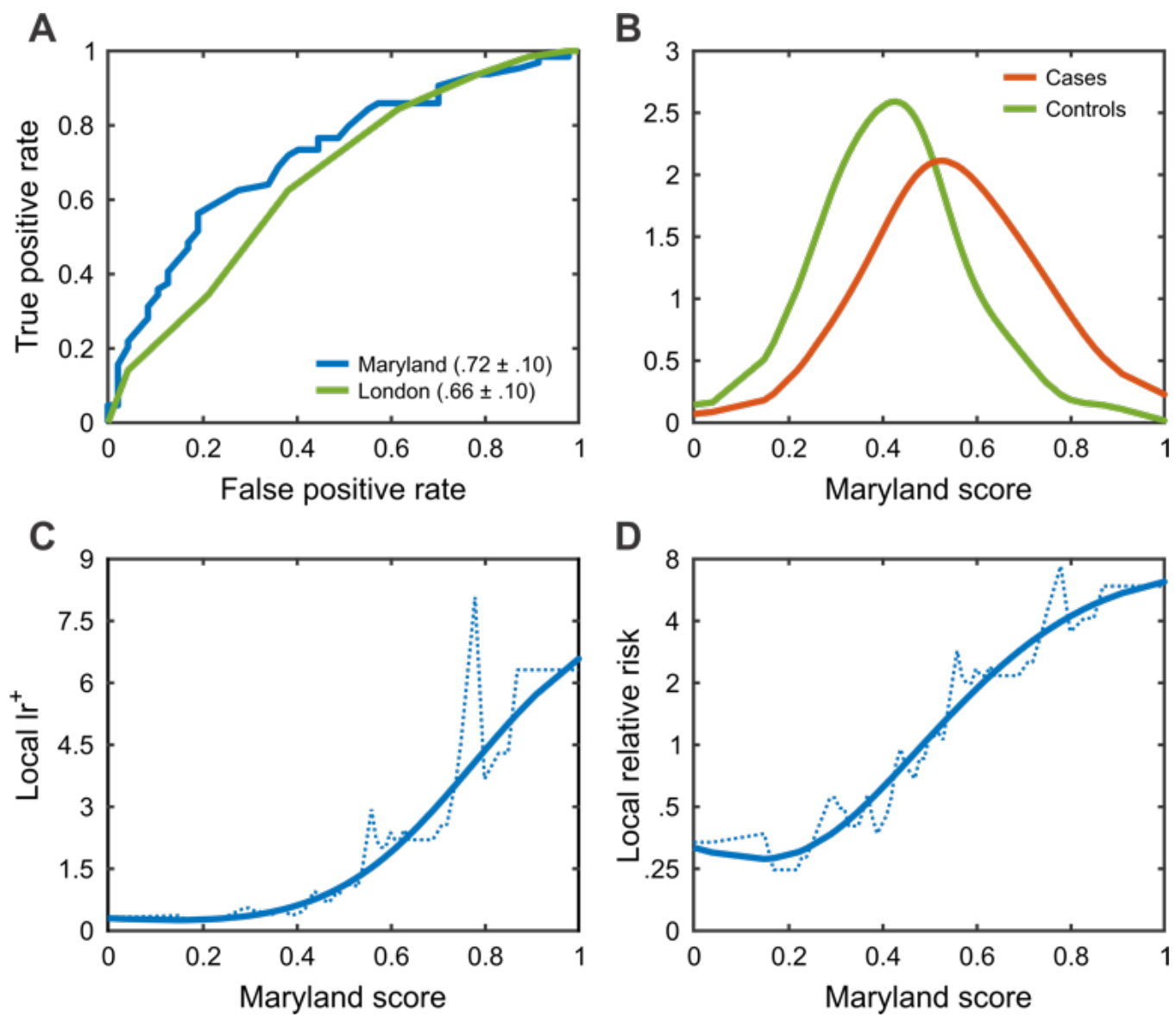

***Figure 6:** Identifying which of a set of individuals are most at risk for Crohn's disease, given exome data.* Examples were selected on the basis of ranking by ROC AUC. (**A**) ROC curves for two selected methods. Statistically significant but relatively low ROC areas are obtained. (**B**) Distributions of disease prediction scores for individuals with the disease (red) and without (green) for the method with the highest AUC (kernel density representation of the data). (**C**) Local positive likelihood ratio ($lr^+$) as a function of prediction score for the method with the highest AUC. (**D**) Relative risk of disease ($log_2$ scale), compared to that in the general population) as a function of prediction score. Individuals with the lowest risk scores have approximately 1/3 the average population risk, while those with the highest scores have risk exceeding six-fold the average, a twenty-fold total range. Depending on the disease, identifying individuals with higher than three-fold the average risk may be sufficient for clinical action.

Performance in matching genomes and disease phenotypes, an additional component in some challenges, was poor. Additional complex trait challenge results are provided in Supplementary Table 7.

**The CAGI Ethics Forum has guided responsible data governance.** Data used in CAGI challenges are diverse in terms of sensitivity (e.g., with respect to participant reidentification risk, potential for stigmatization, potential impact of pre-publication data disclosure), collected under a broad variety of participant consent understandings and protection frameworks, and analyzed by predictors with varying degrees of familiarity with local and international biomedical regulations. This heterogeneity calls for a nuanced approach to data access and the tailored vetting of CAGI experiments. The CAGI Ethics Forum was launched in 2015 to proactively address these concerns. Incorporating input from bioethicists, researchers, clinicians, and patient advocates, it has developed policies for responsible data governance (e.g., assisting in revision of the general CAGI data use agreement, to safeguard human data and also protect all CAGI participants, including data providers for unpublished data), cautioned against overinterpretation of findings (e.g., highlighting the contribution of social and environmental risk factors to disease, and the potential negative consequences, such as stigma, of associating particular disease variants with a specific population), and provided input on a variety of guidelines and procedures, including CAGI's participant vetting process (e.g., how to identify a *bona fide* researcher) and a system of tiered access conditions for datasets, depending on their sensitivity. Future directions include investigating the scalability of current user validation and data access models, exploring implications for family members of unexpected challenge findings, discussing policies to ensure proper credit attribution for constituent primary methods used by metapredictors, and identifying additional means of ensuring accountability options with respect to responsible data sharing.

## DISCUSSION

**Principal findings.** Over CAGI's first decade, five rounds of CAGI challenges have provided a picture of the current state of the art in interpreting the impact of genetic variants on a range of phenotypes and provided a basis for the development of improved methods as well as for more calibrated use in clinical settings.

A key finding is that for most missense challenges it is possible to relate phenotype values to a pathogenicity threshold, and so deduce potential performance in a clinical setting, particularly for rare Mendelian diseases. The results suggest that computational methods are generally more reliable than recognized in the current clinical use guidelines.[26] The Annotate All Missense challenge most directly probed clinical utility, assessing pre-deposited computational annotations of pathogenicity against subsequent clinical annotation. Encouragingly, the results show that of recently classified variants, up to 15% can have computational annotations that are sufficiently reliable to be considered Supporting evidence under the present guidelines,[26] with some reaching Moderate and Strong thresholds of reliability (Figure 3). Together with the recent results from the ClinGen's Sequence Variant Interpretation group,[43] these results suggest that computational methods can play a beneficial larger role in clinical variant interpretation.

The plurality of challenges has required prediction of the impact of missense variants on biochemical properties as varied as protein stability, enzyme activity or cell growth rate. Although most methods were not trained for this type of use, there are some informative overall results albeit with limited accuracy for individual variants (Figure 2, Supplementary Tables 2 and 3). Results are also modestly but consistently better than for an older method trained on smaller datasets. Some challenges are still beyond the current state of the art. Examples are the liver pyruvate kinase challenge, which required prediction of allosteric regulation effects,[61] NPM-ALK for prediction of protein binding changes, and the asthma challenge where comparison of RNASeq data between twins could not be used to assign disease status.

Several challenges have directly assessed the usefulness of missense impact and other methods under clinical conditions. Three have been particularly informative. The two SickKids challenges required participants to identify diagnostic variants using whole-genome sequence data for a set of pediatric rare disease cases where a state-of-the-art pipeline was not successful. For two of these difficult cases, CAGI participants were able to provide interpretations accepted by physicians, and some others may also be correct. The Johns Hopkins pipeline challenge required participants to match gene panel data to a set of heart and lung disease classes that individuals had been diagnosed with. Despite the handicap of not knowing the phenotype/genomic relationships, participants were apparently able to find diagnostic variants for a number of cases where the conventional pipeline had not succeeded, and to suggest that in some cases the initial diagnosis may be incorrect. Not all diagnostic variants in these three challenges are missense and there was limited success with other types of variation. The challenges also highlighted the fact that successful application in the clinic requires integration of the computational methods into a more comprehensive pipeline, creating a barrier to participation for some research groups.

Computational methods for identifying variants that introduce small splicing or gene expression effects have so far been less well represented in CAGI, and issues with data availability and appropriateness have limited insights. However, the results do show that these methods have potential for providing auxiliary evidence for assigning pathogenicity and may also help interpret GWAS results.

Missense and splicing variants also play major roles in cancer, and the challenges have provided insight into potential clinical use there. In addition, CAGI has included a set of challenges probing the ability of computational methods to identify germline variants that predispose to cancer. Results for the two BRCA gene challenges suggest that computational methods are sufficiently accurate to contribute to clinical assessment of variants, but more data are needed.

Up to now, complex traits have proven a difficult area in which to obtain large and robust datasets to test the computational methods in CAGI, partly because of concerns about data privacy (most datasets are subject to strict permission rules) and partly because of limited reliability for early datasets. Nevertheless, the challenges have provoked experimentation with a range of interesting computational approaches.[23] For the one Crohn's disease challenge with clear outcomes, the best methods were able to identify some high-risk individuals (Figure 6). Future challenges may also include other types of omics data.

Most current computational methods for scoring single variants are general; i.e., they are typically applied to any variant of the same type in any gene without adjustment, and are primarily based on sequence conservation, with limited inclusion of functional and structural properties. CAGI results show there is potential for more bespoke approaches together with contributions from functional and structural analysis. Protein structure properties, particularly effects on stability, have helped in several challenges (CDKN2A and TP53), and it appears that knowledge of domain functional properties would have helped in two more (CBS and RAD50). Similarly, although of limited accuracy, methods that identify variants with small splicing and expression effects have potential to strengthen pathogenicity predictions. These and other structural and functional properties provide information that is largely independent of sequence conservation, so there is scope for combination with other properties into more comprehensive models. It is likely that future methods will be more gene-specific and will combine multiple information sources. In general, methods integrating predictions from multiple separate methods in a calibrated manner tend to perform better than single methods. Strong inter-method calibration, usually achieved by machine learning, is key here and simple consensus approaches should be avoided, unless specifically evaluated first.[43]

Participants have sometimes used unanticipated information to improve performance on challenges. For example, in a PGP challenge requiring matching of full genome sequences to extensive phenotype profiles, a participant made use of information in the PGP project blog.[62] Though these sometimes subvert the intended challenge, in some ways this reflects what happens in a clinical setting—all relevant information is used, however obscure.

**Critical assessment for genome interpretation methods.** CAGI relies on the same factors as other critical assessment community experiments: a willingness of the relevant research groups to participate, clearly defined metrics for success, the availability of large enough and accurate enough sets of experimental data to provide a gold standard, and independent objective assessment of the results. Participation in CAGI has been strong in most areas, particularly for missense variants, and a vibrant and interactive community has developed. New researchers have been attracted to the field and new collaborations have resulted in the development of creative algorithms with broad applicability.[63-65]

The regulatory variants research community is smaller but has participated, with encouraging results.[53] Complex trait interpretation methods, particularly PRS,[57] are rapidly growing, so we expect more participation in future, with challenges perhaps including other types of omics data. For clinical challenges such as SickKids and the Johns Hopkins panel, CAGI has not yet succeeded in attracting much participation from professionals in that area. CAGI has been fortunate in its assessors, with many excellent analyses.[12, 13]

CAGI has not committed to any evaluation metric as a primary performance indicator. Instead, methods have been characterized using descriptive and useful metrics for both real-valued and discretized predictions, spanning the interests of the CAGI community and its intended audience. Open questions remain regarding a set of metrics that can give more complete insight into the prediction performance, and new metrics, such as balanced precision,[66] log ROC curves[67] and the LC index[68] are under consideration (Supplementary Figures 1, 2, 3, 4, and 7 provide log ROCs).

The biggest obstacle to clear assessment has been and continues to be data diversity and quality, a key difference between CAGI and related community endeavors. Other initiatives, such as CASP,[69] deal primarily with one type of data (protein structure) and the data are usually of very high quality and directly relevant to the goals of the computational methods. By contrast, CAGI deals with many different settings, including studies of biochemical effects with a broad range of phenotypes, the pathogenicity of variants both germline and somatic, clinical phenotypes, and statistical relationships. Also, while genome variant calling is reliable, it does have limits.[70] For example, in the SickKids challenges, some variants suggested as diagnostic by CAGI participants had been found to be incorrect calls and so eliminated in the clinical pipeline, using sequence validation data CAGI participants did not have access to. Adapting available data to form suitable challenges is difficult, and compromises are sometimes needed to devise a challenge where the results can be objectively assessed. For example, in one of the SickKids and in the Johns Hopkins clinical challenges, assessment hinged on requiring participants to match genomes to phenotypes. But that makes it much harder to identify diagnostic variants than in the real-life situation. Conversely, challenge providers have sometimes benefited from the detailed scrutiny of their data by CAGI participants prior to its publication. In some cases, interpretation of clinical challenge results is also complicated by there being no conclusive diagnoses.

For biochemical effect experiments there has been a trend towards high-throughput methods producing data for 1000s of variants.[14] These surveys of the variant/phenotype landscape are providing major insights.[71] But there are usually biases and compromises in the experimental design, as well as limited evaluation of uncertainty and error that make them less suitable for CAGI. For example, in the SUMO-ligase challenge, there is evidence that different interactions between the human protein in the yeast milieu compared to that of human contributed to odd results for some variants.[17] For the Annotate All Missense challenge, very high ROC areas are obtained by some methods. Since ClinVar and HGMD both contain an unknown level of misannotation,[34] it is not clear how much of the remaining disagreement is due to such factors versus imperfect methods. Similar caveats apply to the high-throughput methods used for regulatory variants which usually use artificial gene constructs that may not fully reflect the genome context. For most challenges, experimental error estimates are almost always based on technical replicates, so likely underestimating true uncertainty. Some disagreement between predicted and experimental values may also be due to the complicated nature of the biochemical effects. For example, a variant may alter a post-translational modification that affects protein abundance.[14] In clinical data provided to CAGI, all phenotypes ascribed to a patient appear to have equal weight, while the physician managing a case will have knowledge of which are major versus minor features.

**Potential and prospects.** Results from the first ten years of CAGI have not only highlighted current abilities and limitations but have pointed to the way forward. As already noted, one of the most important implications is the need for greater utilization of computational methods in the clinic.

Experimental biotechnology platforms have become widely available[71] and the genomic data collection in the clinic has greatly increased. While the next generation of computational tools should benefit from these developments, they also pose new challenges. Deeper characterization of experimental approaches is needed to address data uncertainty and biases. Potential circularity between computation-assisted variant annotation and method assessment also needs to be considered. Future rounds of CAGI will address these issues by using assessment methods that mitigate or eliminate problems with data, by the development and promotion of practices and standards for application of methods, by working with experimental groups to provide sufficiently large and high-quality datasets, and by effectively following up on disagreements at the intersection of high-throughput functional experiments, genetic association studies, and outputs of computational prediction methods. CAGI also plans to increase evaluation of the combinatorial effect of variants, either in single-molecule biochemical assays or in clinical applications with whole-genome studies for both rare and complex phenotypes.

The current performance levels for exonic variation in Mendelian disorders, combined with rapidly accumulating data and a recent breakthrough in protein structure prediction,[69] suggest that upcoming methods should more consistently achieve Strong and Very Strong clinical evidence levels. On the other hand, accurately modeling the biology of variant impact on protein function, such as catalysis, binding and allostery, will probably require the development of a different class of methods capable of generating meaningful scores on a continuous scale. Although this will take time, it may be reasonable to expect improved calibration in specific cases in the short term, which could translate to an increased use of variant interpretation methods in biomedical research.

So far CAGI has had few large-scale challenges for regulatory variants, but it is clear that because of many small effect sizes, future challenges will need very high-quality experimental data to enable a clear assessment. Complex trait challenge data quality has also been a limiting factor but given that data quality and availability are improving, future challenges are expected to resolve this issue. A question still to be addressed with both regulatory and complex trait challenges is that the extent to which the computational methods can provide mechanistic insight rather than just statistical association, and future challenges should provide answers.

Genomic data relevant to disease are increasingly complemented by other data types, particularly expression data, proteomics, and metabolomics. CAGI has already included such multiomics challenges (Figure 1) and will do so more in future.

Improvements in overall variant impact prediction and disease gene prioritization will also lead to the development of new tools with increased automation in clinical diagnostics. This will require careful integration of the component tools and pipelining, including improved standardization and interoperability practices. As of now, CAGI has not evaluated the seamlessness of use of these methods and may consider it as these platforms mature.

Finally, owing to complex institutional and national regulatory policies and practices, obstacles to data sharing remain significant. CAGI has been addressing such issues by developing robust data use agreements and will seek to increase closed-environment assessments.[72] Through the activities of its Ethics Forum, it will also seek to create data use agreement standards and practices to that will simplify responsible data sharing while promoting responsible and respectful research uses.

# METHODS

We describe different evaluation scenarios considered in the Critical Assessment of Genome Interpretation (CAGI), motivate the selection of performance measures, and discuss ways to interpret the results.

**Terminology and notation.** Let $(x, y) \in \mathcal{X} \times \mathcal{Y}$ be a realization of an input-output random vector $(X, Y)$. The input space $\mathcal{X}$ may describe variants, gene panels, exomes, or genomes in different CAGI scenarios. Similarly, the output space $\mathcal{Y}$ can describe discrete or continuous targets; e.g., it can be a binary set {pathogenic, benign} when the task is to predict variant pathogenicity, or a continuous set $[0, \infty)$ representing a percent of enzymatic activity of the wildtype protein (a mutated molecule can have increased activity compared to the wildtype) or a cell growth rate relative to that with the wildtype gene.

Let $s: \mathcal{X} \rightarrow \mathbb{R}$ be a predictor that assigns a real-valued score to each input and $f: \mathcal{X} \rightarrow \mathcal{Y}$ be a predictor that maps inputs into elements of the desired output space; i.e., $f$ is real-valued when predicting a continuous output and discrete when predicting a discrete output. When predicting binary outcomes (e.g., $\mathcal{Y} = \{0, 1\}$), $s(x)$ is often a soft prediction score between 0 and 1, whereas $f(x)$ can be obtained by discretizing $s(x)$ based on some decision threshold $\tau$. Scores $s(x)$ can also be discretized based on a set of thresholds $\{\tau_j\}_{j=1}^{m}$, as discussed later. In a binary classification scenario, $f(x) = 1$ (pathogenic prediction) when $s(x) \geq \tau$ and $f(x) = 0$ (benign prediction) otherwise. We shall sometimes denote a discretized binary model as $f_\tau(x)$ to emphasize that the predictor was obtained by thresholding a soft scoring function $s(x)$ at $\tau$. The target variable $Y$ can similarly be obtained by discretizing the continuous space $\mathcal{Y}$ using a set of thresholds $\{\tau'_k\}$, that are different from $\{\tau_j\}$ used for the scoring function. In one such case, discretizing the continuous space $\mathcal{Y}$ of functional impact of an amino acid variant into {damaging, nondamaging} transforms a regression problem into classification, which may provide additional insights during assessment. With a minor abuse of notation, we will refer to both continuous and the derived discrete space as $\mathcal{Y}$. The exact nature of the target variable $Y$ and the output space will be clear from the context.

Finally, let $\mathcal{D} = \{(x_i, y_i)\}_{i=1}^{n}$ be a test set containing $n$ input-output pairs on which the predictors are evaluated. Ideally, this data set is representative of the underlying data distribution and non-overlapping with the training data for each evaluated predictor. Similarly, we assume the quality of the measurement of the ground truth values $\{y_i\}_{i=1}^{n}$ is high enough to ensure reliable evaluation. While we took multiple steps to ensure reliable experiments and blind assessments, it is difficult to guarantee complete enforcement of either of these criteria. For example, an *in vitro* assay may be an imperfect model of an *in vivo* impact or there might be uncertainty in collecting experimental read-outs. Additionally, the notion of a representative test set may be ambiguous and cause difficulties when evaluating a model that was developed with application objectives different from those used to assess its performance in CAGI.

**Evaluation for continuous targets.** Evaluating the prediction of continuous outputs is performed using three primary measures ($R^2$, Pearson's correlation coefficient, and Kendall's tau) and two secondary measures (root mean square error and Spearman's correlation coefficient). $R^2$ is defined as the difference between the variance of the experimental values and the mean-squared error of the predictor, normalized by the variance of the experimental values. It is also referred to as the fraction of variance of the target that is explained by the model. $R^2 \in (-\infty, 1]$ is estimated on the test set $\mathcal{D}$ as

$$R^2 = 1 - \frac{\sum_{i=1}^{n}(f(x_i)-y_i)^2}{\sum_{i=1}^{n}(y_i-\bar{y})^2}, \tag{1}$$

where $\bar{y} = \frac{1}{n}\sum_{i=1}^{n} y_i$ is the mean of the target values in $\mathcal{D}$ (observe that each value $y_i$ may itself be an average over technical or biological replicates, if available in the experimental data). The $R^2$ values above 0 indicate that the predictor is better than the trivial predictor—one that always outputs the mean of the target variable—and values close to 1 are desired. The values below 0 correspond to models with inferior performance to the trivial predictor. Maximizing the $R^2$ metric requires calibration of output scores; that is, a high correlation between predictions and target values as well as the proper scaling of the prediction outputs. For example, a predictor outputting a linear transformation of the target such as $f(X) = 10 \cdot Y$ or a monotonic nonlinear transformation of the target such as $f(X) = \log Y$ may have a high correlation, but a low $R^2$. $R^2$, therefore, can be seen as the strictest metric used in CAGI. However, this metric can adversely impact methods outputting discretized prediction values. Such methods are preferred by some tool developers as they simplify interpretation by clinicians, experimental scientists, or other users.

In some cases, it may be useful to also report the root mean squared error (RMSE), estimated here as

$$\text{RMSE} = \sqrt{\frac{1}{n}\sum_{i=1}^{n}(f(x_i) - y_i)^2}. \tag{2}$$

RMSE can offer a useful interpretation of the performance and is provided as a secondary measure in CAGI evaluations.

The correlation coefficient between the prediction $f(X)$ and target $Y$ is defined as a normalized covariance between the prediction output and the target variable. Pearson's correlation coefficient $-1 \leq r \leq 1$ is estimated on $\mathcal{D}$ as

$$r = \frac{\sum_{i=1}^{n}(f_i - \bar{f})(y_i - \bar{y})}{\sqrt{\sum_{i=1}^{n}(f_i - \bar{f})^2 \sum_{i=1}^{n}(y_i - \bar{y})^2}}, \quad (3)$$

where $f_i = f(x_i)$ and $\bar{f} = \frac{1}{n}\sum_{i=1}^{n} f_i$ is the mean of the predictions. Pearson's correlation coefficient does not depend on the scale of the prediction, but it is affected by the extent of a linear relationship between predictions and the target. That is, a predictor outputting a linear transformation of the target such as $f(X) = 10 \cdot Y$ will have a perfect correlation. However, a monotonic nonlinear transformation of the target such as $f(X) = \log Y$ may have a relatively low $r$. Although not our main metric, we also explored Spearman's rank correlation as a secondary metric. Spearman's correlation is defined as Pearson's correlation on the rankings.

We also computed Kendall's tau, which is the probability of a concordant pair of prediction-target points linearly scaled to the $[-1, 1]$ interval instead of $[0, 1]$. Assuming that all prediction and target values are distinct, a pair of points $(f(x_i), y_i)$ and $(f(x_j), y_j)$ is concordant if either $(f(x_i) > f(x_j)$ and $y_i > y_j)$ or $(f(x_i) < f(x_j)$ and $y_i < y_j)$. Otherwise, a pair of points is discordant. Kendall's tau was estimated on $\mathcal{D}$ as

$$\tau = \frac{2}{n(n-1)} \sum_{i=1}^{n-1} \sum_{j=i+1}^{n} \mathrm{sign}(f(x_i) - f(x_j)) \cdot \mathrm{sign}(y_i - y_j). \quad (4)$$

It ranges between $-1$ and 1, with 1 indicating that all pairs are concordant, 0 indicating half of the concordant pairs (e.g., a random ordering) and $-1$ indicating that all pairs are discordant. A predictor outputting a linear transformation of the target $f(X) = 10 \cdot Y$ and a monotonic nonlinear transformation of the target $f(X) = \log Y$ will both have a perfect tau of 1. Compared to Pearson's correlation, Kendall's tau can be seen as less sensitive to the scale but more sensitive to the ordering of predictions. Eq. 4 is defined under the assumption that both the predictions and the outputs are unique. However, this assumption is not satisfied by all biological datasets and predictors. To address this issue, we use Kendall's tau-b, a widely accepted correction for ties

$$\tau_b = \frac{\sum_{i=1}^{n-1} \sum_{j=i+1}^{n} \mathrm{sign}(f(x_i) - f(x_j)) \cdot \mathrm{sign}(y_i - y_j)}{\sqrt{\left(\beta(n) - \sum_{i=1}^{T} \beta(u_i)\right)\left(\beta(n) - \sum_{i=1}^{S} \beta(v_i)\right)}}, \quad (5)$$

where $\beta(n) = n(n-1)/2$, $u_i$ $(v_i)$ is the size of the $i^{\text{th}}$ group of ties in the predictions (outputs) and $T$ $(S)$ is the number of such groups in the predictions (outputs).[73]

**Evaluation for binary targets.** Evaluating binary outputs is performed using standard protocols in binary classification.[74] We compute the Receiver Operating Characteristic (ROC) curve, which is a 2D plot of the true positive rate $\gamma = P(f_\tau(X) = 1 | Y = 1)$ as a function of the false positive rate $\eta = P(f_\tau(X) = 1 | Y = 0)$, where $\tau$ is varied over the entire range of prediction scores. The area under the ROC curve can be mathematically expressed as $\mathrm{AUC} = \int_0^1 \gamma \, d\eta$ and is the probability that a randomly selected positive example $x_+$ will be assigned a higher score than a randomly selected negative example $x_-$ by the model.[75] That is, assuming no ties in prediction scores $\text{AUC} = P(f(X_+) > f(X_-))$. In the presence of ties, AUC is given by $P(f(X_+) > f(X_-)) + \frac{1}{2}P(f(X_+) = f(X_-))$.[76] The AUC is estimated on the test set $\mathcal{D}$ using the standard numerical computation that allows for ties.[77] Although AUC does not serve as a metric that directly translates into clinical decisions, it is useful in that it shows the degree of separation of the examples from the two groups of data points (positive vs. negative). Another useful property of the AUC is its insensitivity to class imbalance.

Though AUC is a useful measure for capturing the overall performance of a classifier's score function, it has limitations when applied to a decision-making setting such as the one encountered in the clinic. Typically, clinically relevant score thresholds that determine the variants satisfying Supporting, Moderate or Strong evidence[26] lie in a region of low false positive rate (FPR). A measure well-suited to capture clinical significance of a predictor ought to be sensitive to the variations in the classifier's performance in the low FPR region (when predicting pathogenicity). However, the contribution of the low FPR region to AUC is relatively small. This is because it not only represents a small fraction of the entire curve, but also because the TPR values in that region are relatively small. Thus, AUC is not sensitive enough to the variation in a predictor's performance in the low FPR region. To mitigate this problem, we also provide area under the ROC curve truncated to the $[0, 0.2]$ FPR interval. What constitutes low FPR is not well defined; however, it appears that the $[0, 0.2]$ FPR interval combined with the $[0, 1]$ TPR interval is a reasonable choice in CAGI applications; see Figures 2-3. We normalize the truncated AUC to span the entire $[0, 1]$ range by dividing the observed value by 0.2, the maximum possible area below the ROC truncated at FPR = 0.2.

CAGI evaluation of binary classifiers also involves calculation of the Matthews correlation coefficient.[78] The Matthews correlation coefficient (MCC) was computed as a Pearson's correlation coefficient between binary predictions and target values on the test set $\mathcal{D}$. Efficient MCC estimation was carried out from the confusion matrix.[78]

**Evaluation for clinical significance.** Current guidelines from the American College for Medical Genetics and Genomics (ACMG) and Association for Molecular Pathology (AMP) established a qualitative framework for combining evidence in support of or against variant pathogenicity for clinical use.[26] These guidelines point to five different levels of pathogenicity and (effectively) nine distinct types of evidence in support of or against variant pathogenicity. The five pathogenicity levels involve classifications into pathogenic, likely pathogenic, variant of uncertain significance (VUS), likely benign, and benign variants, whereas the nine levels of evidential support are grouped into Very Strong, Strong, Moderate, and Supporting for either pathogenicity or benignity, as well as indeterminate evidence that supports neither pathogenicity nor benignity.

Richards et al.[26] have manually categorized different types of evidence and also listed twenty rules for combining evidence for a variant to be classified into one of the five pathogenicity-benignity groups. For example, variants that accumulate one Very Strong and one Strong line of evidence of pathogenicity lead to the classification of the variant as pathogenic; variants that accumulate one Strong and two Supporting lines of evidence lead to the classification of the variant as likely pathogenic, etc.[26] The guidelines allow for the use of computational evidence such that a computational prediction of pathogenicity can be considered as the weakest (Supporting) line of evidence. Thus, combined with other evidence, these methods can presently contribute to a pathogenicity assertion for a variant, but in a restricted and arbitrary way.[43] Since Supporting evidence is qualitatively the smallest unit of contributory evidence in the ACMG/AMP guidelines, we refer to any computational model that reaches the prediction quality equivalent of Supporting evidence and higher as a model that provides contributory evidence in the clinic.

Numerically, a variant that is classified as pathogenic should have at least a 99% probability of being pathogenic given all available evidence, whereas a variant that is likely pathogenic should have at least a 90% probability of being pathogenic given the evidence.[26, 45] Variants that cross the 90% probability threshold for pathogenicity are considered clinically actionable.[26] Analogously, variants with sufficient support for benignity will typically be ruled out from being diagnostic in a clinical laboratory. Note that, though the guidelines provide a probabilistic interpretation of the pathogenicity assertions, they do not provide any general quantitative interpretation of the evidence. Consequently, any framework designed to express the evidence levels quantitatively, must tie such quantitative evidential support to the pathogenicity probabilities, mediated by the ACMG/AMP rules for combining evidence.

The possibility of incorporating computational methods into clinical decision making in a properly calibrated manner presents interesting opportunities and unique challenges. In particular, since the evidence levels are only described qualitatively, it is not obvious how to determine what values of a predictor's output score qualify as providing a given level of evidence. Thus, to apply a computational line of evidence in the clinic in a principled manner, and consistent with the guidelines, there is a need for a framework that assigns a quantitative interpretation to each evidence level.

Tavtigian et al.[45] proposed such a framework to provide numerical support for each type of evidential strength for its use in ACMG/AMP guidelines for or against variant pathogenicity. This approach is based on the relationship between prior and posterior odds of pathogenicity as well as on independence of all lines of evidential support for a given variant. We briefly review this approach.

Let $E$ be a random variable indicating evidence that can be used in support of or against variant pathogenicity. The positive likelihood ratio ($\text{LR}^+$) given concrete evidence $e$ is defined as

$$\text{LR}^+ = \frac{\text{posterior odds of pathogenicity}}{\text{prior odds of pathogenicity}} \tag{6}$$

or equivalently

$$\text{LR}^+(e) = \frac{P(Y=1|E=e)}{1-P(Y=1|E=e)} \cdot \frac{1-P(Y=1)}{P(Y=1)}, \tag{7}$$

where the first term on the right corresponds to the posterior odds of pathogenicity given the evidence and the second term on the right corresponds to the reciprocal of the prior odds of pathogenicity. The prior odds of pathogenicity depend solely on the class prior $P(Y=1)$; that is, the fraction of pathogenic variants in the selected reference set. The expression for $\text{LR}^+$ also allows for an easy interpretation as the increase in odds of pathogenicity given evidence $e$ compared to the situation when no evidence whatsoever is available. The likelihood ratio of 2, for example, states that a variant with evidence $e$ is expected to have twice as large odds of being pathogenic than a variant picked uniformly at random from a reference set. As CAGI only considers computational evidence, we will later replace the posterior probability $P(Y=1|E=e)$ by $P(Y=1|f(X)=1)$ for discretized predictors or by $P(Y=1|s(X)=s)$ for the predictors that output a soft numerical score $s$. The probability $P(Y=1|f(X)=1)$ is the positive predictive value (or precision) of a binary classifier, whereas the probability $P(Y=1|s(X)=s)$ can be seen as the local positive predictive value, defined here in a manner analogous to the local false discovery rate.[79]

It can be shown[80] that the positive likelihood ratio can also be stated as

$$\mathrm{LR}^{+}(e)=\frac{P(E=e|Y=1)}{P(E=e|Y=0)} \tag{8}$$

thus clarifying that $\mathrm{LR}^{+}$ can be seen as the ratio of the true positive rate and false positive rate when $P(E=e|Y=1)$ is replaced by $P(f(X)=1|Y=1)$ and $P(E=e|Y=0)$ by $P(f(X)=1|Y=0)$.

Tavtigian et al.[45] give an expression relating the posterior $P(Y=1|E=e)$ to $\mathrm{LR}^{+}$ and the prior $P(Y=1)$ as

$$P(Y=1|E=e)=\frac{\mathrm{LR}^{+}(e)P(Y=1)}{(\mathrm{LR}^{+}(e)-1)P(Y=1)+1} \tag{9}$$

which itself is obtained from Eq. 7. They also present a framework that allows for assigning probabilistic interpretations to different types of evidential strength (Supporting, Moderate, Strong, and Very Strong) and combining them in a manner consistent with the rules listed in Richards et al.[26] and the probabilistic interpretation of likely pathogenic and pathogenic classes. Their formulation is given in terms of the positive likelihood ratio $\mathrm{LR}^{+}$ in an exponential form. We restate this model using a notion of the total (or combined) positive likelihood ratio $\mathrm{LR}_{\mathrm{T}}^{+}$, based on all available evidence, $E_{\mathrm{T}}$, of a variant that is expressed as a product of $\mathrm{LR}^{+}$ factors from different strengths of evidence as

$$\mathrm{LR}_{\mathrm{T}}^{+}=\mathrm{c}^{\frac{n_{\mathrm{su}}}{8}+\frac{n_{\mathrm{mo}}}{4}+\frac{n_{\mathrm{st}}}{2}+\frac{n_{\mathrm{vs}}}{1}}, \tag{10}$$

where $n_{\mathrm{su}}$, $n_{\mathrm{mo}}$, $n_{\mathrm{st}}$, and $n_{\mathrm{vs}}$ are the counts of Supporting (su), Moderate (mo), Strong (st), and Very Strong (vs) lines of evidence present in $E_{\mathrm{T}}$, and $c$ is the $\mathrm{LR}^{+}$ value assigned to a single line of Very Strong evidence. It is easy to show that $\sqrt[8]{c}$ , $\sqrt[4]{c}$, and $\sqrt[2]{c}$ correspond to the $\mathrm{LR}^{+}$ for a single line of Supporting, Moderate, and Strong line of evidence, respectively. In other words, the model from Eq. 10 enforces that if a Very Strong piece of evidence increases $\mathrm{LR}_{\mathrm{T}}^{+}$ by a multiplicative factor of $c$, then a Supporting, Moderate, or a Strong piece of evidence increases $\mathrm{LR}_{\mathrm{T}}^{+}$ by a factor of $\sqrt[8]{c}$, $\sqrt[4]{c}$, and $\sqrt[2]{c}$, respectively. For a reasonable consistency with Richards et al.[26] this model also explicitly encodes that one line of Very Strong evidence is equal to the two lines of Strong evidence, four lines of Moderate evidence, and eight lines of Supporting evidence.

The appropriate value of $c$, however, depends on the class prior. It is the smallest number for which the $\mathrm{LR}_{\mathrm{T}}^{+}$ values computed for the qualitative criteria from the likely pathogenic class in Richards et al.[26] reach $P(Y=1|E_{\mathrm{T}}=e)$ values of at least 0.9 and, similarly, for those in the pathogenic class, reach a $P(Y = 1|E_{\mathrm{T}} = e)$ value of at least 0.99. The dependence on the class prior is due to the conversion between $\mathrm{LR}_{\mathrm{T}}^{+}$ and $P(Y = 1|E_{\mathrm{T}} = e)$ governed by Eq. 9. If the class prior is small, a larger value of $\mathrm{LR}_{\mathrm{T}}^{+}$ will be required to achieve the same posterior level, thereby requiring a larger value of $c$ (Supplementary Figure 14).

Tavtigian et al.[45] also proposed that two rules from Richards et al.[26] be revised; that is, one of the rules was proposed to be "demoted" from pathogenic to likely pathogenic, whereas another rule was proposed to be "promoted" from the likely pathogenic to pathogenic. For a class prior of 0.1 that was selected based on the experience from the clinic, the value $c = 350$ was found to be suitable. This, in turn, suggests that the Supporting, Moderate, and Strong lines of evidence should require the likelihood ratio values of $\sqrt[8]{c} = 2.08$, $\sqrt[4]{c} = 4.32$, and $\sqrt[2]{c} = 18.7$, respectively. However, note again that for different priors, these values will be different; see next Section and Supplementary Figure 14. Moreover, while the level of posterior for the combined evidence (Eq. 13) is required to be at least 0.9 to satisfy the likely pathogenic rule and 0.99 for pathogenic, this does not mean that the posterior level for a single line of evidence is the same for all values of $c$. This is a consequence of the fact that the framework provides intuitive interpretation only at the level of the combined posterior.

When drawing evidence from a pathogenicity predictor, it is necessary to further clarify what evidence is in the first place. At least two options are available: (i) the evidence is the score $s(x)$; that is, a raw prediction of pathogenicity, or (ii) the evidence is a discretized prediction $f_{\tau}(x)$, obtained by thresholding $s(x)$. These approaches, referred to here as local and global, respectively, lead to different interpretations because all evaluation metrics hold only on average, either over all variants with a score $s(x)$ or all variants satisfying $f_{\tau}(x) = 1$; i.e., having a score above $\tau$. When both $s(x)$ and $f_{\tau}(x)$ are available, this leads to difficulties in interpreting the results of the global approach because all scores $s(x)$ that map into $f_{\tau}(x) = 1$ will be treated identically. Unfortunately, this implies that scores $s$ greater than but still close to $\tau$ most likely do not meet the levels of evidential strength for the interval. At the same time, scores close to the high end of the range almost certainly make the levels of evidential strength above the designated level. This means that a clinician seeing a variant with score slightly above $\tau$ would have to interpret this prediction as contributory to pathogenicity, yet this interpretation would almost certainly be incorrect. Based on the recommendations from the ClinGen's Sequence Variant Interpretation group[43] we focus on the local view as well as local performance criteria to define levels of evidential strength and assess whether methods achieve these levels. In the end, however, we also provide global estimates to understand the performance of each tool more comprehensively.

We define the local positive likelihood ratio as

$$\mathrm{lr}^{+}(s) = \frac{p(s|Y=1)}{p(s|Y=0)}, \tag{11}$$

where $p(s|Y = y)$, for $y \in \{0, 1\}$ are class-conditional densities.

We obtain an estimate of the local positive likelihood ratio $\widehat{\mathrm{lr}}^{+}$ from the test data as described in the Section titled "Computing the clinically relevant measures". Now, the threshold to determine the variants with Supporting level of evidence is given as the minimum score above which all variants achieve local positive likelihood value greater than or equal to $\sqrt[8]{c}$; i.e.,

$$\tau_{\mathrm{su}} = \min\{\tau : \forall s \geq \tau, \widehat{\mathrm{lr}}^{+}(s) \geq \sqrt[8]{c}\}, \tag{12}$$

though we note that Pejaver et al.[43] incorporated an additional factor based on the confidence interval for $\widehat{\mathrm{lr}}^{+}(s)$ to result in more stringent recommendations for score thresholding. Similarly, the thresholds for variants with Moderate, Strong and Very Strong evidence are given by $\tau_{\mathrm{mo}} = \min\{\tau : \forall s \geq \tau, \widehat{\mathrm{lr}}^{+}(s) \geq \sqrt[4]{c}\}$, $\tau_{\mathrm{st}} = \min\{\tau : \forall s \geq \tau, \widehat{\mathrm{lr}}^{+}(s) \geq \sqrt[2]{c}\}$ and $\tau_{\mathrm{vs}} = \min\{\tau : \forall \mathrm{s} \geq \tau, \widehat{\mathrm{lr}}^{+}(s) \geq \mathrm{c}\}$.

Once the threshold set $\{\tau_{\mathrm{su}}, \tau_{\mathrm{mo}}, \tau_{\mathrm{st}}, \tau_{\mathrm{vs}}\}$ is determined, we can compute either the global $\mathrm{LR}^{+}$ (e.g., $s \geq \tau_{\mathrm{su}}$) or the $\mathrm{LR}^{+}$ corresponding to an interval of scores (e.g., $\tau_{\mathrm{su}} \leq s < \tau_{\mathrm{mo}}$) by computing the true positive rate and false positive rate for a given set of scores. A global positive predictive value can be similarly estimated once the class prior is known.

In all CAGI evaluations, a predictor is considered to provide contributory evidence in a clinical setting if it reaches any one of the evidence levels according to the ACMG/AMP guidelines, and according to the model by Tavtigian et al.[45] and recommendations by Pejaver et al.[43] Among predictors that reach the desired levels of evidential support, the ones that reach higher levels are generally considered favorably. However, we have not considered any criterion to rank the predictors that reach the same levels of evidential support.

***Selection of class priors for variant pathogenicity.*** Different clinical scenarios require the use of different class priors of variant pathogenicity. We generally distinguish between two clinical situations.

In the first setting, a clinician is presented with a proband with specific phenotypic expression and the objective is to find variants responsible for the clinical phenotype. In certain monogenic disorders with Mendelian inheritance patterns, the fraction of rare variants found to be pathogenic can be as high as 25%, as in the case of the NAGLU challenge. Similarly, Tavtigian et al.[45] report an experience-based prior of 10% based on their work with BRCA genes, which we adopted in this work.

The second setting reflects situations such as screening for potential secondary variants. Here we have used an estimate by Pejaver et al.[34] that up to 1.5% of missense variants in an apparently healthy individual could be disease-causing.

Overall, prior probability of pathogenicity was set to 1% and 10% to demonstrate the distinction in the level of evidential support necessary. These resulted in $c = 8511$ and $c = 351$, respectively (note that $c = 351$ was selected instead of $c = 350$ to avoid rounding errors in finding a $c$ that best models ACMG/AMP rules). In each functional missense challenge, the level of prior probability observed for each gene based on experimental data was further considered. For large class priors such as 50% or above, the Tavtigian et al.[45] framework holds only when an additional rule from Richards et al.[26] is removed; that is, we ignored that two Supporting lines of evidence for benignity assert a likely benign variant.

**Performance measures for clinical application.** ***Diagnostic odds ratio.*** The diagnostic odds ratio (DOR) is commonly used in biomedical sciences to measure the increase in odds of pathogenicity in the presence of evidence $e$ compared to the odds of pathogenicity in the absence of $e$;[80] that is,

$$\mathrm{DOR}(e) = \frac{P(Y=1|E=e)}{1-P(Y=1|E=e)} \cdot \frac{1-P(Y=1|E\neq e)}{P(Y=1|E\neq e)}\,. \tag{13}$$

The difference between Eq. 7 and Eq. 13 is that the prior odds, those governed by the prior $P(Y=1)$ and used in Eq. 7, are replaced by the odds governed by the probability $P(Y=1|E\neq e)$; that is, odds of pathogenicity when the evidence $e$ was not the one that was observed. The quantity $P(Y=0|E\neq e) = 1 - P(Y=1|E\neq e)$ is referred to as the negative predictive value when the observed evidence is $f(X)=0$. $\mathrm{DOR} \in [0,\infty)$ can also be expressed as

$$\mathrm{DOR}(e) = \frac{\mathrm{LR}^{+}(e)}{\mathrm{LR}^{-}(e)}\,, \tag{14}$$

where $\mathrm{LR}^{+}(e)$ is defined in Eq. 8 and

$$\mathrm{LR}^{-}(e) = \frac{P(E\neq e|Y=1)}{P(E\neq e|Y=0)}\,. \tag{15}$$

In contrast to typical studies of variant risk assessment[81] and polygenic risk scores,[57] DOR was calculated without adjustments for usual confounders such as race, ethnicity, etc. that are generally not available in CAGI challenges and, technically, produce conditional odds ratios.[80] However, the DOR values estimated in our experiments have an identical interpretation as the results of logistic regression run with a single independent variable (co-variate) at a time. Glas et al.[80] give a broader coverage of diagnostic odds ratios that further connect some of the quantities discussed here (e.g., AUC vs. DOR).

We only consider DOR with the computational evidence of the "global" type; that is, when $s(x) \geq \tau$. Consequently, DOR at $\tau$ can be expressed as

$$\mathrm{DOR}(\tau) = \frac{\mathrm{LR}^{+}(\tau)}{\mathrm{LR}^{-}(\tau)} = \frac{P(s(X)\geq\tau|Y=1)}{P(s(X)\geq\tau|Y=0)}\frac{P(s(X)<\tau|Y=0)}{P(s(X)<\tau|Y=1)}. \tag{16}$$

Unlike positive likelihood ratio, DOR does not have a "local" version. This is because one cannot define a local negative likelihood ratio.

***Percent of variants predicted as pathogenic.*** In addition to finding whether a method reaches Supporting, Moderate or Strong levels of evidence, it is important to also quantify the proportion of variants in the reference set for which a given evidence level is reached. To this end, for a given score threshold $\tau$, we define the percent of variants in the reference set that the method assigns a score as high as or higher than $\tau$, and refer to it as "probability of pathogenic (positive) predictions", or PPP. Mathematically, it can be expressed as the following probability

$$\mathrm{PPP}(\tau) = P(s(X) \geq \tau). \tag{17}$$

The probability (equivalently, percent) of variants reaching a given level of evidence can now be quantified as $\mathrm{PPP}(\tau)$, where $\tau$ is the score threshold at which a variant is declared to meet the desired evidential support.

***Posterior probability of pathogenicity.*** Given a method, the posterior probability of pathogenicity or the absolute risk for a variant is defined as the probability that the variant is pathogenic based on the score it is assigned by the method. It is expressed as

$$\rho(s) = P(Y = 1|s(X) = s). \tag{18}$$

We also refer to this quantity as a local positive predictive value or local precision.

***Relative risk.*** Given a method, the relative risk (RR) of pathogenicity of a variant is defined as the posterior probability of pathogenicity (based on the score assigned by the method) relative to the prior probability of pathogenicity. It is expressed as the following ratio

$$\mathrm{RR}(s) = \frac{P(Y=1|s(X)=s)}{P(Y=1)}. \tag{19}$$

The prior probability of pathogenicity can also be interpreted as the average of the posterior probability over all variants in the reference set; that is

$$\begin{aligned} \mathbb{E}[P(Y = 1|s(x) = s)] &= \int_{\mathcal{X}} P(Y = 1|s(x) = s)p(x)dx \\ &= \int_{\mathbb{R}} P(Y = 1|s)p(s)\, ds \\ &= \int_{\mathbb{R}} p(s|Y = 1)P(Y = 1)\, ds \\ &= P(Y = 1), \end{aligned} \tag{20}$$

where the last step follows since $p(s|Y = 1)$ is a density function and its integral over $\mathbb{R}$ is 1. Observe that our definition of relative risk is an extension of the "global" version used in clinical applications where the denominator would be $P(Y = 1|s(X) \neq s)$, which effectively equals $P(Y = 1)$ for all predictors outputting continuous scores.

***Computing clinically relevant measures.*** We show here how the measures for evaluation of binary targets and clinically relevant measures are computed from the test data $\mathcal{D}$. It is necessary to be cautious when making decisions on a reference (target) population based on the measures computed on the test set $\mathcal{D}$. Some of the measures computed on $\mathcal{D}$ accurately represent the corresponding values on the target population. However, other measures are biased because the test data set for many challenges is not representative of the target population. In particular, the proportions of positives (e.g., pathogenic variants) in the test set $\alpha_{\mathcal{D}} = P_{\mathcal{D}}(Y = 1)$ may be vastly different from that in the target population $\alpha = P(Y = 1)$. Consequently, the class-prior dependent measures, when estimated directly from the test set, are incorrectly calibrated to the test set class priors.

Fortunately, the class-prior dependent measures can be corrected using an estimate of the target population's class priors if known or if estimated using a principled approach.[82, 83] The correction is derived under the assumption that the reference population and the test set are distributionally identical, except for the differences in class priors. To elaborate, the target distribution of inputs $p(x)$ can be expressed in terms of the class-conditional distributions, $p(x|Y = y)$ for $y \in \{0, 1\}$, and the class priors as follows

$$p(x) = \alpha \cdot p(x|Y = 1) + (1 - \alpha) \cdot p(x|Y = 0). \tag{21}$$

We assume that the test set distribution of inputs might have different class priors, but the same class-conditional distributions as the target population. Precisely,

$$
\begin{aligned}
p_{\mathcal{D}}(x) &= \alpha_{\mathcal{D}} p_{\mathcal{D}}(x|Y=1) + (1-\alpha_{\mathcal{D}}) p_{\mathcal{D}}(x|Y=0) \\
&= \alpha_{\mathcal{D}} p(x|Y=1) + (1-\alpha_{\mathcal{D}}) p(x|Y=0).
\end{aligned} \tag{22}
$$

It is easy to see that any of the clinical and non-clinical measures that only depend on the class-conditional distributions, but not class priors, when computed on the test set is an unbiased estimate of the measure on the target population. However, if a measure also depends on the class priors, it needs to be corrected to reflect the reference population's class prior. All the class-prior independent measures used in this paper can be expressed in terms of class-conditional derived quantities such as the true positive rate (TPR), the false positive rate (FPR) and the local positive likelihood ratio $\mathrm{lr}^+(s)$. The class-prior dependent measures additionally have the class-prior in their expressions.

***TPR, FPR and $lr^+(s)$.*** Formally, TPR is defined as the proportion of positive inputs that are correctly predicted to be positive. Mathematically,

$$
\mathrm{TPR}(\tau) = P(s(x) \geq \tau | Y = 1), \tag{23}
$$

where $s(x)$ is a continuous score function of a classifier and $\tau$ is a threshold such that an input scoring above $\tau$ is predicted to be positive. Similarly, FPR is defined as the proportion of negative inputs that are incorrectly predicted to be positive. Mathematically,

$$
\mathrm{FPR}(\tau) = P(s(x) \geq \tau | Y = 0), \tag{24}
$$

TPR and FPR can be computed from the test data as the proportion of positive and negative test inputs scoring $\geq \tau$, respectively. That is,

$$
\begin{aligned}
\widehat{\mathrm{TPR}}(\tau) &= \frac{\sum_{x \in \mathcal{D}_+} I[s(x) \geq \tau]}{|\mathcal{D}_+|} \\
\widehat{\mathrm{FPR}}(\tau) &= \frac{\sum_{x \in \mathcal{D}_-} I[s(x) \geq \tau]}{|\mathcal{D}_-|},
\end{aligned} \tag{25}
$$

where $\mathcal{D}_+$ and $\mathcal{D}_-$ are the subsets of points in the test set $\mathcal{D}$ labeled as positive and negative, respectively.

Some of the clinically relevant measures used in our study are "local" in nature in the sense that they are derived from a local neighborhood around a score value instead of the entire range of scores above (or below) the threshold. Such measures can be expressed in terms of the local positive likelihood ratio $\mathrm{lr}^+(s)$. To compute $\mathrm{lr}^+(s)$, we exploit its relationship to the posterior probability at score $s$; that is,

$$
\begin{aligned}
P(Y=1|s(X)=s) &= \frac{p(s(X)=s|Y=1)P(Y=1)}{p(s(X)=s)} \\
&= \frac{p(s(X)=s|Y=1)P(Y=1)}{p(s(X)=s|Y=1)P(Y=1)+p(s(X)=s|Y=0)P(Y=0)} \\
&= \frac{\mathrm{lr}^+(s)P(Y=1)}{\mathrm{lr}^+(s)P(Y=1)+P(Y=0)} \\
&= \frac{\mathrm{lr}^+(s)P(Y=1)}{(\mathrm{lr}^+(s)-1)P(Y=1)+1}.
\end{aligned} \tag{26}
$$

Similarly, the test data posterior probability can be expressed as

$$P_{\mathcal{D}}(Y=1|s(X)=s) = \frac{\text{lr}^+(s)P_{\mathcal{D}}(Y=1)}{(\text{lr}^+(s)-1)P_{\mathcal{D}}(Y=1)+1}. \quad (27)$$

Note that since $\text{lr}^+(s)$ only depends on the class-conditional distribution, it does not change when defined on the target population. Unlike the target population's posterior, the test data posterior can be estimated from the test data as described below. Once the test posterior is estimated, the equation above can be inverted to estimate $\text{lr}^+$ as

$$\widehat{\text{lr}}^+(s) = \frac{\hat{P}_{\mathcal{D}}(Y=1|s(X)=s)}{1-\hat{P}_{\mathcal{D}}(Y=1|s(X)=s)} \cdot \frac{1-P_{\mathcal{D}}(Y=1)}{P_{\mathcal{D}}(Y=1)}, \quad (28)$$

where the $\hat{P}_{\mathcal{D}}(Y=1|s(X)=s)$ is an estimate of the test data posterior and $P_{\mathcal{D}}(Y=1)$ is the proportion of positives in the test data, which may differ from the true prior for a randomly picked variant in the gene of interest or another reference sample. Note that though the formula above expresses $\widehat{\text{lr}}^+(s)$ in terms of the prior odds, suggesting a dependence on the class prior, $\widehat{\text{lr}}^+(s)$ is class-prior independent, as discussed earlier. In theory, $P_{\mathcal{D}}(Y=1|s(X)=s)$, is the proportion of pathogenic variants among all variants in $\mathcal{D}$ having a score $s$. Therefore, estimating the local posterior efficiently would require observing the same score many times in the set of variants with known labels. This is unlikely since we only have scores for a finite set of variants and thus the posterior cannot be estimated without making further assumptions. However, assuming that the posterior is a smooth function of the score—similar scores correspond to similar local posterior values—we estimate the posterior as the proportion of pathogenic variants in a small window around the score; that is, $[s-\epsilon, s+\epsilon]$, where $\epsilon$ was selected to be 5% of the range of the predictor's outputs, with the range considered to be an interval between the 5$^{th}$ and 95$^{th}$ percentile of predicted values on the dataset, selected as such to minimize the influence of outliers. In addition, for stable estimates, we required that at least 10% of the variants, up to a maximum of 50 variants, from the data set are within a window; therefore, the final window size was dependent on score $s$ and data set $\mathcal{D}$.

***Measures that do not require correction.*** Among the measures considered in this paper, TPR, FPR, ROC curve, AUC, $\text{LR}^+$, $\text{LR}^-$, DOR and $\text{lr}^+$ do not require correction. Class-prior independence of TPR, FPR and $\text{lr}^+$ is obvious from their definitions as discussed earlier. ROC curve is obtained by plotting TPR against FPR and consequently, it is also class-prior independent. By extension AUC, being the area under the ROC curve, is also class-prior independent. The global positive likelihood ratio $\text{LR}^+$, formulated with the evidence of the type $s(x) \geq \tau$, is given by $\text{TPR}(\tau)/\text{FPR}(\tau)$. Similarly, the global $\text{LR}^-$ is given by $(1 - \text{TPR}(\tau))/(1-\text{FPR}(\tau))$. Since DOR is the ratio of $\text{LR}^+$ and $\text{LR}^-$, it is by extension class-prior independent.

***Measures that require correction.*** Among the measures considered in this paper, probability of pathogenic predictions (PPP), positive predictive value (PPV), posterior probability ($\rho$) and relative risk (RR), being class-prior dependent, require corrections to be properly applied to the target population. To show that the measures are indeed class-prior dependent, we re-formulate them by separating the class-prior from the class-conditional dependent terms.

$$\begin{aligned} \text{PPP}(\tau) &= P(s(X) \geq \tau) \\ &= P(s(X) \geq \tau|Y=1)P(Y=1) + P(s(X) \geq \tau|Y=0)P(Y=0) \\ &= \alpha\text{TPR}(\tau) + (1-\alpha)\text{FPR}(\tau) \end{aligned} \quad (29)$$

$$\begin{aligned}\mathrm{PPV}(\tau) &= P(Y=1|S(X)\geq\tau) \\ &= \frac{P(s(X)\geq\tau|Y=1)P(Y=1)}{P(s(X)\geq\tau)} \\ &= \frac{\alpha\mathrm{TPR}(\tau)}{\alpha\mathrm{TPR}(\tau)+(1-\alpha)\mathrm{FPR}(\tau)}\end{aligned} \tag{30}$$

$$\begin{aligned}\rho(s) &= P(Y=1|s(X)=s) \\ &= \frac{\alpha\mathrm{lr}^{+}(s)}{\alpha(\mathrm{lr}^{+}(s)-1)+1},\end{aligned} \tag{31}$$

where the derivation is the same as that for Eq. 26.

$$\begin{aligned}\mathrm{RR}(s) &= \frac{P(Y=1|s(X)=s)}{\alpha} \\ &= \frac{\mathrm{lr}^{+}(s)}{\alpha(\mathrm{lr}^{+}(s)-1)+1}.\end{aligned} \tag{32}$$

We use the expressions above to correctly calculate class-prior dependent metrics on the target domain by first computing the class-conditional dependent terms (TPR, FPR or $\mathrm{lr}^{+}$) using the test data $\mathcal{D}$ and then using an estimate of the class prior of the target distribution in the corresponding expression.

**Statistical significance and confidence interval estimation.** All p-values and confidence intervals in CAGI evaluations were estimated using bootstrapping with 1,000 iterations.[84]

## The Critical Assessment of Genome Interpretation Consortium

***Analysis and writing***
Shantanu Jain[1,+], Constantina Bakolitsa[2,+], Steven E. Brenner[2,$,*], Predrag Radivojac[1,3,*], John Moult[4,*]
[+]Contributed equally
[$]This author was unable to fully contribute to the paper writing due to an injury and its sequelae
***Corresponding authors (brenner@berkeley.edu; predrag@northeastern.edu; jmoult@tunc.org)

***CAGI organizers***
John Moult[4], Susanna Repo[2], Roger A. Hoskins[2], Gaia Andreoletti[2], Daniel Barsky[2], Steven E. Brenner[2]

***Informatics infrastructure and support***
Ajithavalli Chellapan[5], Hoyin Chu[1,6-7], Navya Dabbiru[5], Naveen K. Kollipara[5], Melissa Ly[2], Andrew J. Neumann[2], Lipika R. Pal[4], Gaurav Pandey[2], Robin C. Peters-Petrulewicz[2], Rajgopal Srinivasan[5], Stephen F. Yee[2], Sri Jyothsna Yeleswarapu[5], Maya Zuhl[4,8]

***Predictors***
Ogun Adebali[9-10], Ayoti Patra[11-12], Michael A. Beer[11], Raghavendra Hosur[13-14], Jian Peng[13], Brady M. Bernard[15-16], Michael Berry[9], Shengcheng Dong[17], Alan P. Boyle[17], Aashish Adhikari[2,18],

Jingqi Chen[2,19], Zhiqiang Hu[2], Robert Wang[2,20], Yaqiong Wang[2,19], Maximilian Miller[21], Yanran Wang[21-22], Yana Bromberg[21], Paola Turina[23], Emidio Capriotti[23], James J. Han[24], Kivilcim Ozturk[24], Hannah Carter[24], Giulia Babbi[23], Samuele Bovo[23], Pietro Di Lena[23], Pier Luigi Martelli[23], Castrense Savojardo[23], Rita Casadio[23], Melissa S. Cline[25], Greet De Baets[26], Sandra Bonache[27-28], Orland Díez[27,29], Sara Gutiérrez-Enríquez[27], Alejandro Moles-Fernández[27,29], Gemma Montalban[27,30], Lars Ootes[31], Selen Özkan[31], Natàlia Padilla[31], Casandra Riera[31], Xavier De la Cruz[31], Mark Diekhans[25], Peter Huwe[32-33], Qiong Wei[32,34], Qifang Xu[32], Roland L. Dunbrack[32], Valer Gotea[35], Laura Elnitski[35], Gennady Margolin[35], Piero Fariselli[36-37], Ivan V. Kulakovskiy[38-39], Vsevolod J. Makeev[38], Dmitry D. Penzar[38,40], Ilya E. Vorontsov[38-39], Alexander V. Favorov[11,38], Julia R. Forman[41-42], Marcia Hasenahuer[43-44], Maria S. Fornasari[43], Gustavo Parisi[43], Ziga Avsec[45], Muhammed H. Çelik[45-46], Thi Yen Duong Nguyen[45], Julien Gagneur[45], Fang-Yuan Shi[47], Matthew D. Edwards[13,48], Yuchun Guo[13,49], Kevin Tian[13,50], Haoyang Zeng[13,51], David K. Gifford[13], Jonathan Göke[52], Jan Zaucha[53-54], Julian Gough[55], Graham R.S. Ritchie[44,56], Adam Frankish[44,57], Jonathan M. Mudge[44,57], Jennifer Harrow[57-58], Erin L. Young[59], Yao Yu[60], Chad D. Huff[60], Katsuhiko Murakami[61-62], Yoko Nagai[61,63], Tadashi Imanishi[61,64], Christopher J. Mungall[65], Julius O.B. Jacobsen[57,66], Dongsup Kim[67], Chan-Seok Jeong[67-68], David T. Jones[69], Mulin Jun Li[70-71], Violeta Beleva Guthrie[11,54], Rohit Bhattacharya[11,72], Yun-Ching Chen[11,73], Christopher Douville[11], Jean Fan[11], Dewey Kim[7,11], David Masica[11], Noushin Niknafs[11], Sohini Sengupta[11,74], Collin Tokheim[6,11,75], Tychele N. Turner[11,76], Hui Ting Grace Yeo[11,52], Rachel Karchin[11], Sunyoung Shin[77-78], Rene Welch[77], Sunduz Keles[77], Yue Li[13,79], Manolis Kellis[7,13], Carles Corbi-Verge[80-81], Alexey V. Strokach[80], Philip M. Kim[80], Teri E. Klein[50], Rahul Mohan[82-83], Nicholas A. Sinnott-Armstrong[50], Michael Wainberg[82,84], Anshul Kundaje[50], Nina Gonzaludo[85-86], Angel C.Y. Mak[85,87], Aparna Chhibber[88-89], Hugo Y.K. Lam[88,90], Dvir Dahary[91], Simon Fishilevich[92], Doron Lancet[92], Insuk Lee[93], Benjamin Bachman[94], Panagiotis Katsonis[94], Rhonald C. Lua[94], Stephen J. Wilson[94-95], Olivier Lichtarge[94], Rajendra R. Bhat[96], Laksshman Sundaram[50], Vivek Viswanath[96], Riccardo Bellazzi[97], Giovanna Nicora[97-98], Ettore Rizzo[98], Ivan Limongelli[98], Aziz M. Mezlini[80], Ray Chang[99], Serra Kim[99], Carmen Lai[99], Robert O'Connor[99-100], Scott Topper[99], Jeroen van den Akker[99], Alicia Y. Zhou[99], Anjali D. Zimmer[99], Gilad Mishne[99], Timothy R. Bergquist[101-102], Marcus R. Breese[85,103], Rafael F. Guerrero[3,104], Yuxiang Jiang[3], Nikki Kiga[101], Biao Li[103,105], Matthew Mort[103,106], Kymberleigh A. Pagel[3], Vikas Pejaver[101,107], Moses H. Stamboulian[3], Janita Thusberg[103], Sean D. Mooney[101], Predrag Radivojac[1,3], Nuttinee Teerakulkittipong[4,108], Chen Cao[4,109], Kunal Kundu[4,110], Yizhou Yin[4], Chen-Hsin Yu[4], Maya Zuhl[4,8], Lipika R. Pal[4], John Moult[4], Michael Kleyman[4,111], Chiao-Feng Lin[20,112], Mary Stackpole[4,113], Stephen M. Mount[4], Gökcen Eraslan[7,114], Nikola S. Mueller[114], Tatsuhiko Naito[115], Aliz R. Rao[30], Johnathan R. Azaria[116-117], Aharon Brodie[116], Yanay Ofran[116], Aditi Garg[118], Debnath Pal[118], Alex Hawkins-Hooker[41,69], Henry Kenlay[41,119], John Reid[41,120], Eliseos J. Mucaki[121], Peter K. Rogan[121], Jana M. Schwarz[122], David B. Searls[123], Gyu Rie Lee[101,124], Chaok Seok[124], Andreas Krämer[125], Sohela Shah[125-126], ChengLai V. Huang[2,127], Jack F. Kirsch[2], Maxim Shatsky[65], Yue Cao[128], Haoran Chen[128-129], Mostafa Karimi[127-128], Oluwaseyi Moronfoye[128], Yuanfei Sun[128], Yang Shen[128], Ron Shigeta[130-131], Colby T. Ford[132], Conor Nodzak[132], Aneeta Uppal[132-133], Xinghua M. Shi[132,134], Thomas Joseph[5], Sujatha Kotte[5], Sadhna Rana[5], Aditya Rao[5], V.G. Saipradeep[5], Naveen Sivadasan[5], Uma Sunderam[5], Rajgopal Srinivasan[5], Mario Stanke[135], Andrew Su[136], Ivan Adzhubey[137-138], Daniel M. Jordan[107,139], Shamil Sunyaev[139], Frederic Rousseau[26], Joost Schymkowitz[26], Joost Van Durme[26], Sean V. Tavtigian[59], Marco Carraro[36], Silvio E. Tosatto[36], Orit Adato[116], Liran Carmel[140], Noa E. Cohen[140], Tzila Fenesh[116], Tamar Holtzer[116], Tamar Juven-Gershon[116], Ron S. Unger[116], Abhishek Niroula[141], Ayodeji

Olatubosun[142], Jouni Väliaho[142], Yang Yang[143], Mauno Vihinen[141-142], Mary E. Wahl[34,139], Billy Chang[144], Ka Chun Chong[144], Inchi Hu[145-146], Rui Sun[144,147], William Ka Kei Wu[144], Xiaoxuan Xia[144], Benny C. Zee[144], Maggie H. Wang[144], Meng Wang[47], Chunlei Wu[136], Yutong Lu[147], Ken Chen[147], Yuedong Yang[148-150], Christopher M. Yates[151-152], Anat Kreimer[2,21], Zhongxia Yan[2,13], Nir Yosef[2], Huying Zhao[150], Zhipeng Wei[153], Zhaomin Yao[153], Fengfeng Zhou[153], Lukas Folkman[149,154], Yaoqi Zhou[149,155]

***Challenge data providers***

Roxana Daneshjou[50], Russ B. Altman[50], Fumitaka Inoue[85,156], Nadav Ahituv[85], Adam P. Arkin[2], Federica Lovisa[36,157], Paolo Bonvini[36,121], Sarah Bowdin[158], Stefano Gianni[159], Elide Mantuano[160], Velia Minicozzi[161], Leonore Novak[159], Alessandra Pasquo[162], Annalisa Pastore[163], Maria Petrosino[164-165], Rita Puglisi[42], Angelo Toto[159], Liana Veneziano[160], Roberta Chiaraluce[164], Mad P. Ball[139,166], Jason R. Bobe[139,167], George M. Church[139], Valerio Consalvi[159], Matthew Mort[103,106], David N. Cooper[106], Bethany A. Buckley[126], Molly B. Sheridan[11], Garry R. Cutting[168], Maria C. Scaini[169], Kamil J. Cygan[110,170], Alger M. Fredericks[170], David T. Glidden[170], Christopher Neil[170-171], Christy L. Rhine[170-171], William G. Fairbrother[170], Aileen Y. Alontaga[172], Aron W. Fenton[172], Kenneth A. Matreyek[101,173], Lea M. Starita[101], Douglas M. Fowler[101], Britt-Sabina Löscher[174], Andre Franke[175], Scott I. Adamson[176], Brenton R. Graveley[176], Joe W. Gray[177], John P. Kane[85], Maria Kousi[178], Nicholas Katsanis[179-180], Max Schubach[122], Martin Kircher[122], Nina Gonzaludo[85-86], Angel C.Y. Mak[85,87], Paul L.F. Tang[85,181], Pui-Yan Kwok[85], Wyatt T. Clark[182], Guoying K. Yu[182-183], Jonathan H. LeBowitz[182], Francesco Benedicenti[184], Elisa Bettella[36], Stefania Bigoni[185], Federica Cesca[36], Isabella Mammi[186], Cristina Marino-Buslje[187], Donatella Milani[188], Angela Peron[189-191], Roberta Polli[36], Stefano Sartori[36,157], Franco Stanzial[184], Irene Toldo[36], Licia Turolla[192], Maria C. Aspromonte[36], Mariagrazia Bellini[36], Emanuela Leonardi[36], Xiaoming Liu[193-194], Christian Marshall[80], Richard McCombie[195], Lisa Elefanti[169], Chiara Menin[169], M. Stephen Meyn[77,196], Alessandra Murgia[36], Kari C.Y. Nadeau[50], Lipika R. Pal[4], John Moult[4], Susan L. Neuhausen[197], Robert L. Nussbaum[126,198], Mehdi Pirooznia[73,199], James B. Potash[168], Dago F. Dimster-Denk[2], Jasper D. Rine[2], Jeremy R. Sanford[25], Michael Snyder[50], Sean V. Tavtigian[59], Atina G. Cote[80,200], Song Sun[80,200], Marta W. Verby[80,200], Jochen Weile[80,200], Frederick P. Roth[80,200], Ryan Tewhey[201], Pardis C. Sabeti[7], Joan Campagna[85], Marwan M. Refaat[85,202], Julianne Wojciak[85], Soren Grubb[203], Jay Shendure[101], Amanda B. Spurdle[204], Dimitri J. Stavropoulos[80], Nephi A. Walton[205-206], Peter P. Zandi[168], Elad Ziv[85]

***Ethics forum***

Wylie Burke[101], Flavia Chen[85,139], Lawrence R. Carr[123], Selena Martinez[123], Jodi Paik[123], Julie Harris-Wai[85], Mark Yarborough[207], Stephanie M. Fullerton[208], Barbara A. Koenig[85]

***Assessors***

Roxana Daneshjou[50], Gregory McInnes[50,209], Russ B. Altman[50], Dustin Shigaki[11], Michael A. Beer[11], Aashish Adhikari[2,18], John-Marc Chandonia[2,65], Mabel Furutsuki[2], Zhiqiang Hu[2], Laura Kasak[2,210], Changhua Yu[2,211], Vikas Pejaver[101,107], Yana Bromberg[21], Castrense Savojardo[23], Rui Chen[50,88], Melissa S. Cline[25], Qifang Xu[32], Roland L. Dunbrack[32], Iddo Friedberg[212], Gad A. Getz[7,138,213], Qian Cong[214], Lisa N. Kinch[214], Jing Zhang[214], Nick V. Grishin[214], Alin Voskanian[215], Maricel G. Kann[215], Wyatt T. Clark[182], Elizabeth Tran[82], Nilah M. Ioannidis[2], Maria C. Aspromonte[36], Mariagrazia Bellini[36], Emanuela Leonardi[36], Jesse M. Hunter[77,216], Rupa Udani[77,217], M. Stephen Meyn[77,196], Binghuang Cai[101], Sean D. Mooney[101], Alexander A.

Morgan[50,218], Lipika R. Pal[4], John Moult[4], Stephen M. Mount[4], Alessandra Murgia[36], Robert L. Nussbaum[126,198], Jeremy R. Sanford[25], Artem Sokolov[25,138], Joshua M. Stuart[25], Shamil Sunyaev[139], Sean V. Tavtigian[59], Marco Carraro[36], Manuel Giollo[36,127], Giovanni Minervini[36], Alexander M. Monzon[36], Silvio E. Tosatto[36], Anat Kreimer[2,21], Nir Yosef[2]

***Advisory board and scientific council***
Russ B. Altman[50], Serafim Batzoglou[18,219], Yana Bromberg[21], Atul J. Butte[50,85], George M. Church[139], Garry R. Cutting[168], Laura Elnitski[35], Marc S. Greenblatt[220], Reece K. Hart[221], Ryan Hernandez[79,85], Tim J.P. Hubbard[42], Scott Kahn[18,222], Rachel Karchin[11], Anne O'Donnell-Luria[7], M. Stephen Meyn[77,196], Sean D. Mooney[101], Alexander A. Morgan[50,218], Pauline C. Ng[52], Robert L. Nussbaum[126,198], John Shon[18,223], Michael Snyder[50], Shamil Sunyaev[139], Sean V. Tavtigian[59], Scott Topper[99], Joris Veltman[224], Justin M. Zook[225]

***CAGI chairs***
John Moult[4], Steven E. Brenner[2]

1. Northeastern University, Boston, Massachusetts, USA; 2. University of California, Berkeley, California, USA; 3. Indiana University, Bloomington, Indiana, USA; 4. University of Maryland, College Park, Maryland, USA; 5. Tata Consultancy Services, Hyderabad, India; 6. Dana-Farber Cancer Institute, Boston, Massachusetts, USA; 7. Broad Institute of MIT and Harvard, Cambridge, Massachusetts, USA; 8. ICF International, Cambridge, Massachusetts, USA; 9. University of Tennessee, Knoxville, Tennessee, USA; 10. Sabanci University, Tuzla, Turkey; 11. Johns Hopkins University, Baltimore, Maryland, USA; 12. Intel Corporation, Santa Clara, California, USA; 13. Massachusetts Institute of Technology, Cambridge, Massachusetts, USA; 14. Encoded Genomics, South San Francisco, California, USA; 15. Institute for Systems Biology, Seattle, Washington, USA; 16. Earle A. Chiles Research Institute, Providence Health & Services, Portland, Oregon, USA; 17. University of Michigan, Ann Arbor, Michigan, USA; 18. Illumina, San Diego, California, USA; 19. Fudan University, Shanghai, China; 20. University of Pennsylvania, Philadelphia, Pennsylvania, USA; 21. Rutgers University, New Brunswick, New Jersey, USA; 22. Genentech, South San Francisco, California, USA; 23. University of Bologna, Bologna, Italy; 24. University of California at San Diego, San Diego, California, USA; 25. University of California at Santa Cruz, Santa Cruz, California, USA; 26. Katholieke Universiteit Leuven, Leuven, Belgium; 27. Vall d'Hebron Institute of Oncology, Barcelona, Spain; 28. Germans Trias i Pujol Hospital, Badalona, Spain; 29. University Hospital of Vall d'Hebron, Barcelona, Spain; 30. Université Laval, Québec, Québec, Canada; 31. Vall d'Hebron Institute of Research, Barcelona, Spain; 32. Fox Chase Cancer Center, Philadelphia, Pennsylvania, USA; 33. Mercer University School of Medicine, Macon, Georgia, USA; 34. Microsoft, Redmond, Washington, USA; 35. National Human Genome Research Institute, Bethesda, Maryland, USA; 36. University of Padova, Padova, Italy; 37. University of Turin, Turin, Italy; 38. Vavilov Institute of General Genetics, Moscow, Russia; 39. Institute of Protein Research, Pushchino, Russia; 40. Lomonosov Moscow State University, Moscow, Russia; 41. University of Cambridge, Cambridge, UK; 42. King's College London, London, UK; 43. Universidad Nacional de Quilmes, Bernal, Argentina; 44. European Molecular Biology Laboratory - European Bioinformatics Institute, Hinxton, UK; 45. Technical University of Munich, Munich, Germany; 46. University of California at Irvine, Irvine, California, USA; 47. Peking University, Beijing, China; 48. Verily Life Sciences, South San Francisco, California, USA; 49. CAMP4

Therapeutics, Cambridge, Massachusetts, USA; 50. Stanford University, Stanford, California, USA; 51. Insitro, South San Francisco, California, USA; 52. Genome Institute of Singapore, Singapore; 53. University of Bristol, Bristol, UK; 54. AstraZeneca, Cambridge, UK; 55. MRC Laboratory of Molecular Biology, Cambridge, UK; 56. Amazon Web Services, Seattle, Washington, USA; 57. Wellcome Sanger Institute, Hinxton, UK; 58. ELIXIR, Hinxton, UK; 59. University of Utah, Salt Lake City, Utah, USA; 60. University of Texas MD Anderson Cancer Center, Houston, Texas, USA; 61. National Institute of Advanced Industrial Science and Technology, Tokyo, Japan; 62. Fujitsu Ltd., Tokyo, Japan; 63. Varinos Inc., Tokyo, Japan; 64. Tokai University School of Medicine, Tokyo, Japan; 65. Lawrence Berkeley National Laboratory, Berkeley, California, USA; 66. Queen Mary University, London, UK; 67. Korea Advanced Institute of Science and Technology, Daejeon, South Korea; 68. Korea Institute of Science and Technology Information, Daejeon, South Korea; 69. University College London, London, UK; 70. University of Hong Kong, Hong Kong; 71. Tianjin Medical University, Tianjin, China; 72. Williams College, Williamstown, Massachusetts, USA; 73. Johnson & Johnson, New Brunswick, New Jersey, USA; 74. PierianDx, Creve Coeur, Missouri, USA; 75. Harvard T.H. Chan School of Public Health, Boston, Massachusetts, USA; 76. Washington University School of Medicine, St. Louis, Missouri, USA; 77. University of Wisconsin-Madison, Madison, Wisconsin, USA; 78. University of Texas at Dallas, Richardson, Texas, USA; 79. McGill University, Montreal, Canada; 80. University of Toronto, Toronto, Ontario, Canada; 81. Cyclica Inc., Toronto, Ontario, Canada; 82. Stanford University School of Medicine, Stanford, California, USA; 83. Dashworks, Boston, Massachusetts, USA; 84. Krembil Centre for Neuroinformatics, Centre for Addiction and Mental Health, Toronto, Ontario, Canada; 85. University of California at San Francisco, San Francisco, California, USA; 86. Pacific Biosciences, Menlo Park, California, USA; 87. CytomX Therapeutics, South San Francisco, California, USA; 88. Personalis, Menlo Park, California, USA; 89. Bristol Myers Squibb, Redwood City, California, USA; 90. HypaHub, Sunnyvale, California, USA; 91. LifeMap Sciences Inc., Alameda, California, USA; 92. Weizmann Institute, Rehovot, Israel; 93. Yonsei University, Seoul, South Korea; 94. Baylor College of Medicine, Houston, Texas, USA; 95. Calm, San Francisco, California, USA; 96. University of Florida, Gainesville, Florida, USA; 97. University of Pavia, Pavia, Italy; 98. enGenome, Pavia, Italy; 99. Color Genomics, Burlingame, California, USA; 100. Syapse, San Francisco, California, USA; 101. University of Washington, Seattle, Washington, USA; 102. Sage Bionetworks, Seattle, Washington, USA; 103. Buck Institute for Research on Aging, Novato, California, USA; 104. North Carolina State University, Raleigh, North Carolina, USA; 105. Gilead, Foster City, California, USA; 106. Cardiff University, Cardiff, UK; 107. Icahn School of Medicine at Mount Sinai, New York City, New York, USA; 108. Burapha University, Chonburi, Thailand; 109. Google LLC, Mountain View, California, USA; 110. Regeneron, Tarrytown, New York, USA; 111. Moderna, Cambridge, Massachusetts, USA; 112. DNANexus, Mountain View, California, USA; 113. EarlyDiagnostics, Los Angeles, California, USA; 114. Helmholtz Zentrum Muenchen, Neuherberg, Germany; 115. University of Tokyo, Tokyo, Japan; 116. Bar-Ilan University, Tel Aviv, Israel; 117. Imperva, San Mateo, California, USA; 118. Indian Institute of Science, Bengaluru, India; 119. University of Oxford, Oxford, UK; 120. Blue Prism, Warrington, UK; 121. University of Western Ontario, London, Ontario, Canada; 122. Charité - Universitätsmedizin Berlin, Berlin, Germany; 123. No affiliation; 124. Seoul National University, Seoul, South Korea; 125. Qiagen, Germantown, Maryland, USA; 126. Invitae, San Francisco, California, USA; 127. Amazon, Seattle, Washington, USA; 128. Texas A&M

University, College Station, Texas, USA; 129. Carnegie Mellon University, Pittsburgh, Pennsylvania, USA; 130. IndieBio, San Francisco, USA; 131. iAccelerate, North Wollongong, Australia; 132. University of North Carolina at Charlotte, Charlotte, North Carolina, USA; 133. Vindara, Orlando, Florida, USA; 134. Temple University, Philadelphia, Pennsylvania, USA; 135. University of Greifswald, Greifswald, Germany; 136. Scripps Research Institute, San Diego, California, USA; 137. Brigham and Women's Hospital, Boston, Massachusetts, USA; 138. Harvard Medical School, Boston, Massachusetts, USA; 139. Harvard University, Cambridge, Massachusetts, USA; 140. Hebrew University of Jerusalem, Jerusalem, Israel; 141. Lund University, Lund, Sweden; 142. University of Tampere, Tampere, Finland; 143. Soochow University, Suzhou, China; 144. Chinese University of Hong Kong, Hong Kong; 145. Hong Kong University of Science and Technology, Hong Kong; 146. George Mason University, Virginia, USA; 147. Sun Yat-sen University, Guangzhou, China; 148. Indiana University-Purdue University Indianapolis, Indianapolis, Indiana, USA; 149. Griffith University, Queensland, Australia; 150. Sun Yat-sen Memorial Hospital, Guangzhou, China; 151. Imperial College, London, UK; 152. Vertex Pharmaceuticals, Boston, Massachusetts, USA; 153. Jilin University, Changchun, China; 154. CeMM Research Center for Molecular Medicine of the Austrian Academy of Sciences, Vienna, Austria; 155. Shenzhen Bay Laboratory, Shenzhen, China; 156. Kyoto University, Kyoto, Japan; 157. Paediatric Research Institute Città della Speranza, Padova, Italy; 158. Addenbrookes Hospital, University of Cambridge, Cambridge, UK; 159. Sapienza University of Rome, Rome, Italy; 160. Institute of Translational Pharmacology CNR, Rome, Italy; 161. University of Rome Tor Vergata, Rome, Italy; 162. ENEA - Frascati Research Centre, Rome, Italy; 163. Crick Institute, London, UK; 164. University of Rome, Rome, Italy; 165. University of Fribourg, Fribourg, Switzerland; 166. Open Humans Foundation, Sanford, North Carolina, USA; 167. PersonalGenomes.org, Sanford, North Carolina, USA; 168. Johns Hopkins University School of Medicine, Baltimore, Maryland, USA; 169. Veneto Institute of Oncology, San Giovanni Rotondo, Italy; 170. Brown University, Providence, Rhode Island, USA; 171. Remix Therapeutics, Cambridge, Massachusetts, USA; 172. University of Kansas Medical Center, Kansas City, Kansas, USA; 173. Case Western Reserve University, Cleveland, Ohio, USA; 174. Kiel University, Kiel, Germany; 175. Christian-Albrechts-University of Kiel, Kiel, Germany; 176. University of Connecticut, Mansfield, Connecticut, USA; 177. Oregon Health and Science University, Portland, Oregon, USA; 178. Third Rock Ventures, Boston, Massachusetts, USA; 179. Northwestern University, Evanston, Illinois, USA; 180. Rescindo Therapeutics Inc., Durham, North Carolina, USA; 181. AccuraGen Inc., Menlo Park, California, USA; 182. BioMarin, Novato, California, USA; 183. Global Blood Therapeutics, South San Francisco, California, USA; 184. Regional Hospital of Bolzano, Bolzano, Italy; 185. Ferrara University Hospital, Ferrara, Italy; 186. Mirano Hospital, Venice, Italy; 187. Fundacion Instituto Leloir, Buenos Aires, Argentina; 188. Milan Polyclinic, Milan, Italy; 189. University of Milan, Milan, Italy; 190. San Paolo Hospital, ASST Santi Paolo e Carlo, Milan, Italy; 191. University of Utah School of Medicine, Salt Lake City, Utah, USA; 192. AULSS 2 Marca Trevigiana, Treviso, Italy; 193. School of Public Health, University of Texas, Dallas, Texas, USA; 194. University of South Florida, Tampa, Florida, USA; 195. Cold Spring Harbor Laboratory, Cold Spring Harbor, New York, USA; 196. Hospital for Sick Children, Toronto, Ontario, Canada; 197. Beckman Research Institute of City of Hope, Duarte, California, USA; 198. Myriad, Salt Lake City, Utah, USA; 199. National Institutes of Health, Bethesda, Maryland, USA; 200. Mount Sinai Health System, New York City, New York, USA; 201. Jackson Laboratory, Bar Harbor, Maine, USA; 202. American University of Beirut Medical Center, Beirut, Lebanon; 203. University of Copenhagen, Copenhagen, Denmark; 204. QIMR Berghofer Medical Research Institute, Brisbane, Australia; 205. Geisinger Genomic Medicine Institute, Danville, Pennsylvania, USA; 206. Intermountain Healthcare Precision Genomics, St. George, Utah, USA; 207. University of California at Davis, Davis, California, USA; 208. University of Washington School of Medicine, Seattle, Washington, USA; 209. Empirico Inc., San Diego, California, USA; 210. University of Tartu, Tartu, Estonia; 211. California Institute of Technology, Pasadena, California, USA; 212. Iowa State University, Ames, Iowa, USA; 213. Massachusetts General Hospital, Boston, Massachusetts, USA; 214. University of Texas Southwestern Medical Center, Dallas, Texas, USA; 215. University of Maryland, Baltimore County, Baltimore, Maryland, USA; 216. Nationwide Children's Hospital, Columbus, Ohio, USA; 217. Sema4, Stamford, Connecticut, USA; 218. Khosla Ventures, Menlo Park, California, USA; 219. Seer.bio, Redwood City, California, USA; 220. University of Vermont, Burlington, Vermont, USA; 221. MyOme Inc, Palo Alto, California, USA; 222. LunaPBC, San Diego, California, USA; 223. SerImmune, Goleta, California, USA; 224. Newcastle University, Newcastle upon Tyne, UK; 225. National Institute of Standards and Technology, Portland, Oregon, USA

## CONFLICT OF INTEREST

Principal authors of this paper participated as predictors in many of the CAGI challenges reported. The unified numerical framework employed for reanalysis of the challenges yields results that are consistent with those obtained by the independent assessors of each challenge and in particular selected methods are the highest ranked in the independent assessments. Nevertheless, while every care was taken to mitigate any potential biases in this work, the authors' participation in CAGI may have affected the presentation of findings, including the selection of challenges, metrics, assessment criteria and emphasis given on particular results.

VBG is a current employee and shareholder of AstraZeneca; RB is a shareholder of enGenome; AJB is a co-founder and consultant to Personalis and NuMedii as well as a consultant to Samsung, Mango Tree Corporation and in the recent past, 10x Genomics, Helix and Pathway; Carles Corbi-Verge is a computational scientist at the drug discovery company; Cyclica INC and is compensated with income and equity; KC is one of the Regeneron authors and owns options and/or stock of the company; DD is Chief Scientist at Geneyx Genomex Ltd; CD is a consultant to Exact Sciences and is compensated with income and equity; GAG receives research funds from IBM and Pharmacyclics, and is an inventor on patent applications related to MSMuTect, MSMutSig, MSIDetect, POLYSOLVER and SignatureAnalyzer-GPU, and is a founder, consultant and holds privately held equity in Scorpion Therapeutics; NG is an employee and stockholder at Pacific Biosciences; RH is a paid consultant for Invitae and Scientific Advisory Board member for Variant Bio; AK is a consultant at Illumina Inc., Scientific Advisory Board member of OpenTargets; KK is one of the Regeneron authors and owns options and/or stock of the company; IL is an employer and stockholder of enGenome; MSM owns stock in PhenoTips; GN is an employee of enGenome; AOD-L is a member of the Scientific Advisory Board of Congenica; ER is a shareholder of enGenome; PKR is the founder of CytoGnomix; FPR is a shareholder in  Ranomics and SeqWell, an advisor for SeqWell, BioSymetrics, and Constantiam BioSciences, and has received research sponsorships from Biogen, Alnylam, Deep Genomics and Beam Therapeutics; PCS is the co-founder and shareholder of Sherlock

Biosciences, a board member and shareholder of Danaher Corporation, and has filed patents related to this work; PLFT is an employer and stockholder in AccuraGen; RT has filed patents related to this work; MHW is a shareholder of Beth Bioinformatics Co., Ltd.; JZ is an employee of AstraZeneca. SEB receives support at the University of California, Berkeley from a research agreement from TCS.

## ACKNOWLEDGMENTS

The authors are grateful for the contributions of Talal Amin, Patricia Babbitt, Eran Bachar, Stefania Boni, Kirstine Calloe, Ombretta Carlet, Shann-Ching Chen, Chien-Yuan Chen, Jun Cheng, Luigi Chiricosta, Alex Colavin, Qian Cong, Emma D'Andrea, Carla Davis, Xin Feng, Carlo Ferrari, Yao Fu, Alessandra Gasparini, David Goldgar, Solomon Grant, Steve Grossman, Todd Holyoak, Rick Lathrop, Xiaolin Li, Quewang Liu, Mary Malloy, Beth Martin, Zev Medoff, Nasim Monfared, Susanna Negrin, Eric Odell, Michael Parsons, Nathan Pearson, Alexandra Piryatinska, Catherine Plotts, Jennifer Poitras, Clive Pulinger, Francesco Reggiani, Melvin M. Scheinman, Nicole Schmitt, George Shackelford, Vasily Sitnik, Fiorenza Soli, Qingling Tang, Nancy Mutsaers Thomsen, Jing Wang, Chenling Xiong, Lijing Xu, Shuhan Yang, Lijun Zhan, and Huiying Zhao.

## FUNDING

The content of this work is solely the responsibility of the authors and does not necessarily represent the official views of the funding agencies. The CAGI experiments and conferences were supported by National Institutes of Health (NIH) awards U41HG007346, U24HG007346, and R13HG006650 to SEB, as well as a Research Agreement with Tata Consultancy Services (TCS) to SEB, and a supplement to NIH U19HD077627 to RLN. NIH U41HG007346 supported ANA, GA, CB, SEB, J-MC, RAH, ZH, LK, BAK, MKL, ZM, JM, AJN, RCP, YW, SFY; NIH U24HG003467 supported CB, SEB, J-MC, ZH, SMF, RCP, PR, SJ; the TCS research agreement supported CAGI activities of ANA, CB, DB, J-MC, ZH, LK, and SFY; NIH U19HD077627 supported RAH and SEB; NIH R13HG006650 provided CAGI travel support to SA, ANA, GA, JRA, CB, DB, MAB, BB, SEB, BAB, YC, EC, MC, HC, J-MC, JC, MSC, RD, DD, RLD, MDE, BF, AWF, IF, SMF, MF, NG, MSG, NVG, JJH, RH, ZH, CLVH, TJPH, SK, RK, MK, PK, DK, BAK, AK, RHL, MKL, DM, GM, MSM, SDM, AAM, JM, SMM, AJN, AO, KAP, LRP, VRP, PR, SR, GS, AS, JMS, SS, RST, CT, AV, MEW, RW, SJW, ZY, SFY, JZ, MZ; NIH U41HG007346 provided travel support to RK, ZM, SDM; UC Berkeley funds additionally provided CAGI travel support for YB, LS. CAGI projects and participants were further supported as follows: OA by an EMBO Installation Grant (No: 4163), TÜBİTAK (Grant IDs: 118C320, 121E365), TÜSEB (Grant ID: 4587); IA by NIH R01GM078598, R35GM127131, R01HG010372; RBA by NIH GM102365; APA by the Office of Science, Office of Biological and Environmental Research of the US Department of Energy, under contract DE-AC02-05CH11231; ZA, MHÇ, JC, JG and TYDN by a Competence Network for Technical, Scientific High Performance Computing in Bavaria KONWIHR, a Deutsche Forschungsgemeinschaft fellowship through the Graduate School of Quantitative Biosciences Munich and an NVIDIA hardware grant; MAB, AP, DS by NIH U01HG009380; SB by the Asociación Española contra el

Cáncer; PB, FL by the Fondazione CARIPARO, Padova, Italy; APB, SD by NIH U24HG009293; WB by the NIH and the Greenwall Foundation; AJB by the Barbara and Gerson Bakar Foundation, and Priscilla Chan and Mark Zuckerberg; KC, SG, NS, NMT by the Danish National Research Foundation Centre for Cardiac Arrhythmia; EC, PT by MIUR 201744NR8S; RC, VC by the 201744NR8S/Ministero Istruzione, Università e Ricerca, PRIN project; MSC by NIH U01CA242954; MM, DNC by Qiagen through a License Agreement with Cardiff University; RD by a Paul and Daisy Soros Fellowship; XdlC by Spanish Ministerio de Ciencia e Innovación (PID2019-111217RB-I00) and Ministerio de Economía y Competitividad (SAF2016-80255-R and BIO2012-40133) and a European Regional Development Fund (Pirepred-EFA086/15); MD by NIH U41HG007234; OD by FIS PI12/02585 and PI15/00355 from the Spanish Instituto de Salud Carlos III (ISCIII) funding, an initiative of the Spanish Ministry of Economy and Innovation partially supported by European Regional Development FEDER Funds; MDE, DKG, YG, KT, HZ by NIH R01HG008363, U01HG007037, U41HG007346, R13HG006650; LE, VG, MSG by a grant from the Intramural Research Program of the NHGRI to LE (1ZIAHG200323-14); WGF by NIH R01GM127472; PF by MIUR 201744NR8S and 20182022D15D18000410001; MSF, MH, GP by the Comision Nacional de Investigaciones Cientificas y Tecnicas (CONICET) [Grant ID: PIP 112201101–01002] ; Universidad Nacional de Quilmes [Grant ID: 1402/15]; DMF by NIH R01GM109110, R01HG010461; AF, JH, JMM by Wellcome Trust grant [098051]; and NIH U54HG004555; AG, DP by intramural funding; NVG by NIH GM127390 and the Welch Foundation I-1505; SG-E by PI13/01711 and PI16/01218 to SGE from the Spanish Instituto de Salud Carlos III (ISCIII) funding, an initiative of the Spanish Ministry of Economy and Innovation partially supported by European Regional Development FEDER Funds. SGE was also supported by the Miguel Servet Progam [CP16/00034]; TI by JSPS KAKENHI Grant Number 16HP8044; JOBJ, CJM by NIH R24OD011883; SK by NIH U01HG007019, RO1HG003747; MK by NIH AG054012, AG058002, AG062377, NS110453, NS115064, AG062335, AG074003, NS127187, AG067151, MH109978, MH119509, HG008155, DA053631; IVK by RSF 20-74-10075; CL, JVdA by Color Genomics; KAM by NIH R35GM142886; GM by NIH T32LM012409; JM by NIH R01GM120364, R01GM104436; LO, SÖ, NP, CR by the Spanish Ministerio de Ciencia e Innovación (PID2019-111217RB-I00) and Ministerio de Economía y Competitividad (SAF2016-80255-R and BIO2012-40133); European Regional Development Fund (Pirepred-EFA086/15); VP by NIH K99LM012992; SDM and PR by R01LM009722 and R01MH105524; PR by Precision Health Initiative of Indiana University; AR by NIH T32HG002536; SR by a Marie Curie International Outgoing Fellowship PIOF-GA-2009-237751; PKR by Natural Sciences and Engineering Research Council of Canada 371758-2009, Canadian Breast Cancer Foundation, Canada Foundation For Innovation, Canada Research Chairs, Compute Canada and Western University; FR, JVD, JS by VIB and KU Leuven; FPR by One Brave Idea Initiative, the NIH HG004233, HG010461, the Canada Excellence Research Chairs, and a Canadian Institutes of Health Research Foundation Grant; PCS by NIH UM1HG009435; JRS by NIH R35GM130361; CS by MUR PRIN2017 2017483NH8_002; CS by NRF-2020M3A9G7103933; YS by NIH R35GM124952; LMS by NIH RM1HG010461; RT by NIH 1UM1HG009435; JMS by NCI R01CA180778; SVT by NCI R01 CA121245; MV by Finnish Academy, Swedish Research Council, Swedish Cancer Society; MEW by NSF GRF, NIH GM068763; MHW by the National Natural Science Foundation of China (NSFC) [31871340, 71974165]; FZ by The Senior and Junior Technological Innovation Team (20210509055RQ), the Jilin Provincial Key Laboratory of Big Data Intelligent Computing (20180622002JC), and the Fundamental Research Funds for the Central Universities, JLU; YZ by the Australia Research Council [DP210101875]; EZ by NCI K24CA169004, California Initiative to Advance Precision Medicine.

## SUPPLEMENTARY MATERIALS

Supplementary materials are available at https://github.com/genomeinterpretation/CAGI50 and the CAGI web site https://genomeinterpretation.org/.

## DATA AND CODE AVILABILITY

CAGI challenge datasets, predictions, and assessment standards are available on the CAGI website for public access or under controlled access for registered CAGI participants, unless ethical concerns about sensitive human data preclude such sharing. Where possible, such sensitive human datasets are available via repositories such as dbGaP and European Genome-Phenome Archive. The code used for the analysis of this work is available at the GitHub repository https://github.com/genomeinterpretation/CAGI50 and the CAGI web site https://genomeinterpretation.org/.

# Supplementary Materials

## Table of contents

## List of figures

## Unified Analysis Framework

We analyzed a number of challenges in order to provide a unified framework when presenting results. These experiments were performed on all biochemical effect challenges that we considered to be high-quality challenges, the Annotate All Missense challenge, the high-throughput splicing challenges, the eQTL challenges, and the third Crohn's challenge. Here we summarize all data processing pipelines and evaluation approaches. First, we give a high-level description of the commonalities and differences between different datasets and their evaluation.

The datasets were evaluated on regression and/or binary classification type tasks. The goal of the regression type task is to measure the performance of a method on predicting a continuous measurement (functional activity, growth, etc.) made in the experiment that directly or indirectly measures a quantity of interest. For the purpose of the description that follows we will refer such continuous measurements as *experimental values*. Scatter plots and measures of R-squared, RMSE, Pearson's correlation, Spearman's correlation and Kendall's $\tau$ were used for evaluating the prediction of experimental values; see Methods. In many challenges the data was generated with experimental replicates that allowed recording the standard deviation of the experimental values. When available, the standard deviation was used to generate the predictions of a positive control that gives an upper limit on the method performance given the experimental variability. We refer to the positive control as the *Experimental-Max* (sec. Implementation Details).

In many challenges, the experimental values were further used to define classes for the classification task. This was based on either thresholding the experimental values using reasonable thresholds or a combination of threshold, confidence or statistical significance used by previous assessors. Although some of the challenges originally defined more than one class from the experimental values, in such cases we either removed or merged one or more classes to create a binary classification dataset. By convention, the numeric values 1 and 0 are used to represent the positive and negative class, respectively. For some non-coding challenges, we created two binary classification datasets; e.g., one where over-splicing is used as the positive class and the other where under-splicing is used as the positive class. ROC curve, AUC and local $\mathrm{lr}^+$ curve were used for evaluating classification tasks. Additionally, for many challenges we also provide a clinically relevant classification analysis with Truncated AUC, log-log AUC (see Implementation Details), along with the corresponding ROC curves, local posterior probability of pathogenicity ($\rho$) curve and measurements PPP, TPR, FPR, $\mathrm{LR}^+$ (global), $\mathrm{LR}^-$, DOR, MCC and PPV at clinically relevant evidence (Supporting, Moderate and Strong) thresholds; see Methods. Additionally, for complex traits dataset on Crohn's disease, we also provide a relative risk (RR) curve.

In many challenges (mostly functional challenges), the predictions for the experimental values also serve as the continuous method score for binary classification. In some of these cases, where a low experimental value corresponds to the positive class, the predictions are negated (multiplied by $-1$) to create the method score. This ensures that a high method score corresponds to the positive class. This transformation is necessary to correctly compute the classification measures. However, note that when the results are presented in figures (posterior probability and $\mathrm{lr}^+$ curves) the unnegated values are used for consistency with the scatter plots. As a side-effect of this presentation, in some figures the evidence thresholds represent a region towards its left, whereas in other cases it represents a region towards its right depending on whether the predicted experimental values required negation or not, respectively. As a rule of thumb, for a method that captures some signal for the classification task, the evidence thresholds represent the region towards the direction of increasing $lr^+$ or posterior curve. An Experimental-Max baseline for the classification measures is also reported if the standard deviation of the experimental values is available.

In some challenges (most expression and splicing challenges) the participants were asked to submit a method score for the classification task in addition to the prediction for the experimental values. In such cases the method score for the classification is used for the classification related analysis and experimental value predictions are used in the regression related analysis.

The CAGI cancer challenges and the Annotate All Missense dataset analyzed in this paper use curated ground truth variants; that is, they do not fall in the framework of running experimental assays and, consequently, do not have an experimental value to be predicted. Only classification-related analysis is provided for these datasets.

Some biochemical effect challenges, in spite of measuring an experimental value, only required participants to submit discrete labels for the classification task. Here we only report the classification related analysis. The predicted binary class labels are converted to a numeric method score: 1 for a positive prediction and 0 for a negative prediction. Since the resultant method score is binary and discrete, local $lr^+$ and posterior probability of pathogenicity are only defined at the two values taken by the method score, instead of being a continuous curve. Furthermore, the local $lr^+$ and the global $LR^+$ coincide in this case.

# Analyzed challenges

## Coding Challenges

### NAGLU

The NAGLU dataset contains missense variants responsible for the production of N-acetyl-glucosaminidase (NAGLU, NP_000254.2). Deficiency of NAGLU is indicated in neurodegenerative diseases Mucopolysaccharidosis IIIB (MPS IIIB) or Sanfilippo B disease (OMIM #252920). BioMarin functionally assessed the enzymatic activity of 165 novel missense mutations in the ExAC dataset by transfecting plasmids containing cDNAs encoding each of the mutant proteins into HEK293 cells. The activity levels were normalized to represent the fraction of wildtype NAGLU activity.[1] A value of 0 represents no activity, 1 represents wildtype-level of activity and >1 represents an activity greater than the wildtype activity. For example, 0.7 means 70% of wildtype activity and 1.3 means 130% of wildtype activity. Each mutant was assayed in at least three independent transfection experiments. The results from these three determinations were averaged to give the mean activity and the standard deviation was also calculated.

In the NAGLU challenge of CAGI4, participants were asked to submit predictions for the fractional enzymatic activity of the variants. A total of 10 teams participated in the NAGLU challenge, submitting predictions from 17 different models.

We first compared the predictions to experimental fractional activity in a regression type analysis. For a classification type analysis, each variant was assigned either "pathogenic" or "benign" label based on its experimental fractional activity. Variants having activity value below 0.15 were deemed "pathogenic" and the remaining variants were considered to be "benign". The threshold is consistent with that observed from previously identified pathogenic mutations as described by Clark et al,[1] and also used in the CAGI5 assessment paper for NAGLU.[2] Out of the 165 variants, the experimental values could not be measured for 2 variants. Out of the remaining 163 variants, 40 were assigned pathogenic label, giving an overall 25% of pathogenic variants (data prior). As stated in the main text, this appears to be a good estimate of the prior probability of pathogenicity in a diagnostic setting. For the clinical analysis, we used the data prior to compute the clinically relevant thresholds and all class-prior dependent measures. The results of the clinical analysis with the general diagnostic (0.1) and screening (0.01) priors are also reported. The predicted fractional activity was also used as the method score for classification and clinical analysis. Our evaluation for the NAGLU challenge is summarized in Figure 2 (Main text) and Figure 7 (Supplementary text) and Table 2 (Supplementary Excel file).

### PTEN and TPMT

The gene PTEN (Phosphatase and TEnsin Homolog) encodes for a protein that is an important secondary messenger molecule promoting cell growth and survival. Thiopurine S-methyl transferase (TPMT) is a key enzyme involved in the metabolism of thiopurine drugs. A library of 4002 PTEN and 3952 TPMT mutations was assessed to measure the stability of the variant protein using a multiplexed Variant Stability Profiling (VSP) assay. The VSP assay exploits a fluorescent reporter system to measure steady-state abundance of missense protein variants. Here, each cell expresses a protein variant fused to EGFP. The stability of the variant protein dictates the abundance of the fusion protein and thus the EGFP level of the cell. As a reporter of transcriptional abundance, mCherry is either co-transcriptionally or co-translationally expressed from the same construct. Cells are flow-sorted into bins according to their EGFP/mCherry ratio, and deep sequencing is used to quantify each variant's frequency in each bin. The EGFP/mCherry ratios of cells harboring each library spanned the range previously characterized for the wildtype (WT) and known destabilized variants. Finally, a stability score is calculated as the relative protein abundance based on the bin wise frequency. The relative protein abundance was computed, with 0 meaning unstable, 1 meaning wildtype stability, and >1 meaning more stable than the wildtype protein. The mean, standard deviation, upper and lower confidence intervals (CI) of the relative abundance was also recorded for each variant using replicates.

In the CAGI5 challenges for PTEN and TPMT, participants were asked to predict the relative protein abundance of the variants. A total of 8 teams participated in the challenge, submitting 16 different predictors.

In our analysis, we separately evaluated the performance on the PTEN and TPMT variants. For each gene, the predictions were compared against the experimental relative protein abundance means in a regression type analysis. Wildtype, synonymous and nonsense variants were removed from the dataset to limit our assessment to the missense variants only. Variants with mean relative abundance below 0, being outside the interpretable range, were also excluded from our analysis. For the classification type analysis, each variant was assigned either "pathogenic/destabilizing" or "benign/wildtype stabilizing" label based on its mean relative abundance. Variants having a value below 0.4 and between 0.4 (inclusive) and 1.2 were deemed "pathogenic" and "benign", respectively. The scores above 1.2 were interpreted as more stabilizing than the wildtype and were not considered in the classification analysis. The previous assessment paper for the two challenges, used a different threshold based on the lower CI, upper CI and the mean.[3] However, in our analysis, we use a simpler threshold based on the mean only. It is obtained after modeling the distributions of functional and nonfunctional variants using a multi-sample Gaussian mixture model (MSGMM) as proposed by Jain et al.[4] We further refer the reader to Figure 1 from Pejaver et al.[3] for the details of MSGMM modeling. Out of the total number of PTEN (TPMT) variants assayed, 3,716 (3,484) variants remained after removing the non-missense variants and those for which experimental values could not be measured. Out of these variants, 3,537 (3,228) variants made it to the classification set for PTEN (TPMT), having a proportion of 0.21 (0.12) pathogenic variants (data prior). For the clinical analysis, we used the data prior to compute the clinically relevant thresholds and all class-prior dependent measures. Although we have no evidence that the data prior probability reflects a population-based prior (as in the NAGLU challenge, where variants were selected from ExAC), this is still a useful quantity because the reference set of variants is essentially a set of all possible variants for the two genes. The results of the clinical analysis with the general diagnostic (0.1) and screening (0.01) priors are also reported. The predicted relative protein abundance was also used as a method score for the classification and clinical analysis. Our evaluation for the PTEN challenge is summarized in Figure 2 (Main text), Figure 7 (Supplementary text) and Table 2 (Supplementary Excel file); for TPMT in Figure 1 (Supplementary text) and Table 2 (Supplementary Excel file).

## GAA

Acid alpha-glucosidase (GAA) is a lysosomal enzyme involved in the breakdown of glycogen. Some mutations in GAA cause Pompe disease (Glycogen Storage Disease II), a rare autosomal recessive metabolic disorder. BioMarin has functionally assessed the enzymatic activity of the 356 rare and novel missense mutations in from ExAC with transfected cell lysates. The fractional enzyme activity of each mutant protein compared to the wildtype enzyme was recorded such that a score of 0 means no activity, 1 means a wildtype-level of activity and >1 means greater than wildtype activity. Each mutant was assayed in at least three independent transfection experiments. The results from these determinations were averaged, and the standard deviation of experimental read-outs was calculated.

In the CAGI5 challenge for GAA, participants were asked to submit predictions for the fractional enzymatic activity of the variants. A total of 8 teams participated in the challenge, submitting predictions for 26 different models. The predictions were compared against the experimentally determined activity values in a regression type analysis. The experimental activity values were divided by 100 to scale them between 0 and 1 for consistency with other challenges. The predictions were already scaled between 0 and 1. In the most severe case zero GAA activity is observed in the patient's fibroblasts; however, in less severe cases the activity is in the range of 25-30%. Taking a more conservative approach, we considered a variant "pathogenic" if its experimentally measured activity was less than 0.1, otherwise it was considered "benign". The choice of the threshold is consistent with one of the thresholds used in the previous assessment of a submitted method for GAA.[5] Based on the threshold of 0.1, 18% of the variants were labeled as pathogenic (data prior). For the clinical analysis, we used the data prior to compute the clinically relevant thresholds and all class-prior dependent measures. The results of the clinical analysis with the general diagnostic (0.1) and screening (0.01) priors are also reported. The predicted fractional activity was also used as the method score for the classification and clinical analysis. Our evaluation for the challenge is summarized in Figure (Supplementary text) and Table 2 (Supplementary Excel file).

## CBS

CBS is a vitamin-dependent enzyme involved in cysteine biosynthesis. The human CBS requires two cofactors for function, vitamin B6 and heme. Homocystinuria due to CBS deficiency (OMIM #236200) is a recessive inborn error of sulfur amino acid metabolism. More than 90 different disease-associated mutations have been identified in the CBS gene.[6] About one half of homocystinuric patients respond to high doses of pyridoxine and several alleles are clearly pyridoxine remediable: A114V, R266K, R366H, K384E, L539S and the frequent I278T which accounts for 20% of all CBS mutant alleles.

Jasper Rine's lab at UC Berkeley collected 51 synthetic single amino acid variants for the CAGI1 challenge and 84 variants that had been observed in patients with homocystinuria for the CAGI2 challenge. The functionality of the variants was tested in an *in vivo* yeast complementation assay. The level of mutant human CBS function is measured in terms of the yeast growth in the assay. The rates are normalized as a percentage relative to wildtype (human) growth with the same amount of exogenous pyridoxine supplementation, plus and minus the standard deviation. Two concentrations of pyridoxine, high (400 ng/ml) and low (2 ng/ml), were used. A value of 0 indicates no growth, whereas 1 indicates a wildtype growth rate. In both CAGI1 and CAGI2 challenges for CBS, participants were asked to submit predictions for the effect of the variants in the function of CBS both in high co-factor (pyridoxine) concentration (400ng/ml) and in low co-factor concentration (2ng/ml). A total of 19 different predictors were submitted by 12 teams for the CAGI1 challenge and the same number of predictors were submitted by 9 teams for the CAGI2 challenge.

For our analysis, we created four datasets, CBS1Low, CBS1High, CBS2Low and CBS2High, separating CAGI 1 and 2 and the two co-factor concentration levels. Each of the four datasets was analyzed separately. For the regression type analysis, we evaluated the predictions against the experimentally measured normalized growth rates. The experimental and predicted growth rates were divided by 100 to scale them between 0 and 1 for consistency with other challenges. For the classification type analysis, a variant was interpreted to have "no growth" (positive class) if its experimental normalized growth rate was below 0.2, otherwise it was interpreted have "growth detected" (negative class). The threshold of 0.2 was determined as the lowest average growth observed in remediation variants in the CAGI 1 CBS dataset.[7] Out of the 51 CAGI 1 CBS variants, 33% (39%) fraction were labeled "no growth" in presence of high (low) pyridoxine concentration. Out of the 84 CAGI 2 CBS variants, measurements could only be made on 78, out of which 51% (68%) were labeled "no growth" in presence of high (low) pyridoxine concentration. These percentages are the data priors for the four datasets. For the clinical analysis, we used the data prior to compute the clinically relevant thresholds and all class-prior dependent measures. The results of the clinical analysis with the general diagnostic (0.1) and screening (0.01) priors are also reported. The predicted relative growth rate was also used as the method score for the classification and clinical analysis. Our evaluation for the challenge is summarized in Figure 2A and 2B (Supplementary text) and Table 2 (Supplementary Excel file).

## SUMO ligase

The human genome encodes several small ubiquitin-like modifier proteins (SUMOs) that collectively 'tag' and modulate the functions of hundreds of proteins, including proteins implicated in cancer, neurodegeneration, and other diseases. As the only human SUMO-conjugating protein (SUMO E2 ligase), UBE2l is solely responsible for identifying target proteins and covalently attaching SUMO,[8] thereby serving a very important function.

The Roth Lab has generated a library of over 6,000 UBE2I clones. These clones collectively express nearly 2,000 unique amino acid changes in various combinations, including several single substitutions. They have also implemented a yeast-based complementation assay in which expression of human UBE2I in *S. cerevisiae* rescues a temperature-sensitive mutant version of yeast UBC9. A library expressing mutant human UBE2I clones in yeast is grown competitively and quantified via DNA barcode sequencing to assess the functional impact of individual UBE2I variants. The growth scores are normalized relative to the growth of the clone considered to be wildtype. A growth score of 0 means that the mutant clone was completely ineffective, whereas a score of 1 means that it was as effective as the wildtype clones. The final growth score was obtained as the mean of the technical replicates for each variant; standard deviation was also recorded.

In the CAGI4 challenge for the SUMO ligase, participants were asked to submit predictions for the effect of the variants on the competitive growth. To help participants calibrate their numeric values appropriately, the experimental distribution of numeric growth scores was also provided. The variants were divided in three subsets; Subset 1: the high-accuracy subset of 219 single amino acid variants for which at least three independent barcoded clones are represented, providing internal replicates of the experiment; Subset 2: the remaining 463 (of 682 total) single amino acid variants; Subset 3: the additional 4,427 alleles corresponding to clones containing two or more amino acid variants. Participants were allowed to submit predictions for multiple subsets. A total of 16 different predictors were submitted by 9 teams for subset 1, 13 predictors from 7 teams for subset 2 and 12 predictors from 6 teams for subset 3.

We ran separate analyses on subsets 1 (SUMO1), 2 (SUMO2) and 3 (SUMO3). For each set, the predictions were assessed against the experimentally determined normalized growth rates. The previous assessment paper for the SUMO challenge[9] labeled a variant as "deleterious", if its growth rate was below 0.3; "intermediate", if between 0.3 and 0.7; "wildtype", if between 0.7 and 1.3 and "advantageous", if above 1.3. In our assessment, we merged the "intermediate", "wildtype" and "advantageous" variants into a single class "non deleterious". In effect, if a variant had growth rate below 0.3, it was labeled "deleterious" (positive), otherwise it was labeled "non deleterious" (negative). Since the experimental growth rate could not be measured for many variants across the three subsets, the effective dataset size for the three subsets was 215 (subset 1), 410 (subset 2) and 3880 (subset 3). The proportion of "deleterious" variants was 41%, 48% and 67%, respectively (data prior). For the clinical analysis, we used the data prior to compute the clinically relevant thresholds and all class-prior dependent measures. The results of the clinical analysis with the general diagnostic (0.1) and screening (0.01) priors are also reported. The predicted growth rate was also used as method score for the classification and clinical analysis. Our evaluation for the challenge is summarized in Figure (Supplementary text) and Table 2 (Supplementary Excel file).

## CALM1

Calmodulin is a calcium-sensing protein that modulates the activity of a large number of proteins in the cell. It is involved in many different cellular processes and is especially important for neuron and muscle cell function. Variants that affect calmodulin function have been found to be causally associated with two cardiac arrhythmias.

A team in Fritz Roth's Lab at the University of Toronto and Lunenfeld Tanenbaum Research Institute (Sinai Health Systems), led by Jochen Weile and Song Sun, has assessed a large library of calmodulin variants using a high-throughput yeast complementation assay. The variants are assessed based on their ability to rescue a yeast strain carrying a temperature-sensitive allele of the yeast calmodulin orthologue CMD1.[10] A fitness score was computed for each variant as competitive growth rate on a log scale and then normalized relative to the wildtype and nonsense variant scores such that a score of 0 means no growth at a restrictive temperature, whereas a score of 1 means wildtype growth. Technical replicates were used to measure the standard deviations.

In the CAGI 5 challenge for CALM1, participants were asked to submit predictions for the competitive growth scores of 1,813 variants. To help participants calibrate their numeric values appropriately, the experimental distribution of numeric growth scores was also provided. A total of 7 different predictors were submitted by 4 teams. The predictions were compared against the experimental growth scores in a regression type analysis. In the previous assessment paper for CALM1, a variant was labeled as "deleterious", if its growth rate was below 0.3; "intermediate", if between 0.3 and 0.8; and "neutral", if above 0.8.[11] In our analysis, we dropped the "intermediate" variants to create a binary classification dataset. In effect, the variants having growth rate below 0.3 were labeled "deleterious" (positive) and those with growth rate above 0.8 were labeled “neutral” (negative), respectively. Based on this criteria, 1,284 variants, were selected in the classification set, 21% of which were labeled “deleterious” (data prior). For the clinical analysis, we used the data prior to compute the clinically relevant thresholds and all class-prior dependent measures. The results of the clinical analysis with the general diagnostic (0.1) and screening (0.01) priors are also reported. The predicted growth rate was also used as a method score for the classification and clinical analysis. Our evaluation for the challenge is summarized in Figure 1 (Supplementary text) and Table 2 (Supplementary Excel file).

## Frataxin

Frataxin is a highly conserved protein found in prokaryotes and eukaryotes that is required for efficient regulation of cellular iron homeostasis. Humans with a frataxin deficiency have the cardio- and neurodegenerative disorder Friedreich’s Ataxia. The role of frataxin in cancer is still ambiguous; studies have shown that frataxin protects tumor cells against oxidative stress and apoptosis, but also acts as a tumor suppressor.[12, 13]

A library of eight missense variants, selected from the Catalog of Somatic Mutations in Cancer (COSMIC) database, were assessed by near and far-UV circular dichroism and intrinsic fluorescence spectra to determine thermodynamic stability at different concentration of denaturant. These data were used to calculate a $\Delta\Delta GH_20$ value, the difference in unfolding free energy $\Delta\Delta GH_20$ between the variant and wildtype proteins for each variant measured in kcal/mol.

In the CAGI5 challenge for Frataxin, participants were asked to submit predictions for the $\Delta\Delta GH_20$ values of the 8 variants. A total of 10 different predictors were submitted by 6 teams. The predictions were compared against the experimental $\Delta\Delta GH_20$ values in a regression type analysis. For the classification type analysis, variants having an experimental $\Delta\Delta GH_20$ score $\leq -1.0$ kcal/mol were labeled “destabilizing” (positive class), whereas those having score $> -1.0$ were labeled “neutral/stabilizing” (negative class). The choice of the threshold is consistent with the previous assessment paper for the challenge.[14] Five out of the eight variants were labeled “destabilizing” in this manner. The predicted $\Delta\Delta GH_20$ was also used as a method score for the classification type analysis. A clinical analysis was not performed for the challenge due to the small dataset size, which makes binning based local $\mathrm{lr}^+$ and posterior estimates noisy. Our evaluation for the challenge is summarized in Figure 6 (Supplementary text) and Table 2 (Supplementary Excel file).

## PCM1

The PCM1 (Pericentriolar Material 1) gene is a component of centriolar satellites occurring around centrosomes in vertebrate cells. Several studies have implicated PCM1 variants as a risk factor for schizophrenia. Ventricular enlargement is one of the most consistent abnormal structural brain findings in schizophrenia.

The Katsanis lab assessed 38 missense mutations within PCM1 in a zebrafish model. Native zebrafish embryo PCM1 protein was suppressed by injecting morpholino (MO). The brain ventricle formation was measured for three groups suppressed PCM1 embryos: (1) injected with the human variant (MO+Var) or (2) injected with the wildtype mRNA (MO+WT) or (3) not injected with any mRNA (MO). The p-values for statistically different volume of brain ventricle between pairs of conditions were obtained using a Student's t-test. When the p-value is:

- not statistically different from MO, but statistically significantly different from MO+WT, the variant is "pathogenic".
- statistically different from MO, but not from MO+WT, the variant is "benign".
- Statistically different from MO, and at the same time statistically significantly different from MO+WT, the variant is "hypomorphic" (partial loss of function).

In the CAGI5 challenge for PCM1 the participants were asked to submit the predictions for the p-values comparing ventricle size of MO+Var to MO and MO+WT groups. Additionally, the participants were also asked to submit the predictions for the class labels: pathogenic, benign and hypomorphic. A total of 6 different predictors were submitted by 5 teams.

Since no predictions for the relative change in the brain volume were solicited from the participants, a regression type analysis was not performed. For the classification type analysis, the p-value predictions were ignored since it is not obvious how two combine the two p-values into a single continuous method score for classification. Binary class labels were created from the three classes by merging the hypomorphic and pathogenic classes into a single pathogenic class and retaining the benign class as is. The merging was performed for both the true class labels derived from the experiment and the predicted class labels submitted by the participants. The merging resulted in 22 out of the 38 mutations (data prior: 58%) being assigned the pathogenic class based on the experiment. The evaluation was performed using the numeric class labels: 1 for pathogenic and 0 for benign. The numeric class label corresponding to the predicted class was interpreted as the method score. For the clinical analysis, we used the data prior to compute the clinically relevant thresholds and all class-prior dependent measures. The results of the clinical analysis with the general diagnostic (0.1) and screening (0.01) priors are also reported. Since the method scores were binary and discrete, local $lr^+$and posterior probability of pathogenicity are only defined at the two values taken by the method score, instead of being a continuous curve. Furthermore, the local $lr^+$ and the global $LR^+$ coincide in this case. Our evaluation for the challenge is summarized in Figure 4 (Supplementary text) and Table 2 (Supplementary Excel file).

## L-PYK

Pyruvate kinase (PYK) catalyzes the last step in glycolysis and is regulated by allosteric effectors. Defects in the glycolytic pathway due to PYK deficiency is a known cause for anemia. Isozymes of PYK expressed in the red blood cells (R-PYK) and the liver (L-PYK) are expressed from the same genes (pklr). The difference between R-PYK and L-PYK is minor and it appears to have no effect on enzyme function and regulation. However, L-PYK is easier to study in *E. coli* since 50% of R-PYK expressed in *E. coli* is truncated, whereas L-PYK is not similarly truncated. Several non-synonymous variants of R/L-PYK observed in PYK deficient patients fall in or near the allosteric effector binding sites. Therefore, modifications in allostery seem to be sufficient to cause disease. Two sets of variants were created by Aron Fenton at University of Kansas Medical Center. The first set of 113 variants were created by substituting the residues at nine sites in or near to the binding of the negative allosteric regulator, alanine. The second part of the challenge consisted of mutations to alanine at 430 sites throughout the protein. The variants were assayed in *E. coli* extracts for the effect on allosteric regulation of enzyme activity. The enzyme activity was recorded as a binary variable indicating presence (1) or absence (0). Allosteric effect was measured as the ratio (Qax) of apparent affinity in absence versus saturating presence of the effectors alanine and Fru-1,6-BP.

In the CAGI4 challenge for L-PYK the participants were asked to submit the predictions on the effect of mutations from the two sets on L-PYK enzyme activity and allosteric regulation. The prediction for enzymatic activity was interpreted as the probability of retaining enzymatic activity (a continuous score). A total of 5 different predictors were submitted by 4 teams for both the sets.

In our analysis, we disregarded the allosteric regulation predictions and only evaluated the predictors for the classification task of predicting enzymatic activity. We ran two separate classification analysis (LPYK1 and LPYK2) on the two sets. To make the interpretation of the positive class consistent with the other challenges we flipped the numeric labels so that 1 and 0 represents "absence" (positive) and "presence" (negative) of enzymatic activity, respectively. Presence or absence of enzymatic activity could not be measured for four variants from the second set, effectively reducing its size to 426. The percent of variants labeled as "absence" in the first and the second set was 20% and 10%, respectively. These percentages are the data priors for the two datasets. For the clinical analysis, we used the data prior, to compute the clinically relevant thresholds and all class-prior dependent measures. The results of the clinical analysis with the general diagnostic (0.1) and screening (0.01) priors are also reported. Our evaluation for the challenge is summarized in Figure 4 (Supplementary text) and Table 2 (Supplementary Excel file).

## p16

Coded by the CDKN2A gene, p16 is a tumor suppressor protein that acts as cyclin-dependent kinase inhibitor and is essential for regulating the cell cycle. Constitutional and inactivating p16 mutations are common in malignant melanoma. Saturation mutagenesis experiments were carried out to measure the cell proliferation rate on a few pivotal positions in the p16 protein and mutants at these positions were assayed along with some proband-related missense mutations (total 10 mutations). The proliferation rates of the mutation-like (positive) control cells was set as 100%. The proliferation rate for p16 wildtype (negative) control cells was approximately 50%. In this CAGI3 challenge, predictors were asked to assess the 10 p16 VUS for their ability to block cell proliferation. The challenge attracted 22 submissions from 10 groups.

In our analysis, we evaluated the methods on the regression-type prediction of the proliferation rates. Note that the original proliferation rates and their predictions were divided by 100 to give values between 0 and 1. For the classification type analysis, the variants with proliferation rates above 0.75 were labeled "pathogenic" and the remaining variants were labeled "benign". The choice of the threshold is consistent with one of the three thresholds (0.65, 0.75 and 0.9) suggested by the data providers and used in the previous assessment paper for the challenge.[15] Based on the threshold of 0.75, 5 out of the 10 variants were labeled "pathogenic". The predicted proliferation rate was also used as the method score for the classification analysis. A clinical analysis was not performed for the challenge due to the small dataset size, which makes binning based local $lr^{+}$ and posterior estimates noisy. Our evaluation for the challenge is summarized in Figure 6 (Supplementary text) and Table 2 (Supplementary Excel file).

## p53 rescue

Known as the guardian of the genome, p53 is a central tumor suppressor protein that controls DNA repair, cell cycle arrest, and apoptosis (programmed cell death). Mutations in the p53 gene are the most recurrent genetic alterations in human cancers. Most of these alterations are of the missense type and show a very distinct distribution, localizing to the DNA-binding domain, and include several hotpots. Interestingly, a second mutation in p53 can in some cases rescue its function by altering a second amino acid that likely provides a structural change that compensates for the initial mutation. The aim of this challenge was to predict which second amino-acid change will rescue the p53 function. The dataset consisted of the exhaustive testing all 3,667 possible single amino acid change mutations in the entire core domain of p53 (194 amino acids from codon 96 to 289), in four different initial hit contexts; M237I, R248Q, R282W, Y220C. This amounts to a total of 14,668 mutations. The effect of the cancer rescue mutants was measured by wet-lab experimental assays of p53 function in yeast and/or human cell lines. A training set of 16,772 functionally characterized p53 mutants was also provided. In general, there are very few rescuing mutations–six mutations for M237I, one for R282W, one for Y220C and none for R248Q.

There were 8 predictions submitted from 5 different groups. Each submission provided a probability for rescue for each of the 14,668 mutations. We performed a classification type analysis to separate the "rescue mutations" (positives) from the remaining mutations (negatives). SWITCH was the best performing method with an AUC of 0.8. Of note, the approach for SWITCH uses both structural and conservation considerations as well as a more physics-based approach, which calculates stability of p53 by estimating the $\Delta\Delta G$ of the mutant vs. wildtype, looking for the changes that regain p53 stability. Regression and clinical analysis were not performed for the challenge. Our evaluation for the challenge is summarized in Figure 5 (Supplementary text) and Table 2 (Supplementary Excel file).

## BRCA

Mutations in the BRCA1 and BRCA2 genes increase the risk of breast and ovarian cancer. Myriad Genetics created the BRACAnalysis test in order to assess a woman's risk of developing hereditary breast or ovarian cancer based on detection of mutations in the BRCA1 and BRCA2 genes. This test has become the standard of care in identification of individuals with hereditary breast and ovarian cancer (HBOC) syndrome. Myriad Genetics makes one of the following four classifications for a mutation:

1. Deleterious
2. Benign
3. Genetic Variant, Favor Polymorphism (VFP)
4. Variant of Unknown Significance (VUS).

These designations are based on a database of patient testing, including frequency of the variants in populations and segregation of variants with disease in families. Precisely, how Myriad Genetics assigns these designations, and their complete database of assignments, is proprietary. Nevertheless, using the BRACAnalysis test results from clinics, it was possible to determine these assignments for a set of 100 variants observed in patients. These variants and associated pathogenicity assessment were not found in the public domain.

In the CAGI 3 BRCA challenge, participants were asked to predict the probability that Myriad Genetics classified a variant to be deleterious for the 100 variants in the dataset. There were 14 predictions submitted from 5 different groups.

In our evaluation, we only considered deleterious and benign missense variants, i.e., variants labeled as VFP and VUS were removed, additionally, indels, truncated variants and intron mutations were also removed. The resulting set had 10 missense variants, 5 of which were deleterious and the other 5 were benign. Only a classification type analysis was performed. The top performing method[16] had an AUC of 0.88. Our evaluation for the challenge is summarized in Figure 10B (Supplementary text) and Table 5 (Supplementary Excel file).

## ENIGMA

Breast cancer is the most prevalent cancer among women worldwide. The association between germline mutations in the BRCA1 and BRCA2 genes and the development of cancer has been well established, with mutations in these genes found in 1-3% of breast cancer cases. Testing for variation in these genes has emerged as a standard clinical practice, helping women to better understand and manage their heritable risk of breast and ovarian cancer. However, the increased rate of BRCA1/2 testing has led to an increasing number of variants of uncertain significance (VUS), and the rate of VUS discovery currently outpaces the rate of clinical variant interpretation. ENIGMA consortium (https://enigmaconsortium.org) is an international consortium focused on determining the clinical significance of sequence variants in BRCA1, BRCA2 and other known or suspected breast cancer related genes, providing expert input to global database and classification initiatives.

In the CAGI5 ENIGMA challenge, participants were asked submit predictions on 326 newly interpreted variants from the ENIGMA Consortium. Variants included in the dataset were classified according to the IARC 5-tier classification scheme using multifactorial likelihood analysis. The procedure assesses clinically calibrated bioinformatics information and clinical information (pathology, segregation, co-occurrence, family history, case-control) for each variant to produce a likelihood of pathogenicity. Likelihood values were calibrated against the features of known high-risk cancer-causing variants in BRCA1/2.[17, 18] Each mutation was assigned to one of five classes depending in the pathogenicity likelihood, as shown in the table. A combination of public and unpublished information was used to arrive at the final classifications, and all the classifications provided in the dataset for this challenge were either new or improved compared to what is in the public domain.[19]

| Class | Probability of Pathogenicity |
|---|---|
| 5: Pathogenic | >0.99 |
| 4: Likely pathogenic | 0.95-0.99 |
| 3: Uncertain | 0.05-0.949 |
| 2: Likely not pathogenic | 0.001-0.049 |
| 1: Not pathogenic | <0.001 |

Twelve predictions from 6 participating teams were submitted for this challenge. Four metrics were chosen for the assessment: ROC AUC, precision/recall AUC, precision and recall.[20] The rank order was largely consistent between metrics. The best-performing method used feature categories including splice predictions, population frequencies, conservation scores, and clinical observation data, such as personal and family history and covariant information.[21] The population frequencies, leveraged from gnomAD, were instrumental in many accurate predictions, as was the splicing information, a feature also used successfully by another team.

In our analysis for the challenge, we derived a binary class label from the original class label. Classes 5 and 4 were merged to create a single "pathogenic' (positive) class, classes 1 and 2 were merged to create a single "benign" (negative) class, and the variants from class 3 were not included in the analysis. The resulting dataset had 321 variants, out of which 17 were labeled "pathogenic". In absence of any continuous experimental measurement, we only perform the classification type analysis. Our evaluation for the challenge is summarized in Figure 10A (Supplementary text) and Table 5 (Supplementary Excel file).

## Annotate All Missense

dbNSFP is a database of human nonsynonymous single nucleotide variants (nsSNVs) and their functional predictions and annotations.[22-25] Version 3.5 compiles 18 functional prediction scores and 6 conservation scores, as well as other related information including allele frequencies observed in different large datasets, various gene IDs from different databases, functional descriptions of genes, gene expression and gene interaction information.

For this CAGI5 challenge, a large list of possible SNVs based on the human reference sequence was created from dbNSFP. This resulted in 81,084,849 possible protein-altering variants. Predictors were asked to predict the functional effect of each of these coding SNVs. For the vast majority of these missense and nonsense variants, the functional impact is not known, but experimental and clinical evidence is accruing rapidly. Rather than drawing upon a single discrete dataset as typical with CAGI, predictions were assessed by comparing with experimental or clinical annotations made available after the prediction submission date. If predictors provided their assent, predictions would also be incorporated into dbNSFP.

A test dataset of newly annotated missense variants was constructed from ClinVar and HGMD databases, considering only variants added to these databases between June 2018 (after the close of the annotate all missense CAGI5 challenge) and December 2020. In particular, the June 3, 2018 and December 26, 2020 releases of ClinVar were obtained and variants annotated as missense SNVs were extracted. Similarly, 2019.1 (first quarter of 2019) and 2020.4 (last quarter of 2020) releases of HGMD were obtained and restricted to missense SNVs. A set of newly annotated ClinVar missense variants were obtained by subtracting from the December 26, 2020 release any variants present in the June 3, 2018 release, except those with a clinical significance annotation of "Uncertain significance" in the June 3, 2018 release, as well as any variants present in HGMD 2019.1. Any variant with a review status of "no assertion provided", ''no assertion criteria provided", "no interpretation for the single variant" and "conflicting interpretations" was removed from the set. A set of newly annotated HGMD variants was generated by subtracting from the HGMD 2020.4 release any variants present in HGMD 2019.1, as well as any variants present in the June 3, 2018 ClinVar release, except those with a clinical significance annotation of "Uncertain significance." Subtraction was done based on the chromosome and position of the variant in each case. In total, there were 3,309 pathogenic (P+LP) ClinVar variants, 2,732 of which were likely pathogenic (LP), 10,677 disease mutations (DM; 1,141 of which DM?) from HGMD, and 23,096 benign variants (B+LB); 11,078 of which likely benign, (LB) from ClinVar. The newly annotated ClinVar and newly annotated HGMD variants were combined to generate two test sets, 1) AAM1All: containing only variants with confident assertions ("pathogenic" or "benign" in ClinVar or "DM" in HGMD) and 2) AAM2All: additionally, containing variants with less confident assertions ("pathogenic," "pathogenic/likely_pathogenic," or "likely_pathogenic" in ClinVar; "DM" or "DM?" in HGMD; "benign", "benign/likely_benign," or "likely_benign" in ClinVar). We used these test sets to evaluate performance of the four submitted predictors for this challenge. Additionally, all tools (functional predictions and conservation scores) with results deposited in dbNSFP v3.5 were also evaluated, as a set of commonly available tools that could not have been trained on the test variants, since dbNSFP v3.5 was released in 2017. In all, the following tools were included in the evaluation: VEST, Turkey, Bologna, Condel, SIFT, PROVEAN, PolyPhen-2, LRT, MutationTaster, MutationAssessor, FATHMM, CADD v1.4, fitCons, DANN, MetaSVM, MetaLR, GenoCanyon, Eigen, M-CAP, REVEL, MutPred, GERP++, phyloP, phastCons and SiPhy. Some of these tools have multiple prediction scores, included in the analysis; for example, VEST has versions 3 and 4 predictors and FATHMM has version 2.3 and fathmm-mkl.

Only classification and clinical analysis was performed for this challenge. For the clinical analysis, we used the general diagnostic (0.1) and screening (0.01) priors, to compute the clinically relevant thresholds and all class-prior dependent measures. In addition to AAM1All and AAM2All, evaluation was performed on the following six data subsets created from the two sets. (1) AAM1BiClass: containing variants restricted to the "bi-class" genes that have both pathogenic and benign variants in AAM1All, (2) AAM2BiClass: containing variants restricted to the "bi-class" genes that have both pathogenic and benign variants in AAM2All, (3,4) AAM1CV and AAM2CV: composed of subsets of AAM1All and AAM2All data, respectively, restricted to the variants from ClinVar only and (5,6) AAM1HGMD and AAM1HGMD: containing subsets of AAM1All and AAM2All data, respectively, with the pathogenic variants only coming from HGMD. A total of 22,131 variants from 6,482 genes were present in the AAM1All dataset, out of which 7,429 variants came from 1,022 bi-class genes. A total of 37,082 variants from 7,723 genes were present in the AAM2All dataset, out of which 21,423 variants came from 2,074 bi-class genes. Since all the datasets created from AAM2All, include less confident variant classes, they are more difficult to predict compared to those created from AAM1All. Within all AAM1All (and AAM2All) generated datasets the bi-class gene dataset is the most difficult to predict. The evaluation for the challenge is summarized in Figure 3 (Main text), Figure 7 and 9 (Supplementary text) and Table 4 (Supplementary Excel file).

## Expression and Splicing Challenges

### Vex-seq

In the CAGI5 challenge, Vex-seq, a barcoding approach, Variant exon sequencing (Vex-seq), was applied to assess the effect of around 2,000 natural single nucleotide variants and short indels on splicing of a globin mini-gene construct transfected into HepG2 cells. The results are expressed as $\Delta\Psi$ (delta PSI, or Percent Spliced-In), between the variant $\Psi$ and the reference $\Psi$. If $\Delta\Psi$ is calculated from a reference exon that is always spliced in ($\Psi$(reference) = 100), and a variant exon that is only spliced-in in half of the transcripts observed for that variant ($\Psi$(variant) = 50), the $\Delta\Psi$ would be 50. $\Delta\Psi$ is bounded by −100 and 100. A training set of around 1,000 variants containing the $\Delta\Psi$ values were provided to the participants. Participants were asked to predict the ΔΨ values for the remaining 1,098 test variants. A total of six groups participated in the challenge.

In our analysis, for this challenge, we created two datasets: one for over-splicing (VexSeq1) and the other for under-splicing (VexSeq2). For VexSeq1, a variant was assigned an "over-splicing" (positive) label, if its experimental ΔΨ value was more than one standard deviation (13.56) above the mean ΔΨ (−1.88) value, the remaining variants were considered as negatives. Similarly, for VexSeq2, a variant was assigned an "under-splicing" (positive) label, if its experimental ΔΨ value was more than one standard deviation below the mean ΔΨ value, the remaining variants were considered as negatives. Creating the class labels, using the mean and the standard deviation was done based on the previous assessment paper for the challenge.[26] A regression type analysis was performed, comparing the predicted ΔΨ value to the experimental values. The data for the regression analysis was identical for VexSeq1 and VexSeq2. Some participants submitted predicted ΔΨ in the range from −1 to 1. We multiplied the predictions by 100 in such cases. 6.5% and 7.5% of the variants were labeled as over-splicing and under-splicing in VexSeq1 and VexSeq2, respectively. The predicted ΔΨ was also used as the method score for the classification type analysis. A clinical analysis was not performed for the challenge. Our evaluation for the challenge is summarized in Figure 11 (Supplementary text) and Table 6 (Supplementary Excel file).

## MaPSy

In the CAGI 5 challenge, the Massively Parallel Splicing Assay (MaPSy) approach was used to screen sets of 4,964 and 797 reported exonic disease mutations using a mini-gene system, assaying both *in vivo* via transfection in tissue culture, and *in vitro* via incubation in cell nuclear extract. The loss or gain of splicing efficiency was measured in terms of $log_2$ allelic skew ratio computed from the read counts of input DNA and correctly spliced cDNA for the variant and the wildtype. The $log_2$ ratios are expected to be 0 for a neutral mutation, negative for under-splicing and positive for over-splicing. The variants were categorized as exonic splicing mutations (ESMs) if they both changed the allelic ratio by 1.5-fold or more and passed a two-sided Fisher's exact test with a false discovery rate (FDR) of 5% both *in vivo* and *in vitro*. The set of 4,964 variants along with the measurements of the $log_2$ allelic skew ratios (*in vivo* and *in vitro*) and the ESM class labels were provided to the participants for training their models. The challenge was to predict the $log_2$ allelic skew ratios (*in vivo* and *in vitro*) and the ESM class labels on the test set of 797 variants. The participants were asked to submit their predictions for the two $log_2$ allelic ratios and the variants' probability of being an ESM. A total of five groups participated in the challenge.

For this challenge, we created three datasets, MaPSy1, MaPSy2 and MaPSy3, to be analyzed separately. MaPSy1 and MaPSy2 included the $log_2$ allelic skew ratios measured in vivo and in vitro, respectively, whereas MaPSy3 included the ESM class labels. A regression type analysis was performed on MaPSy1 and MaPSy2 for predicting the *in vivo* and *in vitro* $log_2$ allelic skew ratios, respectively. A classification type analysis was performed on MaPSy3 for the predicting the ESM class label. The *in vivo* and *in vitro* allelic ratios did not agree (implying opposite effects on splicing) for some of the variants. The challenge assessors performed a "consistent" analysis where the disagreeing variants were relabeled "non ESM". Following that approach, the ESM class label for such variants, if originally set to 1 (ESM), were changed to 0 (non ESM).[26]

In this manner, labels were changed for 19 variants. Only 30 out of the 797 variants were finally labeled ESM in MaPSy3. The predicted probability of ESM was used as the method score for the classification task. A clinical analysis was not performed for the challenge. Our evaluation for the challenge Is summarized in Figure 5A (Main text) and Table 6 (Supplementary Excel file).

## eQTL

Genome-wide association studies (GWAS) suggest that much of the variation underlying common traits and diseases maps within regions of the genome that do not encode protein. However, identifying the causal alleles responsible for variation in expression of human genes has been particularly difficult. In the CAGI4 eQTL causal SNPs challenge, a massively parallel reporter assay (MPRA) was applied to thousands of single nucleotide polymorphisms (SNPs) and small insertion/deletion polymorphisms in linkage disequilibrium (LD) with cis-expression quantitative trait loci (eQTLs). The results identify variants showing differential expression between alleles. The challenge is to identify the regulatory sequences and the expression-modulating variants (emVars) underlying each eQTL and estimate their effects in the assay.

The CAGI4 eQTL challenge comprised two parts. In the first, 3,006 potential regulatory sequences and variants (2,811 SNVs and 195 indels) associated with a distinct subset of 1,050 eQTLS were provided. Participants were asked to predict the level of transcriptional activity for each allele and to determine for each variant whether at least one of the alleles is a "regulatory hit" (positive class), based on significant activation of reporter gene expression. 12% of the variants were labeled as regulatory hit. A sample dataset of 3,044 variants associated with 1,052 eQTLs was provided for training. In the second part of the challenge, 401 regulatory sequences (370 SNVs and 31 indels) associated with a third distinct subset of 1,055 eQTL were provided. Participants were given variants that were confirmed regulatory hits and asked to predict the difference between the transcriptional activity of the two alleles, both quantitatively as the $\log_2$ allelic skew (the $\log_2$ ratio of expression level of the alternative allele relative to the reference allele) and qualitatively as expression-modulating variant, "emVar" (positive class). 26% of the variants were labeled as emVar. Seven groups participated in this challenge, submitting 20 predictions for the first part, and 13 submissions for the second. The prior assessment of this challenge identified chromatin accessibility and transcription factor binding as features leading to the most accurate results.[27]

For the first part of the challenge (eQTL1), we evaluated the methods only on the classification task of predicting if a variant is a regulatory hit, using the predicted probability of regulatory hit as the method score. For the second part of the challenge (eQTL2), we evaluate the methods on the regression task of predicting the $\log_2$ allelic skew and the classification task of predicting if a variant is an emVar. The predicted probability of emVar (different from the predictions for $\log_2$ allelic skew) is used as the method score for the classification task. A clinical analysis was not performed for the challenge. Our evaluation for the challenge is summarized in Figure 12A and 12B (Supplementary text) and Table 6 (Supplementary Excel file).

## Regulation-Saturation

Gene regulatory variants are known to play an important role in a number of common human diseases, including diabetes, neuropsychiatric disorders, autoimmune disorders, cardiovascular disease, and cancer. These variants modulate the strength of interactions between enhancers and promoters and the transcription factors (TFs) that bind them and alter the cell-specific transcriptional control of gene regulatory networks central to the proper development and functioning of human cells and tissues. Although we have a good basic understanding of the general molecular mechanisms of these interactions, quantitative and predictive models of cell-specific enhancer and promoter function are currently under active development.

In this CAGI5 challenge, 17,500 single nucleotide variants (SNVs) in 5 human disease associated enhancers (IRF4, IRG6, MYC, SORT1, ZFAND3) and 9 promoters (TERT, LDLR, F9, HBG1, PKLR, MSMB, HBB, HNF4A, GP1BB) were assessed in a saturation mutagenesis massively parallel reporter assay (MPRA) in different cell lines; see Kircher et al.[28] for a more detailed description of the MPRA experimental methods. Promoters were cloned into a plasmid upstream of a tagged reporter construct, and reporter expression was measured relative to the plasmid DNA to determine the impact of promoter variants. Enhancers were placed upstream of a minimal promoter and assayed similarly. A multiple linear regression model fitting the reporter's log expression level with binary (dependent) variables, one corresponding to each variant, is used to estimate the contribution of a variant as its (fitted) coefficient. A confidence score was derived for each coefficient after scaling its p-value on a log 10 scale and normalizing to a 0-1 range. Effectively, a confidence score of 1 corresponds to a p-value of ≤10-50, 0.5 to 10-25 and 0 to a p-value of 1. The coefficient served as a continuous measurement capturing a variant's effect on expression. A ternary class variable was derived for each variant taking value −1 (decrease expression) or 1 (increase expression), if the coefficient is negative or positive, respectively, and the confidence is greater than 0.1 (p-value of $10^{-5}$), otherwise the variable was set to 0 (no effect on expression). Participants were given the impact of the variants (coefficients and ternary labels) in selected subsets from each region to train their models, consisting of around 25% of the variants. The remaining variants were used for evaluation. The challenge is to predict the functional effects of these variants (coefficients and the ternary labels) in the regulatory regions as measured from the reporter expression. The participants were only required to submit discrete values: −1 (decrease expression), 0 (no effect on expression) and 1 (increase expression) for each variant, along with a probability that the discrete values are correctly assigned. However, we were, additionally, able to procure the continuous prediction for a variant's effect on expression, obtained during the previous assessment of the challenge for the best methods from the top three performing groups.[29]

According to the data providers, the dataset incorrectly included positions 20bp up and downstream of each construct due to technical reasons. These variants were identified by the positions listed ahead were removed from the training and test datasets. F9:X 138612621; GP1BB:22 19710788; HBG1:11 5271309; HNF4A:20 42984159; IRF6:1 209989736; MSMB:10 51548987; PKLR:1 155271656; SORT1:1 109817273; HBB:11 5248439; IRF4:6 396142; LDLR:19 11199906; MYCrs6983267:8 128413073; ZFAND3:6 37775274.

Seven groups participated, with a total of 23 submissions. All top performing models for variant impact prediction used machine learning based ANN (or gkm-SVM) DNA sequence features trained on chromatin accessibility or chromatin state data.[29] These models consistently outperformed models using sequence features derived from other sources (evolutionary conservation, kmers, or more generic sequence features, e.g., GC content). The machine learning-based models also outperformed models using chromatin accessibility, chromatin state, or TF ChIP-seq data without using epigenomic data to derive DNA sequence-based models.

High prediction accuracy was obtained when machine learning-based DNA sequence features were combined with proper importance weighting derived from another layer of machine learning on a subset of the mutation data used as training for each cell type.

In our analysis, the classification tasks corresponded to predicting a set of binary labels derived from the ternary labels and the regression analysis corresponded to the prediction of the fitted coefficients, measuring a variants' effect on expression. We evaluated the performance on the enhancer and promoter variants separately. From a total of 13,790 variants available in the test set, 6,295 and 6,868 corresponded to the enhancers and promoters, respectively. From each set, we created two binary classification datasets to separately evaluate the performance on predicting the increase and decrease in expression. For the increase in expression the variants with the ternary class label 1 were considered as positives and the remaining variants as negatives. Similarly, for decrease in expression, the variants with the ternary class label −1 were considered as the positives (relabeled as 1) and the remaining variants as negatives. In total we perform four separate analyses: RegSatEnh1 (increase), RegSatEnh2 (decrease), RegSatProm1 (increase) and RegSatProm2 (decrease), each with both regression and classification type evaluation. Given the enhancers or the promoters, the regression type task in both the increase and decrease of expression analysis is the same: evaluate the predictions for a variant's contribution to the reporter's expression level. The two analyses differ only w.r.t to the classification task since only the class labels are different. The proportion of positives for the increase in expression were 0.09 and 0.06 for the enhancers and promotors, respectively. The proportion of positives for the decrease in expression were 0.14 and 0.13 for the enhancers and promotors, respectively. The discrete predictions were used for both the regression and classification type analyses for most methods, except the three methods, for which continuous predictions were procured. A clinical analysis was not performed for the challenge. Our evaluation for the challenge is summarized in Figure 5B (Main text) and Table 6 (Supplementary Excel file).

# Complex disease challenges

This work re-assessed only a single complex disease challenge.

## Crohn's disease

Crohn's disease (CD; MIM #266600) is a chronic inflammatory bowel disease (IBD) characterized by relapsing inflammation that can involve any part of the gastrointestinal tract and also extra-intestinal manifestations. It is caused by the complex interplay between an overly active immune system and environmental triggers in genetically susceptible individuals. Results from twin and familial aggregation studies,[30] as well as evidence from GWAS,[31, 32] have shown that genetic factors play an important role in CD etiology. To date, 163 genetic susceptibility loci have been identified for IBD with 30 loci exclusive to CD, 23 to ulcerative colitis (UC), and 110 shared by the two.[32] Early-onset cases of IBD, with an age of onset before 10, often show a more severe disease course with a higher risk of complications, and genetic factors likely play a larger role in these individuals.[33]

Three successive iterations of this challenge were performed. The 2011 (CAGI2) dataset had 56 exomes (42 cases, 14 controls), all of German ancestry.[34] During assessment, substantial batch effect was discovered in the data as a side effect of sample preparation and sequencing.[35] The 2013 (CAGI3) dataset had 66 exomes (51 cases, 15 controls). Although these samples were also of German ancestry, cases were selected from pedigrees of German of families with multiple occurrences of Crohn's disease. As such, some of these cases were related. This led to a substantial difference in clustering between cases and controls, suggesting the presence of sampling bias.[35] The 2016 (CAGI4) challenge had 111 unrelated German ancestry exomes (64 cases, 47 controls). For the CAGI4 challenge, submitting groups were allowed to use the data from the CAGI2 and CAGI3 Crohn's challenges for training. In all iterations of the challenge, groups were asked to report a probability of Crohn's disease (between 0 and 1) for each individual and a standard deviation representing their confidence in that prediction. For the CAGI4 challenge, teams were also asked to predict whether age of onset was greater or less than 10 years of age. The problems with batch effects and sampling bias were no longer present in the CAGI4 Crohn's challenge.[35] This was the challenge selected for further analysis in this study.

We analyzed the CAGI4 Crohn's data on classification problem of separating the cases (positives) from the controls (negatives). In addition to the ROC curve, AUC and the local $\text{lr}^+$ curve, we also give the RR (relative risk) curve and the kernel density estimation-based distribution of the method scores for the cases and controls. A class-prior of 1.3% is used to compute the RR curve with Equation 31 (see Methods). The choice of the prior was based on the recent data on the prevalence of inflammatory bowel disease in US adults.[36] Our evaluation for the challenge is summarized in Figure 6 (Main text) and Table 6 (Supplementary Excel file).

## Non-analyzed challenges

Summaries of challenges that were not analyzed for this work are presented below in alphabetical order.

**Asthma twins (CAGI2).** The dataset includes whole genomes of 8 pairs of discordant monozygotic twins (randomly numbered from 1 to 16) that is, in each pair identical twins one has asthma and one does not. In addition, RNA sequencing data for each individual is provided. One of the twins in each pair suffers from asthma while the other twin is healthy.

There were 6 submissions from 6 groups. All predicted the correct twin pairs but the asthma correction rate was 63%, no better than random. In the genomic data, the number of errors was greater than the number of variants. This sequencing error rate might have masked the differences between the twins in the genomic data. Further, the RNA sequencing data appeared to correlate with the twins, rather than with the disease status. The experimental dataset remains unpublished, and thus the results are not further discussed here.

**Bipolar exomes (CAGI4).** This challenge involved the prediction of which of a set of individuals have been diagnosed with bipolar disorder, given exome data. 500 of the 1000 exome samples were provided for training. Nine groups participated in this challenge, providing 29 submissions. No participant was very successful, with the highest AUC being 0.64.[35] Although not impressive, the best-scoring method is interesting. While most participants used similar approaches to those deployed for Crohn's, this one used linear genotype status as an input vector to a three-layer neural network, trained on the data provided, and used no information about the disease or known GWAS loci. The result suggests that more sophisticated machine learning approaches have potential in this area. A caveat is that the case and control data were from different sources, so it is possible that the method identified some underlying sequence features not related to the disease.

**Breast cancer pharmacogenomics (CAGI2).** Cancer tissues are specifically responsive to different drugs. For this experiment, predictors are asked to predict the response of each of 54 breast cancer cell lines to a panel of 54 drugs. Data about the tissues include transcriptional profiling, SNP data and copy number profiles measured for cells grown in the absence of any treatment. The prediction requested was GI50 values with standard deviation.

Three groups participated, providing 3 submissions. The assessors used RMSE and Kendall's tau to evaluate predictions. RMSE was used to measure how well each submitted prediction estimated the overall level of sensitivity to a particular drug, while Kendall's tau was used to measure the quality of cell line rankings from least to most sensitive, as reported by each method. All three submissions performed significantly better than random on Tamoxifen, Bortezomib, and Iressa. Additionally, submission SID#16A had the lowest RMSE on 10 of the 15 drugs. Kendall's tau was low overall (<0.3) and no algorithm was able to accurately rank the cell lines from least to most sensitive.

**CHEK2 (CAGI1, CAGI5).** Cell-cycle-checkpoint kinase 2 (CHEK2; OMIM #604373) is a protein that plays an important role in the maintenance of genome integrity and in the regulation of the G2/M cell cycle checkpoint. CHEK2 has been shown to interact with other proteins involved in DNA repair processes such as BRCA1 and TP53. These findings render CHEK2 an attractive candidate susceptibility gene for a variety of cancers. The challenges in both CAGI1 and CAGI5, involved classifying variants as occurring in breast cancer cases or controls.

In CAGI1, predictors were provided with 41 rare missense, nonsense, splicing, and indel CHEK2 variants. Ten groups participated in this challenge, making a total of 16 submissions. Assessment showed several methods performing better than the baseline method (Align-GVGD), which had been trained on this dataset. Functional contribution to the predictions is particularly helpful when evolutionary information is not discriminative enough. Participants tended to not properly consider the likely distribution of neutral mutations. A probability of 0.5 would indicate that the mutation is neutral (equal in both populations) while a probability of less than 0.5 would be indicative of a variant that is actually protective.

In CAGI5, data involved the targeted resequencing of CHEK2 from approximately 1000 Latina breast cancer cases and 1000 ancestry matched controls. Fifty-three variants in the list, observed between 1 and 20 times, were provided for this challenge. Eight groups participated, with a total of 18 submissions. While group 5, submission 1, appeared to do best overall, it had many false positives. Additionally, most of the variants were found in cases, and methods that favored this performed better.[37]

**Clotting disease (DVT or PE) exomes (CAGI5).** African Americans have a 30-60% higher incidence of developing venous thromboembolisms (VTE), which includes deep vein thrombosis (DVT) and pulmonary embolism (PE) than people of European ancestry.[38] The risk factors for VTE are complex and include environmental risk factors (e.g., vessel injury; and blood stasis) and genetic risk factors, including common and/or rare variants that predispose to hypercoagulation.[39] In this challenge, participants are provided with exome data and clinical covariates for a cohort of African Americans who have been prescribed Warfarin, an anticoagulant, either because they had experienced a VTE event or had been diagnosed with atrial fibrillation (which predisposes to clotting). The challenge requested participants to distinguish between these conditions.

Seven groups participated in this challenge, providing 16 submissions. Assessment was complicated by two factors. First, the warfarin doses of study individuals were known to participants, and VTE patients are usually given high doses, providing a strong predictive signal that a number of participants exploited. In hindsight, as noted by the assessors, given the strong genetic relationship between warfarin dosage and genetics, it may have been better for the challenge to not provide warfarin dosage to the participants and to remove genes related to warfarin pharmacokinetics and pharmacodynamics from the exomes.[40] Second, unlike Crohn's disease and Bipolar disorder, studies in Europeans in the UK Biobank have calculated the heritability on the liability scale in Europeans to be 0.14 and disease prevalence to be 2%, which indicate that the theoretical maximum AUROC that could be achieved in predicting VTE from coding regions is a low 0.62.[41, 42] The best AUC from methods that did not appear to use warfarin dose information was 0.65, while a previously published baseline method developed for European populations[43] did better than any submitted method that did not use warfarin dose in their predictions, with an AUC of 0.71.

**FCH (CAGI3).** Familial combined hyperlipidemia (FCH; OMIM 14380) the most prevalent hyperlipidemia, is a complex metabolic disorder characterized by variable occurrence of elevated low-density lipoprotein cholesterol (LDL-C) level and high triglycerides (TG)—a condition that is commonly associated with coronary artery disease (CAD).

The challenge involved exome sequencing data for 5 subjects in an FCH family and comprised two parts. In the first, participants were given which family members have elevated LDL and asked to predict which variant(s) confer the elevated LDL phenotype. In the second part of the challenge, the task was to predict which individuals have abnormally high TG and which individuals have abnormally high HDL levels.

There were 21 submissions from 11 groups. The assessor considered the first part of the challenge very simple, as the presumably causal mutation is listed in OMIM and most people would check LDLR. Several submissions did well in this task, with three groups uniquely predicting the most probable diagnostic mutation. Two submissions only listed the correct mutation, while a third listed two others but with negligible confidence placed on them.

The second part of the challenge was considered hard, and there were no correct submissions. The assessor commented that predicting the abnormal father's TG and HDL is very hard, so mostly required that the unaffected daughter would be predicted correctly. There were three reasonable submissions. Combining two sub-challenges was difficult. Only one submission did reasonably in both cases. Judging solely on the first sub-challenge, three submissions did well.

**HA (CAGI3).** Hypoalphalipoproteinemia (HA; OMIM #604091) is characterized by severely decreased serum high-density lipoprotein cholesterol (HDL-C) levels and low apolipoprotein A-1 (APOA1). Low HDL-C is a risk factor for coronary artery disease. The dataset for this challenge comprises of exome sequencing data for 4 subjects from the same family, one of which has HA.

The challenge attracted 18 submissions from 12 groups. While 3 submissions confidently identified the proband, no group provided the right answer with high confidence. One group had good correlation between probability of illness and actual disease state, by making a series of 'bets' about likely priors. Here again, data quality complicated interpretation and assessment.

**Johns Hopkins clinical panel (CAGI4).** The Johns Hopkins challenge, provided by the Johns Hopkins DNA Diagnostic Laboratory (http://www.hopkinsmedicine.org/dnadiagnostic), comprised of exonic sequence for 83 genes associated with one of 14 disease classes, including 5 decoys.[44] Participants were tasked with identifying the disease class for each of 106 patients; 43 of these patients had received a molecular diagnosis in the clinical pipeline.

Five groups participated, providing a total of 5 submissions. The most successful CAGI method correctly matched 36 of the previously diagnosed patients to their disease class. More interestingly, 39 of the 63 undiagnosed cases were successfully matched by at least one participating group, indicating successful identification of causative variants. Some of these may have been highlighted in the John Hopkins pipeline, but did not have strong enough evidence to meet the ACMG/AMP guidelines.[45] Guidelines also require the official pipeline only search for causative variants in genes consistent with the specific disease a physician requested a test for. For a number of undiagnosed cases, CAGI participants found high-confidence deleterious mutations in genes that were not in the selected panel, suggesting physicians may have misdiagnosed the symptoms.[44] However, because of the IRB guidelines under which the pipeline operates, it has not been possible to further investigate or even publicly report these cases. Ensuring appropriate consents and approvals are in place in advance of a challenge, could maximize the use of clinical data, and allow for a more in-depth and critical analysis of challenge results.

**Intellectual disability panel (CAGI5).** In the ID challenge, 150 individuals with ID and/or Autism Spectrum Disorder (ASD) were assessed through sequencing of 74 genes involved in ID with or without autistic features. Predictors were tasked with matching patients with one or more of 7 phenotypes and identifying causative or contributing variants for each patient.

Five groups participated in this challenge, submitting a total of 15 predictions. The phenotype matching in this challenge had a poor overall performance, with the top method achieving 0.78 for the ID phenotype. While the Hopkins panel was testing for monogenic diseases with Mendelian inheritance, the ID challenge addressed complex disease. The genetic bases of neurodevelopmental disorders (NDDs) are not fully understood but are characterized by strong clinical comorbidity as well as genetic heterogeneity, involving the interplay of de novo, rare, and many common variants which affect phenotype variability and disease severity. Despite these difficulties, some groups made plausible predictions on novel variants that had not been reported to the patient due to being predicted neutral by the majority of standard computational methods.[46] In two of these cases, the proband phenotype was partially consistent with the clinical observations, and segregation analysis showed the variants to be absent from the mother and healthy sibling, suggesting they might be transmitted *in cis* from the other parent. However, the father was not available for further investigation.

**Mouse exomes (CAGI2).** The challenge involved identifying the causative variants leading to one of four morphological phenotypes arising spontaneously in inbred mouse lines (L11Jus74, Sofa, Frg and Stn. Predictors were given SNVs and indels found from exome sequencing. Causative variants had been identified for the L11Jus74 and Sofa phenotypes by the use of traditional breeding crosses,[47] and the predictions were compared to these results, which were unpublished at the time of the CAGI submissions. The L11Jus74 phenotype is caused by two SNVs (chr11:102258914A>T and chr11:77984176A>T), whereas a 15-nucleotide deletion in the Pfas gene is responsible for the Sofa phenotype.[9] The predictions for Frg and Stn phenotypes could not be compared to experimental data, as the causative variants could not successfully be mapped by linkage.

There were 2 submissions from 2 groups. The approach of the first team consisted of two steps: (1) searching the JAX Mice Database (https://mice.jax.org) for chromosome regions where the phenotypes are known to map; (2) examining the effects of missense variants located in the found regions with the help of MutPred[16] and SNPs&GO.[48] In addition, variants in the proximity of splice sites, determining a frameshift or a stop gain/loss in the coding sequence were included in the submission. Altogether the group submitted seven candidate variants for L11Jus74, and four variants for the Sofa phenotype. None of the predicted variants coincided with the causative variants identified. The second team also utilized the JAX Mice Database and assigned predicted effects for all types of variants. For the L11Jus74 phenotype, they only considered variants in chromosome 11, assuming that the phenotype name implied a causal variant in that region. Their submission included 82 candidate variants for L11Jus74, and 31 for the Sofa phenotype. They were able to identify the linkage matching variants for both phenotypes and assigned these with the highest probabilities. However, because of the large number of possible variants included, the absolute probabilities are low ($p = 0.0195$ and 0.0586 for L11Jus variants; $p = 0.0541$ for the Sofa variant). These limited results indicate that causative monogenic-like variants can be identified with current methods, though perhaps not unambiguously.

**MR-1 (CAGI2, CAGI3).** *Shewanella oneidensis* strain MR-1 (formerly known as *S. putrefaciens*) is a model organism for studying metal reduction, as MR-1 can utilize a wide range of metal ions and solid metals as electron acceptors and also grows aerobically. MR-1 is in the same division of bacteria as *E. coli* (the γ-Proteobacteria), but they are not closely related. Of the ~4,500 proteins in MR-1, only about a third have orthologs in *E. coli*.

The Arkin Lab at UC Berkeley created a large number of *S. oneidensis* MR-1 transposon insertions with known location and with a known tag or barcode. These insertions were pooled together into two pools, and the pools grown under a given (stress) condition for ~6-8 generations. The abundance of each tagged strain was measured with microarray at the beginning and at the end of the experiment. The fitness of the strain is the log2 ratio of these abundances. (This is not the same scale as fitness in population genetics.) The data is normalized so that the median strain has a fitness of 0. The fitness value of a gene is computed as the average of the values for the insertions in that gene. In this experiment it is assumed that the insertions of a given gene deactivate that gene. A study of MR-1 gene-phenotype relationships for 121 conditions has already been published.[49] The CAGI challenge involved predicting results under eight more conditions.

Despite being offered in two successive rounds of CAGI, this challenge did not attract any submissions.

**MRN complex (CAGI3).** Genomes are subject to constant threat by damaging agents that generate DNA double-strand breaks (DSBs). The Mre11–Rad50–Nbs1 (MRN) complex plays important roles in detection and signaling of DSBs, as well as in the repair pathways of homologous recombination and non-homologous end-joining. The importance of Mre11-Rad50-Nbs1 complex in the cellular response to DNA double-strand breaks was initially revealed by ataxia telangiectasia-like disorder and Nijmgen breakage syndrome.

In this challenge, mutation screening of MRE11 and NBS1 genes was conducted from a series of approximately 1,300 breast cancer cases and 1,100 controls. There were 42 mutations listed for MRE11, and 44 mutations for NBS1, with more added during a short (one week) window in an optional second challenge (9 variants for MRE11, 1 for NBS1). Thirteen groups participated in the primary challenge, making a total of 23 submissions. Additional 17 submissions were made from 9 groups for the optional challenge. Assessment employed a logistic regression likelihood ratio test of the status of each subject (case/control) against the predicted probability of pathogenicity of the variant(s) that they carried. Predictions were also be assessed by calculated odds ratios and ROC areas.

Assessment revealed that method performance differed sharply on the two proteins, even though they were part of the same complex. Additionally, participants tended to not properly consider the likely distribution of neutral ($p = 0.5$) or protective ($p < 0.5$) mutations, which formed the majority in this challenge. Furthermore, a recent study for breast-cancer risk genes in over 113,000 women, revealed no significant association between NBS1 (also known as NBN) or MRE11 and breast cancer.[50]

**NPM-ALK (CAGI4).** Nucleoplasmin-anaplastic lymphoma kinase (NPM-ALK) is an oncogenic fusion protein found exclusively in a specific type of T-cell lymphoid malignancy, namely ALK-positive anaplastic large cell lymphoma (ALCL).[51] Aberrantly activated NPM-ALK, specifically constitutive activation of the ALK tyrosine kinase, causes cell transformation through activation of several biological pathways related to cell proliferation, cell-cycle control and apoptosis. Small-molecule inhibitors of ALK are among the most promising drugs in several high-risk cancers, since ALK activation by mutation, amplification, or gene rearrangement is highly oncogenic. However, inhibitor efficacy can be hampered by several resistance mechanisms including point mutations in ALK.[52, 53] An alternative approach involves inhibiting the molecular chaperone Hsp90, required for ALK folding, stability and/or activity.

In this experiment, NPM-ALK constructs with mutations in the kinase domain were assayed in extracts of transfected Hek-293T cells. ALK kinase activity was assessed by Western blotting using site-specific antibodies against phosphorylated ALK (Tyr1604) and STAT3 (Tyr705), one of ALK's downstream targets. Binding to Hsp-90 was assessed by immunoblotting and measured as the interaction density (band density) of each mutant relative to wildtype NPM-ALK. 23 variants (single amino acid, multiple amino acid substitutions and deletions) were assessed this way. The challenge involved predicting both the kinase activity and the Hsp90 binding affinity of the mutant proteins relative to the reference (wildtype) NPM-ALK fusion protein.

There were 4 submissions from 4 groups for this challenge. Assessment showed that predictors performed better than baseline tools, with the effect of short deletions being easier to predict than other mutation types.

**PGP (CAGI1-CAGI4).** A rather different class of challenge using large scale genome data used information from the Personal Genome Project (PGP). Participants in the project make their full sequence and phenotypic profile data publicly available. The four CAGI challenges were based on prerelease samples from this resource. The first two challenges, in 2010 and 2011, asked CAGI participants to predict which of 32 binary traits each individual has, given complete genome sequence. Using precision as the metric, results were quite poor, although for unclear reasons, the AUC values were much better, with a best AUC over 0.8. The second pair of challenges required matching each genome to a set of 239 self-reported binary phenotypes. Here results were slightly better than random (6 correct matches in the first round and 5 out of 23 in the second), but this is clearly a difficult task. Although the full 239 traits were available, participants seem to have gained most from a few strong genome/phenotype relationships, such ancestry, rare blood type and eye color.[54] There were also some discrepancies observed between provided, self-reported traits and information from genomic data. Although the results are not the impressive, these challenges inspired one group to develop an interesting comprehensive Bayesian approach that may have broader application.

**RAD50 (CAGI2).** RAD50 is a component of the MRN (MRE11-NBS1-RAD50) complex, which plays a central role in double-strand break repair, DNA recombination, maintenance of telomere integrity and meiosis. RAD50 may be required to bind to DNA and hold the other two protein components of MRN in close proximity. Mutations in RAD50 are observed in a variety of cancers (stomach, intestinal, endometrial), and it is considered a candidate intermediate-risk breast cancer susceptibility gene. For this challenge, predictors are provided with a list of 69 RAD50 variants observed from sequencing RAD50 gene in about 1,400 breast cancer cases and 1,200 ethnically matched controls. These variants were observed between 1 and 20 times. The challenge is to predict the probability of the variant occurring in a case individual.

Eight groups participated in this challenge, submitting a total of 14 predictions. Assessment revealed no evidence in favor of pathogenicity from truncating variants, posing a problem for evaluating RAD50 pathogenicity and the quality of these predictions. However, limiting analysis to rare missense variants in the RAD50 DNA-binding domain (P-loop hydrolase and Zn hook) significantly improved predictor performance (AUROC for the top-performing methods increased by 20-25%), suggesting that incorporating gene-specific information could substantially improve results over typical methods that train on genome-wide mutations data. Furthermore, a recent study for breast-cancer risk genes in over 113,000 women, revealed no significant association between RAD50 and breast cancer.[50]

**riskSNPs (CAGI2, CAGI3).** Data from genome wide association studies (GWAS) are providing extensive information on the relationship between genetic variation and the risk of complex trait disease, such that there are now over 1000 reliable associations between the presence of SNPs at a particular locus and risk of a common disease (https://www.ebi.ac.uk/gwas/). Each association implies that a variant in that locus influences a molecular process affecting disease risk. The goal of these challenges is to investigate the community's ability to identify underlying molecular mechanisms, given GWAS results. Since the correct mechanisms are unknown, there is no ranking of accuracy. In this sense, the challenge is different from the others in CAGI, aiming to assess the value of crowdsourcing in solving a pressing problem in data interpretation. Specifically, the goals are to ascertain which mechanisms appear most relevant, the degree of consensus between methods, and what fraction of loci can be assigned plausible mechanisms. Participants were provided with candidate SNPs for disease associated loci discovered in the Wellcome Trust Case Control Consortium (WTCCC1)[55] and follow-up studies of seven complex diseases.

In all, SNPs in 178 loci were included in the CAGI2 challenge. Participants were asked to consider whether each candidate SNP might influence disease risk via any of the following mechanisms: missense SNPs (those that result in an amino acid substitution in a protein, thus potentially affecting *in vivo* function) - 4 predictor groups contributed using a total of 7 methods; expression (altering RNA level by a variety of possible mechanisms - two groups submitting), splicing (two groups submitting), and microRNA binding sites on messenger RNA (1 submission). In the second iteration of this challenge (CAGI3), 6 groups participated submitting a total of 13 predictors.

Missense methods are also used in a number of other CAGI challenges. Broadly, the methods use information on relative conservation of amino acid type at the substituted position, analysis of the effect of the substitution on protein structure and function, or a combination of both approaches. Methods for identifying expression effects either made use of information on known functional sites such as transcription factor binding positions or information from large scale studies of the association between the presence of SNPs and altered levels of expression. For splicing effects, one group restricted predictions to direct impact on splice junctions, yielding a very small number of possible mechanism SNPs. The second group used a more comprehensive approach, including possible enhancer sites and effects on cryptic splicing sites. The single microarray binding site method also makes use of database information.

There are two primary conclusions from this crowd-sourcing experiment. First, the results suggest both missense effects and changes in gene expression levels play a substantial role in molecular level mechanisms underlying these diseases. The exact extent is not yet clear, both because of limited expression predictions, and because the precise numbers depend on which set of SNPs are considered candidates. More data are also needed to assess the relative roles of splicing and microarray binding, as well as other factors. Although there is considerable variation in the missense results, a consensus view is encouraging. Consistency is not the same as accuracy, and the results suggest that large scale testing and benchmarking of missense analysis methods is needed to establish accuracy measures. Results for expression are intriguing in that the two methods used are based on different principles and produce rather different results.

**SCN5A (CAGI2).** Brugada syndrome (BrS) (OMIM #601144) is a rare clinical condition characterized by atypical right bundle branch block (RBBB) and elevated ST-segments in right precordial leads in the absence of structural heart disease. Though most individuals with BrS are asymptomatic, the disease manifests at young age (20-40 y) and men are more likely affected than women. Common symptoms are syncope and cardiac arrest or sudden death at rest or during sleep. BrS is inherited as an autosomal- dominant trait, with incomplete penetrance. Mutations in nine genes encoding ion channel subunits or gene products affecting ion channel function have been associated with BrS or proposed as risk factors (SCN5A, SCN1B, SCN3B, CACNA1C, CACNB2, KCND3, KCNE3, GPD1L, and MOG1). Mutations in SCN5A represent the majority with about 300 mutations in SCN5A linked to the syndrome. On a functional level, BrS mutations in SCN5A lead to loss of $Na^+$ current through several mechanisms.

In this study, novel mutations in SCN5A were identified in two independent families with BrS and their effects on Nav1.5 channel function were investigated.

The mutant proteins were generated in the laboratory, heterologously expressed in CHO-K1 cells and analyzed using the patch-clamp technique. In these experiments, parameters such as current densities and channel kinetics (activation, inactivation, recovery from inactivation) were analyzed, comparing mutant channels to wildtype channels. Thus, the change induced by the mutant as a percentage change as compared to the wildtype channel was measured.

The challenge involved predicting the effect of 3 mutants on Nav1.5 function with respect to current densities, expressed as the percentage of current density reduction compared to the wildtype channel with a standard deviation. The predictions were assessed against the values obtained for each mutation in the patch-clamp experiments.

Four groups participated in this challenge, submitting a total of 7 predictions. However, the dataset (3 variants) was too small to draw any significant conclusions regarding performance.

**SickKids clinical genomes (CAGI4, CAGI5).** In the SickKids4 challenge, participants were provided full sequence data and phenotypic profiles in the form of Human Phenotype Ontology (HPO) terms[56] for 25 undiagnosed patients from the SickKids Genome Clinic Project (https://www.sickkids.ca/), and asked to identify diagnostic variants and also to provide secondary findings – putative variants relevant to other diseases other than the reason for clinical presentation. Proposed rare disease causative mutations for two of the cases were deemed diagnostic by the referring physicians. This was the first instance of the CAGI community directly contributing in the clinic. Four groups participated in this challenge, submitting a total of 4 predictions. Prioritized variants were located in genes that had partial overlap with the clinical phenotype and were successfully validated. In two instances, the patient's referring clinical geneticist re-assessed the patient in light of the proposed disease gene and concluded that it was a good fit for the patient's phenotype. This was the first instance of CAGI participants providing a clinical diagnosis.[57]

SickKids5 contained WGS data and associated phenotypic profiles for 24 undiagnosed SickKids patients, and additionally required that participants match each genome to a patient phenotype list. This situation is artificially more difficult than encountered in the clinic but has the advantage of providing a clearer test of the methods – genome/phenotype matches above random must be due to correct identification of causative variants. Eight groups participated in this challenge, submitting a total of 9 predictions. No group did better than random in this assignment. However, two of the nominated diagnostic variants predicted with the highest probability for correct genome-patient matches, while not meeting ACMG criteria, were considered reasonable candidates for phenotype expansions, again potentially resolving previously intractable cases. In three of those cases, the referring physician accepted the proposed variants as causative, again resolving previously intractable cases. However, as seen in the other similar challenges, while these and other proposed variants may be correct, they are not supported by additional evidence as required by the ACMG/AMP guidelines for clinical assignment.[45]

**TP53 Splicing (CAGI3).** This challenge involved 3 TP53 splicing mutations implicated in cancer, assayed using minigene constructs to experimentally determine if each mutation influences splicing fidelity in HEK293T cells. The aim of the challenge was to predict the outcome of these experiments.

Five groups participated in this challenge, submitting a total of 5 predictions. The best-performing method, with an accuracy of 67%, used VEP (Variant Effect Predictor),[58] followed by manual inspection.

**Warfarin exomes (CAGI4).** Warfarin is the most commonly prescribed anticoagulant for preventing thromboembolic events. Warfarin has a twenty-fold inter-individual dose variability and a narrow therapeutic index, and it is responsible for a third of adverse drug event hospitalizations in older Americans.[59] Both clinical modifiers and genetic polymorphisms are known to affect an individual's stable therapeutic warfarin dose.[60] A dose estimator (IWPC) based on the status of SNPs in two genes as well as age, height, weight, race and two other prescribed drugs has been developed for Europeans; however, these algorithms are less predictive in diverse populations.[61] In practice, physicians probably utilize a trial and adjust approach.

In this challenge, exome data and clinical covariates were provided for 53 African American individuals, with the aim to predict the therapeutic dose of warfarin. Exomes from an additional 50 African including warfarin doses were provided for use in training. Three groups participated with a total of 9 submissions. Results here were disappointing, with a maximum $R^2$ value of 0.25, compared with 0.35 obtained with the Caucasian standard IWPC method.[60] The small sample size was likely a limiting factor in method performance.[35]

# Implementation Details

## Experimental-Max

When experimental assays use replicates, the uncertainty in the measurements can be quantified as the standard deviation of the replicates. The mean of the replicates is used as the experimental value, against which the predictions are evaluated. The uncertainty in the measurements poses an upper limit on the performance of the methods. To quantify this upper limit, we generate predictions reflecting the experimental uncertainty by sampling from a Gaussian distribution with the replicate mean and standard deviation as the parameters. Predictions are generated by this simulation 1000 times and the mean performance over these predictions is reported as the Experimental-Max performance. In the case where a method's experimental value prediction is also used to create a score for the classification task, the Experimental-Max estimate of the classification performance measures is also computed. A summary of the estimates of a measure, in terms of mean, standard deviation, median, $5^{th}$ and $95^{th}$ percentile are stored, to be used in the figures and tables. In particular, the confidence intervals are derived from the $5^{th}$ and $95^{th}$ percentile and the AUC values in the figures are reported along with the 1.96 standard deviation.

## ROC from discrete class labels

In a few cases the methods submit a discrete class label instead of a continuous score. To plot the ROC curve in this case, we convert the predicted class labels to numeric scores: 0 (negative) and 1 (positive). The ROC curve is constructed from the numeric scores using the standard corrections for scores containing ties (see Handling ties in scores).

## log-log AUC

As argued in Methods, AUC, as a measure, is not calibrated appropriately for assessing a predictor's performance in the clinical context. We use Truncated AUC as the main measure to this end; see Methods. Additionally, we use log-log AUC as another measure in the clinical context. Loosely, we define it as the area under the curve formed by plotting log TPR against log FPR. Intuitively, log-scale stretches the small values of TPR and FPR and consequently, enhancing the contribution of low TPR and FPR regions in the area computation. This is not to mean that a low TPR at a given FPR value contributes to improved log-log AUC. As one would expect, a higher TPR at a given value of FPR still contributes to a larger area. However, a small increase in TPR at low TPR values leads to a higher increase in the area compared to the same increase at high TPR values. Similarly, a small decrease in FPR at small FPR values leads to a higher increase in the area compared to the same decrease at high FPR values.

There are some difficulties in practically computing log-log AUC that stem from the fact that log function maps the interval, $[0,1]$, the range of TPR and FPR, to an unbounded interval $(-\infty, 0]$. This would lead to an infinite area under the log-scaled curve, unless the TPR and FPR values are both bounded above 0. We bound the TPR and FPR values around the lowest positive values that they can take on a given dataset. For a dataset with $n_+$ positives and $n_-$ negatives, $1/n_+$ and $1/n_-$ are the smallest positive values that TPR and FPR can take at any given threshold. We restrict the range of log TPR and log FPR to $\log 1/n_+$ and $\log 1/n_- - \log 2$, respectively; i.e., if TPR at a given threshold is $0$, then log TPR is assigned the value $\log 1/n_+$ and similarly, if FPR at a given threshold is $0$, then log FPR is assigned the value $\log 1/n_- - \log 2$. The factor of $\log 2$ is subtracted for computing the minimum log FPR to account for the area under the standard ROC curve between FPR values of 0 and $1/n_{-}$, which is non-zero when a positive TPR is achieved at 0 FPR. When converted to log-scale this region has infinite length. Instead of completely removing this region from the area computation, we bound this region to a length of $\log 2$ to the left of $\log 1/n_{-}$. The choice $\log 2$ comes from the distance between the smallest positive value of FPR and the second smallest positive value of FPR $(2/n_{-})$ on a log-scale; i.e., $\log 2 = \log 2/n_{-} - \log 1/n_{-}$.

Note that to correctly compute the area under the log-scaled curve, we use log TPR = $\log 1/n_{+}$ as the reference line, instead of log TPR = 0; in other words, the area is computed between the log-scaled curve and the line log TPR = $\log 1/n_{+}$. In order to ensure that the measure has a range of $[0,1]$ (similar to the standard AUC), we normalize the area thus computed by dividing by the maximum area for the perfect classifier. The maximum area is given by the $(\log n_{+}) \times (\log 2n_{-}) = (\log 1 - \log 1/n_{+}) \times (\log 1 - (\log 1/n_{-} - \log 2))$. For convenience, we use the log function with base 10. For example, this allows us to convert the log TPR value of -2 as a TPR value of 0.01.

Unlike AUC and truncated AUC, the log-log AUC for the random classifier is not a constant. It is a function of the number of positives and negatives in the dataset. The width of the log TPR and log FPR axes is $\log n_{+}$ and $\log 2n_{-}$, respectively. If $n_{+} > 2n_{-}$, the log-log ROC curve of a random classifier intersects the log TPR axis. In this case the unnormalized log-log AUC is $0.5(\log 2n_{-})^2 + \log 2n_{-} (\log n_{+} - \log 2n_{-})$. After normalization it is $1 - 0.5(\log 2n_{-}) / (\log n_{+})$, which evaluates to a value above 0.5. However, if $n_{+} < 2n_{-}$, the log-log ROC curve of a random classifier intersects the log FPR axis. In this case the unnormalized log-log AUC is $0.5(\log n_{+})^2$. After normalization it is $0.5 (\log n_{+})/(\log 2n_{-})$, which evaluates to a value below 0.5. In the case when $n_{+} = 2n_{-}$, the log-log AUC of the random classifier is the same as that standard AUC, i.e., 0.5.

In comparison to Truncated AUC, log-log AUC is more aggressive in enhancing the importance of very small FPR and TPR range. Furthermore, the calibration of the log scaled curve is dependent on the number of positives and negatives in the dataset. Smaller FPR and TPR ranges get further enhanced with a larger dataset size. The appropriate level of calibration of the ROC curve in the clinical context requires further research. Potential approaches include using a power function with fractional power such as square root, cube root or the fourth root to scale FPR and TPR.

## Bootstrap

To estimate the variability of the performance measures, 1000 bootstrap samples were generated from the dataset. For all the classification and clinical measures, the positive and negative variants were resampled separately and then combined to create a single bootstrap sample. A summary of the bootstrap estimates of the measures in terms of mean, standard deviation, median, 5$^{th}$ and 95$^{th}$ percentile are stored, to be used in the figures an tables. In particular, the confidence intervals are derived from the 5$^{th}$ and 95$^{th}$ percentile and the AUC values in the figures are reported along with the 1.96 standard deviation.

## Handling infinity and indeterminate values

Some of the classification score dependent measures, considered in our analysis can take infinite value in theory and when computed on real data (e.g., $LR^+$, $LR^-$, DOR, local $lr^+$). To obtain finite bootstrap summaries, we replace the (positive) infinite values by 1000, as a conservative approximation. If a finite value greater than 1000 was achieved by the measure on any other score, the infinite values of the measure are replaced by the maximum finite value, instead of 1000.

Furthermore, measures such as $LR^-$ and DOR when computed from real data achieve indeterminate value of 0/0 at the smallest score. This is because TPR and FPR at the smallest score are both equal to 1 and consequently, FNR and TNR are both equal to 0. Since the ratio FNR/TNR appears in the formulation of $LR^-$ and DOR, the two measures are indeterminate at the smallest score. In this case, we replace the indeterminate value at the smallest score by the value of the measure computed at the second smallest score (strictly smaller than the smallest score) to enforce continuity.

## Handling ties in scores

Ties in scores, predictions and observed experimental values need to be handled explicitly, while computing some measures. This includes Spearman's correlation, Kendall's tau, TPR, FPR, posterior probability of pathogenicity ($\rho$), local $lr^+$, RR and other measures that are derived from these measures. Spearman's correlation is defined as the linear correlation between the observed and predicted values' ranks. As per the standard approach to handle ties, the initial ranks of the tied scores are modified before calculating the correlation. The modified rank of a given set of tied predictions (or observations), is obtained by averaging their original ranks. The standard correction of Kendall's tau for ties is given by Kendall's tau-b (formula in Methods). An efficient algorithm ($O(n \log n)$) to compute Kendall's tau-b was implemented.[62]

As per the standard approach, post-correction TPR (FPR) at a tied score value, is given by the maximum of the pre-correction TPR (FPR) values computed at all data points tied at that score. The pre-correction TPR (FPR) value at a data point is obtained by first ordering all data points in the descending order of its score value and then counting the number of positive (negative) labels encountered traversing from left to right on reaching the given data point and dividing by the total number of positive (negative) labels in the dataset. Correcting TPR and FPR for ties, gives the corrected ROC curve. The AUC, $LR^+$, $LR^-$, DOR, PPP, PPV and MCC are computed from the corrected TPR and FPR.

Computation of local $lr^+$, class prior adjusted posterior and RR, relies on the computation of the unadjusted posterior reflecting the data prior (see Methods). Consequently, correcting the unadjusted posterior for ties also corrects these measures. The unadjusted posterior at a score value is computed as the average class label (proportion of positive labeled points), where 1 and 0 are the positive and negative class labels, respectively, lying within a window around the given score. To correct for ties, the class labels for a set of data point with equal scores are replaced by a soft class label equal to the average class label in the set. Using the new class labels to compute the unadjusted posterior corrects for ties. Note that the correction is only required if a window around a score can potentially contain a non-trivial subset of a set of tied scores, which is indeed the case when a minimum number of data points, in addition to window width, is used to construct the window.

## Supplementary Figures

The evaluation for each challenge is contained in one or more sheets in Tables 2, 4, 5, 6 and 7; see Supplementary Tables and Analyzed Challenges. A summary of each sheet is visually presented as a column of plots in the figures below and in the main text. A description of the contents of each plot type is given below. In the following text, we use the term "selected methods" to refer to a subset of methods used to summarize the performance achieved on a given challenge. Loosely, the selected methods are obtained by first ranking each submitted method based on one or more measures from (Pearson's correlation, Kendall's Tau, AUC and Truncated AUC), depending on the type of analyses applicable to the challenge. Then each method's final rank is calculated as its average rank on the applicable measures; see Supplementary Tables. The selected methods are picked as the top two methods based on the final ranking. In case of Annotate all Missense challenge figures, four selected methods are picked: two from the meta predictors and two from the non-meta predictors. The top-ranking method, among the selected methods is referred to as the "primary selected method" and the others are collectively referred to as the "secondary selected methods".

The plots displayed for a challenge depend on the type of analysis applicable to the challenge; see Supplementary Tables and Analyzed Challenges. The confidence intervals and the 1.96 standard deviations displayed in the figures are computed with 1000 bootstrap samples. The prior used for the clinical analysis, corresponds to the data prior (see Supplementary Tables) for all the Biochemical effect challenges. In case of Annotate all Missense, figures are displayed with two priors: 0.1 and 0.01 corresponding to the diagnostic and screening setting, respectively.

**Scatter plot:** The scatter plot is displayed for any challenge with regression analysis. It displays the continuous experimental values (y-axis) measured in a challenge against the predicted values (x-axis) by the primary selected method. A light grey, solid perfect prediction line, $x = y$, is drawn for an easy comparison. If a classification type analysis is also applicable to a challenge and the ground truth class labels are defined by a class boundary separating the experimental measurements, the class boundary is displayed as a horizontal grey dashed line. The points on either side of the line correspond to the positive (purple) and negative (yellow) class. The scatter plot may include points (grey) not included in the classification analysis. To show how the predictions would separate the positives from the negatives a vertical grey dashed line is also displayed at the class boundary. Note that, though it is natural to use the class boundary as a threshold for the predictions, it is not the only possible threshold. When performing a clinical analysis, the evidence thresholds are more relevant. In all challenges with a clinical analysis, we display the reachable evidence thresholds and their 90% confidence interval below the scatter plot. Since the evidence thresholds are computed w.r.t. a class prior, the prior is also displayed.

**Correlation bar plot:** A correlation bar plot is displayed for any challenge with a regression analysis. Kendall's tau and Pearson's correlation is displayed for all selected methods and, if available, the baseline method and the Experimental-Max. 90% confidence interval are displayed for the correlations. The correspondence between the bar colors and the methods is given in the ROC curve legend.

**ROC plots:** An ROC plot is displayed for any challenge with classification analysis. ROC curves are displayed for all selected methods and, if available, the baseline method and the Experimental-Max. The corresponding AUC value along with 1.96 standard deviation is displayed in the legend. If clinical analysis is applicable to the challenge, Truncated ROC curves (see Methods) and log-log ROC curves (see Implementation Details) are also displayed in two separate plots. The corresponding Truncated AUCs and the log-log AUCs, along with 1.96 standard deviation, are also given in the legend. The dashed grey line in all ROC plots corresponds to the random classifier. For the primary selected method, points corresponding to the (FPR, TPR) values at the evidence thresholds, reached by the method, are displayed on the Truncated ROC curve and are annotated as Sup(porting), Mod(erate) or Str(ong). Since the evidence thresholds, and consequently the (FPR TPR) values they achieve, are computed w.r.t. a class prior, the class prior is also displayed in the plot.

**Posterior and $lr^+$plot:** For all the challenges with a clinical analysis, a plot with posterior probability of pathogenicity (red) and the local $lr^+$(blue) curves of the primary selected method is displayed. If a clinical analysis in not applicable, but classification is, only the $lr^+$ curve is displayed. Both curves are plotted as a function of the predictions by the method. A smoothed version of the curves, derived by fitting a neural network with two hidden neurons, is also provided in cases where the original curve is jagged. For all the biochemical effect challenges, the posterior curve is computed w.r.t. the data prior; see Supplementary Table. In case of Annotate all Missense, two posterior curves corresponding to the diagnostic and screening priors are displayed; see Methods. To prevent the figure from being too crowded the $lr^+$ curve is displayed on a separate plot in that case. For the Biochemical challenges, where the two curves are displayed in the same plot, the posterior values are read w.r.t. the left axis and the $lr^+$ values, w.r.t to the right axis. The reached evidence thresholds along with their confidence intervals are displayed on the posterior curve. The posterior value at which an evidence threshold is displayed gives the minimum value of the posterior required to achieve that level of evidence. The $lr^+$value where a vertical line drawn from an evidence threshold intersects the $lr^+$curve gives the minimum value of $lr^+$ required to achieve that level of evidence. For most Biochemical challenges the posterior and $lr^+$curves increase in the left direction because, in those cases, low prediction scores correspond to the positive class. For other challenges, with curves increasing to the right, high prediction scores correspond to the positive class. This has implication on how the evidence threshold should be interpreted. When the curves are increasing to the left the all the variants having a prediction below an evidence threshold meet that level of evidence. Whereas if the curves increase to the right the variants with a prediction above an evidence threshold meet that level of evidence. The percentage listed next to an evidence threshold is the PPP value at that threshold computed w.r.t. the same prior as that used for computing the threshold and the posterior; see Methods. This value gives the percent of variants reaching that evidence level, assuming the prior used in the computation.

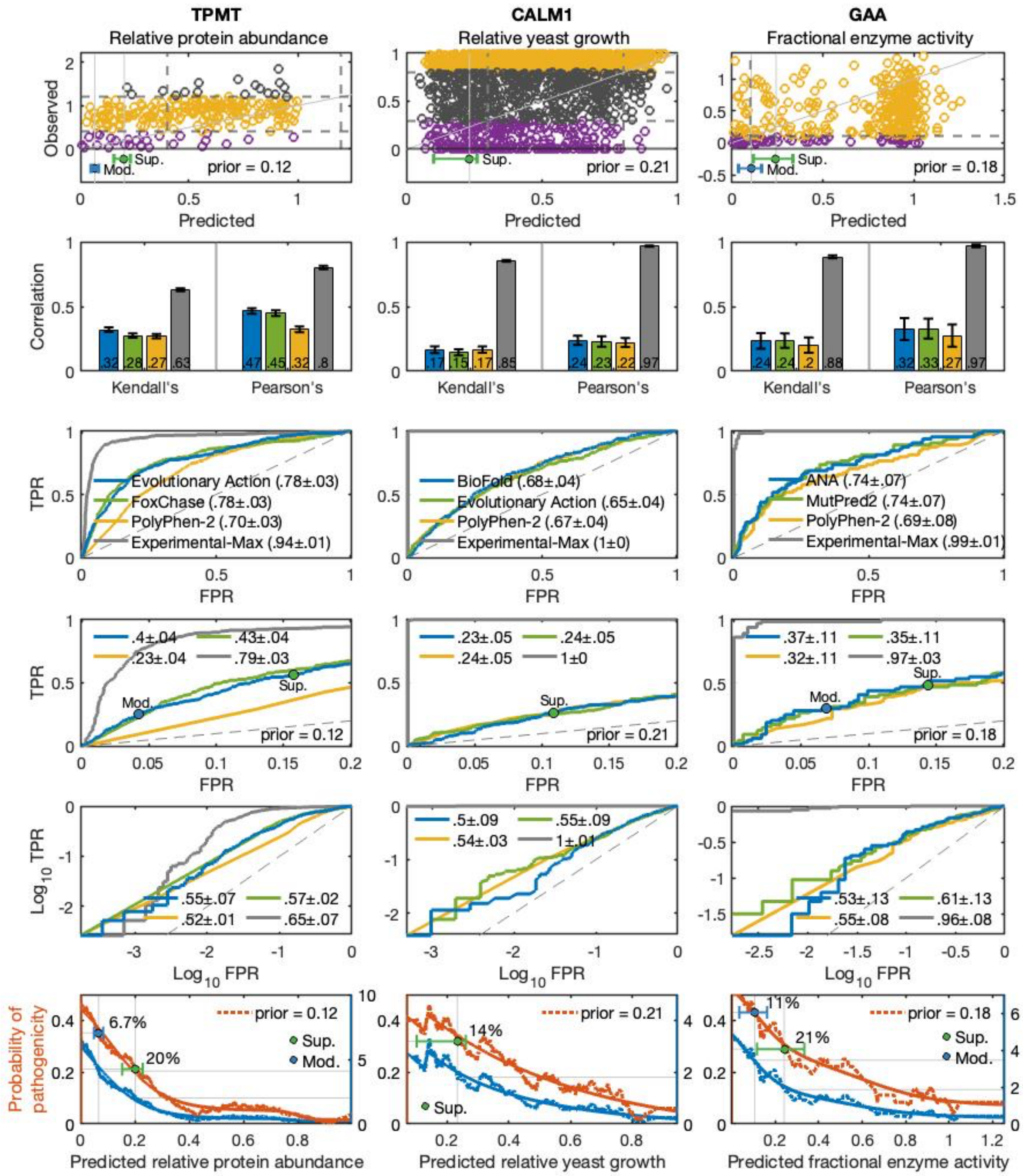

**Figure 1:** Summary of evaluation for TPMT (left), CALM1 (middle) and GAA (right) sheets in Table 2. The selected methods are picked based on Pearson's correlation, Kendall's Tau, AUC and Truncated AUC. (1) Row 1 contains the scatter plots of the experimental values versus their predictions by the primary selected method. The horizontal grey dashed line is the class boundary separating the experimental values into positives (purple) and negatives (yellow). In case of CALM1 there are two class boundaries to additionally separate the neutrals; see Analyzed challenges. A vertical grey dashed line is also drawn at the class boundary. A solid, light grey perfect prediction line, $x = y$, is drawn for easy comparison. Below the scatter plot, the thresholds for the clinical evidence levels, reachable by the primary selected method, along

with their 90% confidence intervals are displayed. The prior used to determine the thresholds is also displayed. (2) Row 2 displays Kendall's $\tau$ and Pearson's correlation for the selected methods (primary: blue, secondary: green), PolyPhen-2 and Experimental-Max. (3-5) Rows 3, 4 and 5 contain the ROC, Truncated ROC, and log-log ROC, respectively, for the selected methods, PolyPhen-2, Experimental-Max and the random classifier (dashed, grey line). The corresponding AUC, Truncated AUC and log-log AUC values, along with their 1.96 standard deviation, are also displayed. For the primary selected method (blue), the FPR, TPR values corresponding to the reachable clinical evidence thresholds are displayed as points on its Truncated ROC curve. The corresponding class prior is also displayed. (6) Row 6 contains the posterior probability of pathogenicity (red) and the local $\mathrm{lr}^+$ (blue) curve of the primary selected method. A smoothed version of both curves is also displayed. The posterior curve is read w.r.t. the left y-axis, whereas $\mathrm{lr}^+$ is read w.r.t. the right y-axis. The reachable evidence thresholds and their 90% confidence intervals are displayed on the posterior curve. The PPP value at an evidence threshold, giving the proportion of variants satisfying the threshold, is displayed as a percentage. The prior used to compute the posterior curve, evidence thresholds and PPP is also displayed.

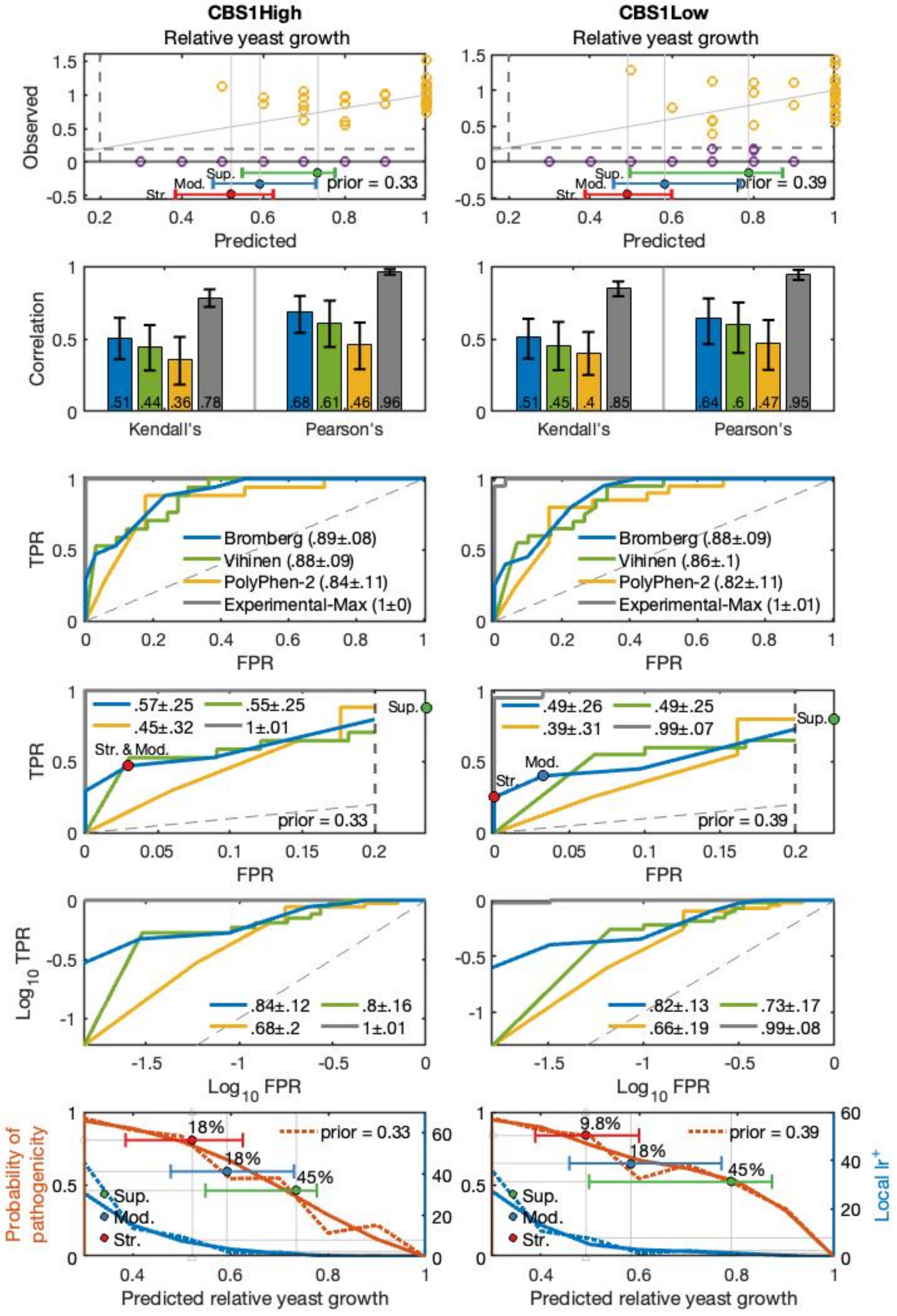

**Figure 2A:** Summary of evaluation for CBS1High (left) and CBS1Low (right) sheets in Table 2 for the CAGI 1 CBS challenge. The selected methods are picked based on Pearson's correlation, Kendall's Tau, AUC and Truncated AUC. (1) Row 1 contains the scatter plots of the experimental values versus their predictions by the primary selected method. The horizontal grey dashed line is the class boundary separating the experimental values into positives (purple) and negatives (yellow). A vertical grey dashed line is also drawn at the class boundary. A solid, light grey perfect prediction line, $x = y$, is drawn for easy comparison. Below the scatter plot, the thresholds for the clinical evidence levels, reachable by the primary selected method, along with their 90% confidence intervals are displayed. The prior used to determine the thresholds is also displayed. (2) Row 2 displays Kendall's $\tau$ and Pearson's correlation for the selected methods (primary: blue, secondary: green), PolyPhen-2 and Experimental-Max. (3-5) Rows 3, 4 and 5 contain the ROC, Truncated ROC, and log-log ROC, respectively, for the selected methods, PolyPhen-2, Experimental-Max and the random classifier (dashed, grey line). The corresponding AUC, Truncated AUC and log-log AUC values, along with their 1.96 standard deviation, are also displayed. For the primary selected method (blue), the FPR, TPR values at the reachable clinical evidence thresholds are displayed as points on its Truncated ROC curve. The corresponding class prior is also displayed. (6) Row 6 contains the posterior probability of pathogenicity (red) and the local $\mathrm{lr}^{+}$ (blue) curve of the primary selected method. A smoothed version of both curves is also displayed. The posterior curve is read w.r.t. the left y-axis, whereas $\mathrm{lr}^{+}$ is read w.r.t. the right y-axis. The reachable evidence thresholds and their 90% confidence intervals are displayed on the posterior curve. The PPP value at an evidence threshold, giving the proportion of variants satisfying the threshold, is displayed as a percentage. The prior used to compute the posterior curve, evidence thresholds and PPP is also displayed.

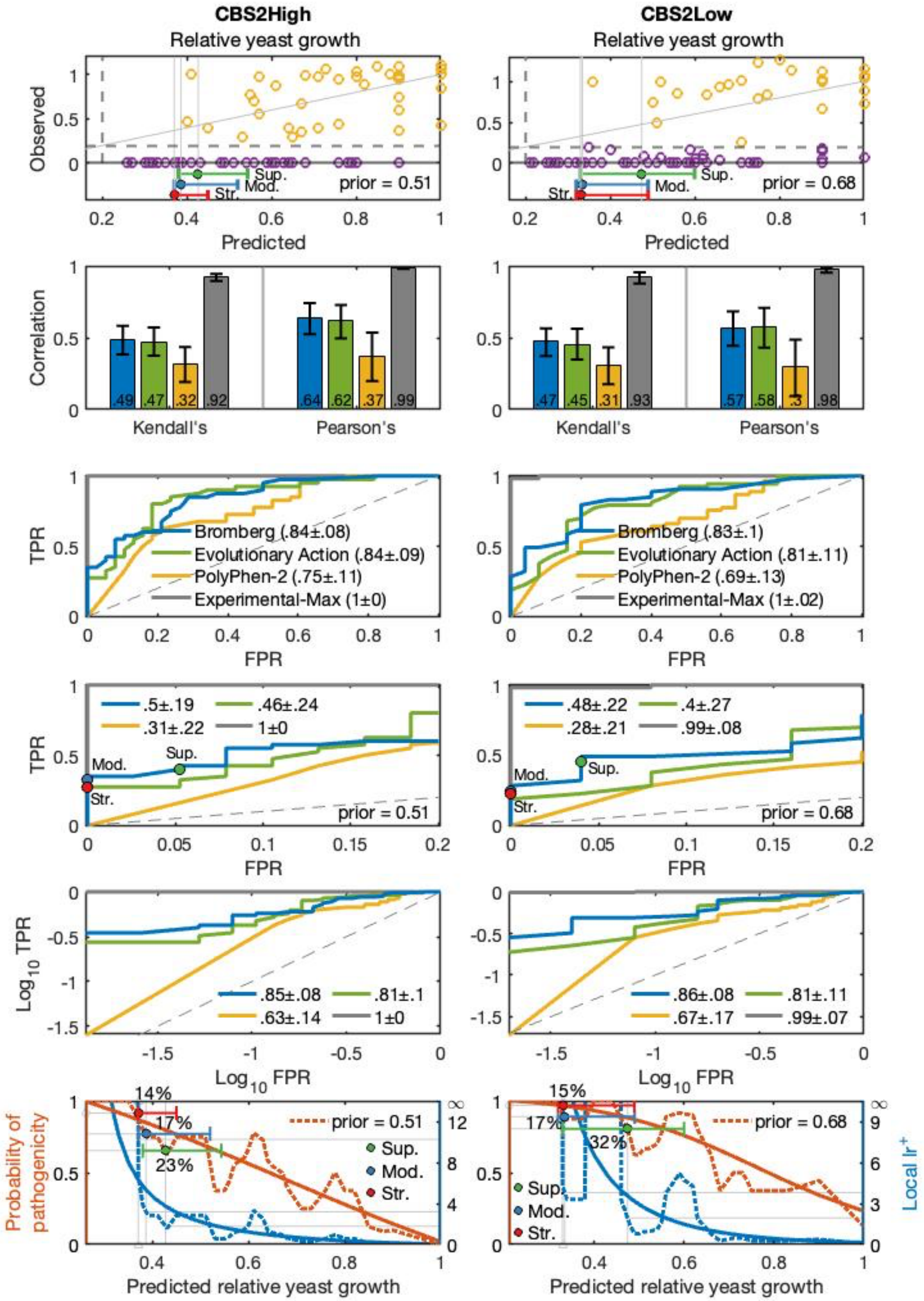
CBS2High
Relative yeast growth
Observed
Predicted
Sup.
Mod.
Str.
prior = 0.51
CBS2Low
Relative yeast growth
prior = 0.68
Correlation
Kendall's
Pearson's
.49 .47 .32 .92 .64 .62 .37 .99
.47 .45 .31 .93 .57 .58 .3 .98
TPR
FPR
Bromberg (.84±.08)
Evolutionary Action (.84±.09)
PolyPhen-2 (.75±.11)
Experimental-Max (1±0)
Bromberg (.83±.1)
Evolutionary Action (.81±.11)
PolyPhen-2 (.69±.13)
Experimental-Max (1±.02)
.5±.19
.46±.24
.31±.22
1±0
.48±.22
.4±.27
.28±.21
.99±.08
Log10 TPR
Log10 FPR
.85±.08
.81±.1
.63±.14
1±0
.86±.08
.81±.11
.67±.17
.99±.07
Probability of pathogenicity
Local lr+
14%
17%
23%
15%
17%
32%
Predicted relative yeast growth

**Figure 2B:** Summary of evaluation for CBS2High (left) and CBS2Low (right) sheets in Table 2 for the CAGI 2 CBS challenge. The selected methods are picked based on Pearson's correlation, Kendall's Tau, AUC and Truncated AUC. (1) Row 1 contains the scatter plots of the experimental values versus their predictions by the primary selected method. The horizontal grey dashed line is the class boundary separating the experimental values into positives (purple) and negatives (yellow). A vertical grey dashed line is also drawn at the class boundary. A solid, light grey perfect prediction line, $x = y$, is drawn for easy comparison. Below the scatter plot, the thresholds for the clinical evidence levels, reachable by the primary selected method, along with their 90% confidence intervals are displayed. The prior used to determine the thresholds is also displayed. (2) Row 2 displays Kendall's $\tau$ and Pearson's correlation for the selected methods (primary: blue, secondary: green), PolyPhen-2 and Experimental-Max. (3-5) Rows 3, 4 and 5 contain the ROC, Truncated ROC, and log-log ROC, respectively, for the selected methods, PolyPhen-2, Experimental-Max and the random classifier (dashed, grey line). The corresponding AUC, Truncated AUC and log-log AUC values, along with their 1.96 standard deviation, are also displayed. For the primary selected method (blue), the FPR, TPR values at the reachable clinical evidence thresholds are displayed as points on its Truncated ROC curve. The corresponding class prior is also displayed. (6) Row 6 contains the posterior probability of pathogenicity (red) and the local $\mathrm{lr}^+$ (blue) curve of the primary selected method. A smoothed version of both curves is also displayed. The posterior curve is read w.r.t. the left y-axis, whereas $\mathrm{lr}^+$ is read w.r.t. the right y-axis. The reachable evidence thresholds and their 90% confidence intervals are displayed on the posterior curve. The PPP value at an evidence threshold, giving the proportion of variants satisfying the threshold, is displayed as a percentage. The prior used to compute the posterior curve, evidence thresholds and PPP is also displayed.

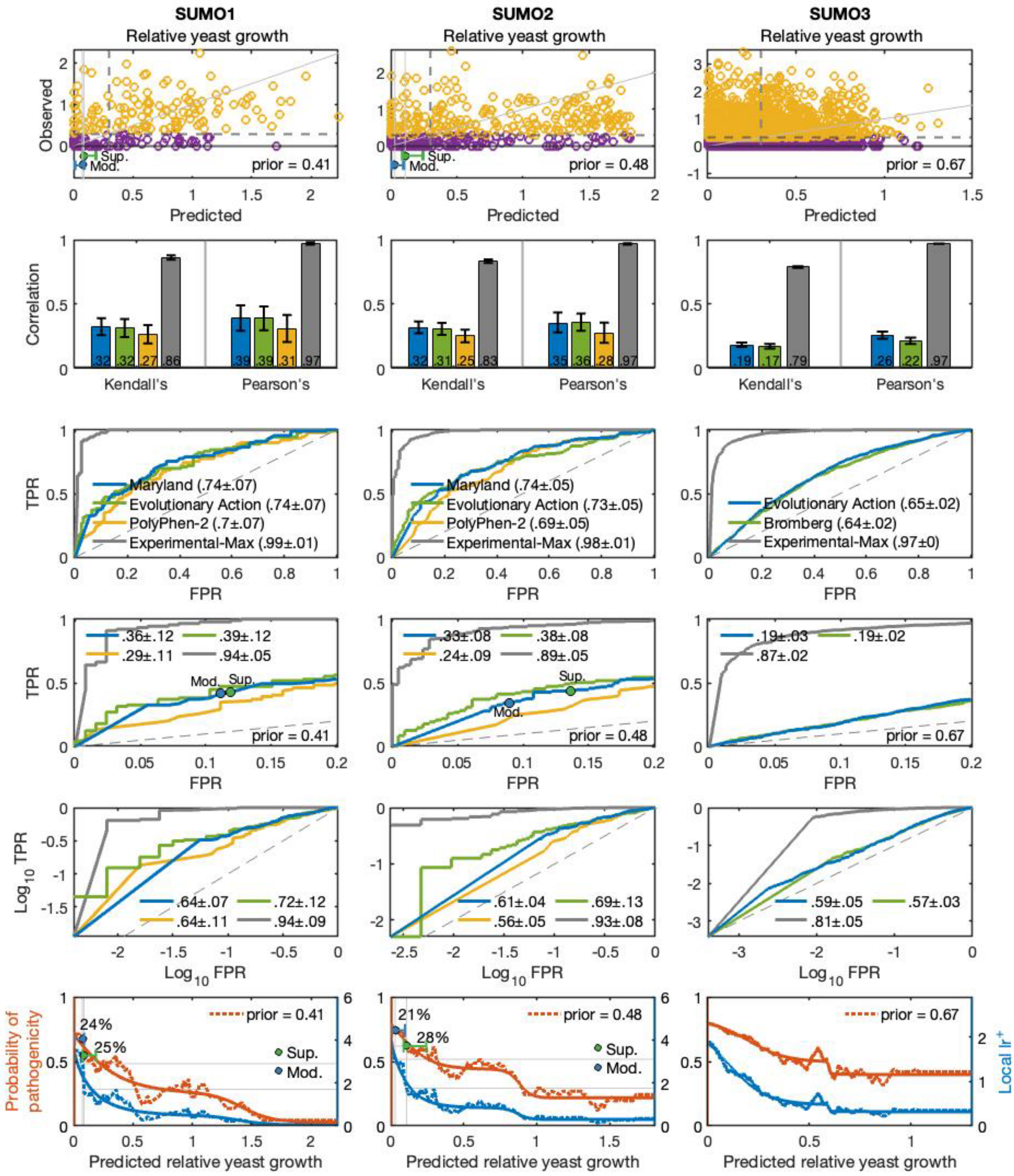

**Figure 3:** Summary of evaluation for SUMO1 (left), SUMO2 (middle) and SUMO3 (right) sheets in Table 2 for the three datasets in the SUMO challenge. The selected methods are picked based on Pearson's correlation, Kendall's Tau, AUC and Truncated AUC. (1) Row 1 contains the scatter plots of the experimental values versus their predictions by the primary selected method. The horizontal grey dashed line is the class boundary separating the experimental values into positives (purple) and negatives (yellow). A vertical grey dashed line is also drawn at the class boundary. A solid, light grey perfect prediction line, $x = y$, is drawn for easy comparison. Below the scatter plot, the thresholds for the clinical evidence levels, reachable by the primary

selected method, along with their 90% confidence intervals are displayed. The prior used to determine the thresholds is also displayed. (2) Row 2 displays Kendall's $\tau$ and Pearson's correlation for the selected methods (primary: blue, secondary: green), PolyPhen-2 and Experimental-Max. (3-5) Rows 3, 4 and 5 contain the ROC, Truncated ROC, and log-log ROC, respectively, for the selected methods, PolyPhen-2, Experimental-Max and the random classifier (dashed, grey line). The corresponding AUC, Truncated AUC and log-log AUC values, along with their 1.96 standard deviation, are also displayed. For the primary selected method (blue), the FPR, TPR values at the reachable clinical evidence thresholds are displayed as points on its Truncated ROC curve. The corresponding class prior is also displayed. (6) Row 6 contains the posterior probability of pathogenicity (red) and the local $\mathrm{lr}^{+}$ (blue) curve of the primary selected method. A smoothed version of both curves is also displayed. The posterior curve is read w.r.t. the left y-axis, whereas $\mathrm{lr}^{+}$ is read w.r.t. the right y-axis. The reachable evidence thresholds and their 90% confidence intervals are displayed on the posterior curve. The PPP value at an evidence threshold, giving the proportion of variants satisfying the threshold, is displayed as a percentage. The prior used to compute the posterior curve, evidence thresholds and PPP is also displayed.

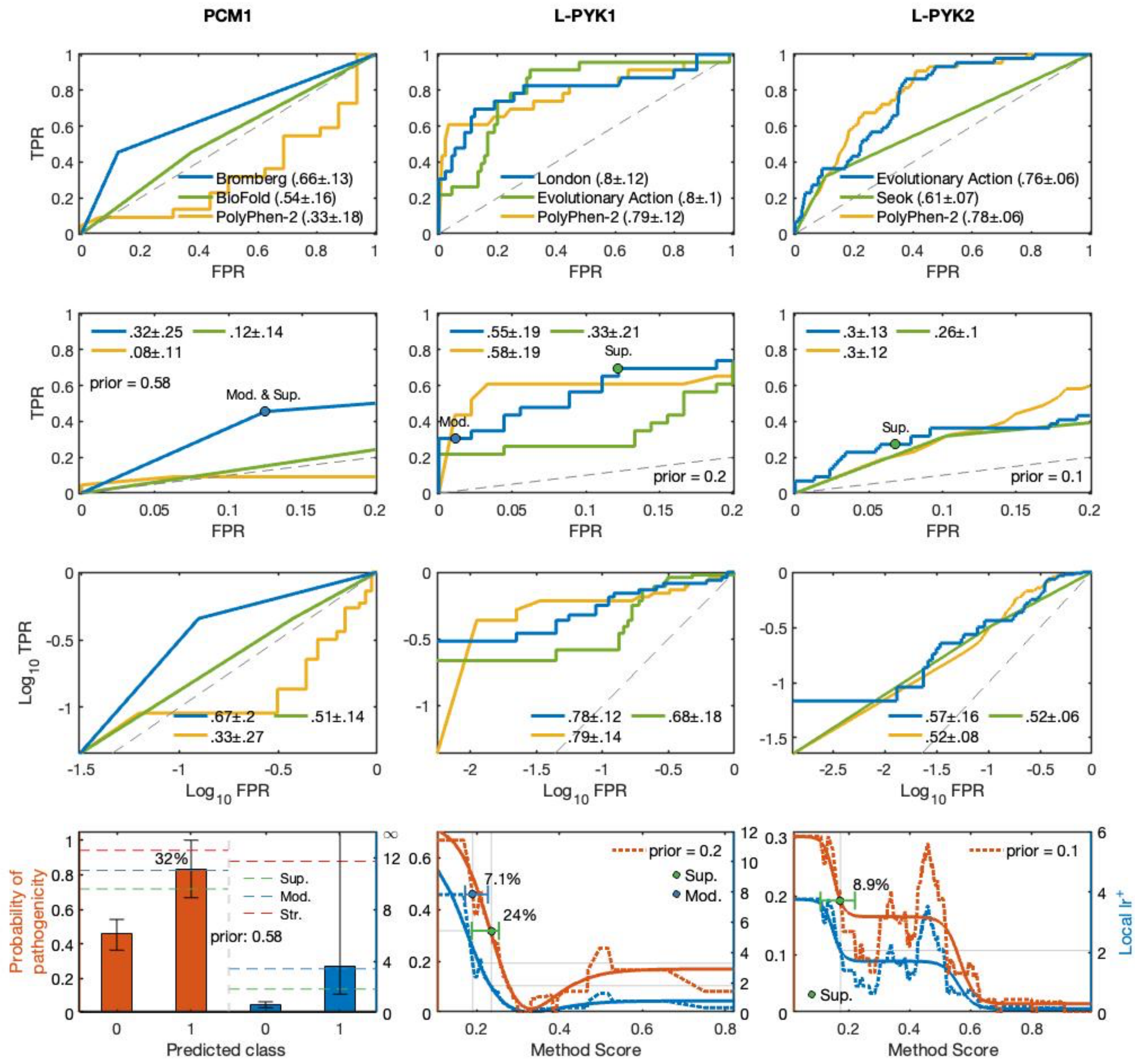

**Figure 4:** Summary of evaluation for PCM1 (left), LPYK1 (middle) and LPYK2 (right) sheets in Table 2 for the PCM1 challenge and the two datasets from the L-PYK challenge. The selected methods are picked based on AUC and Truncated AUC. (1-3) Rows 1, 2 and 3 contain the ROC, Truncated ROC, and log-log ROC, respectively, for the selected methods (primary: blue, secondary: green), PolyPhen-2 and the random classifier (dashed, grey line). The corresponding AUC, Truncated AUC and log-log AUC values, along with their 1.96 standard deviation, are also displayed. For the primary selected method (blue), the FPR, TPR values at the reachable clinical evidence thresholds are displayed as points on its Truncated ROC curve. The corresponding class prior is also displayed. (4) For LPYK1 and LPYK2, row 4 contains the posterior probability of pathogenicity (red) and the local $\mathrm{lr}^+$ (blue) curve of the primary selected method. A smoothed version of both curves is also displayed. Since continuous scores for the classification task in the PCM1 challenge were not available, the posterior and the $\mathrm{lr}^+$ values are displayed as barplots at the predicted class labels. Note that the local $\mathrm{lr}^+$ and $\mathrm{LR}^+$ are the equivalent in this case. The posterior curve is read w.r.t. the left y-axis, whereas $lr^+$ is read w.r.t. the right y-axis. For LPYK1 and LPYK2, the reachable evidence thresholds and their 90% confidence intervals are displayed on the posterior curve. For PCM1, in absence of continuous scores, there are only two possible values for the evidence thresholds: 0 and 1. Consequently, a confidence interval for the evidence threshold does not make sense. To see the clinical relevance of the variants predicted to be positive, we display the posterior and $lr^+$ cutoffs for the supporting, moderate and strong evidence levels. The variants do attain posterior and $lr^+$ values above the respective cutoffs for supporting and moderate evidence. However, the 90% confidence intervals for the posterior and $lr^+$ indicates that the evidence levels might not always be attained. The PPP value at an evidence threshold, giving the proportion of variants satisfying the threshold, is displayed as a percentage in all the plots. The prior used to compute the posterior curve, evidence thresholds and PPP is also displayed.

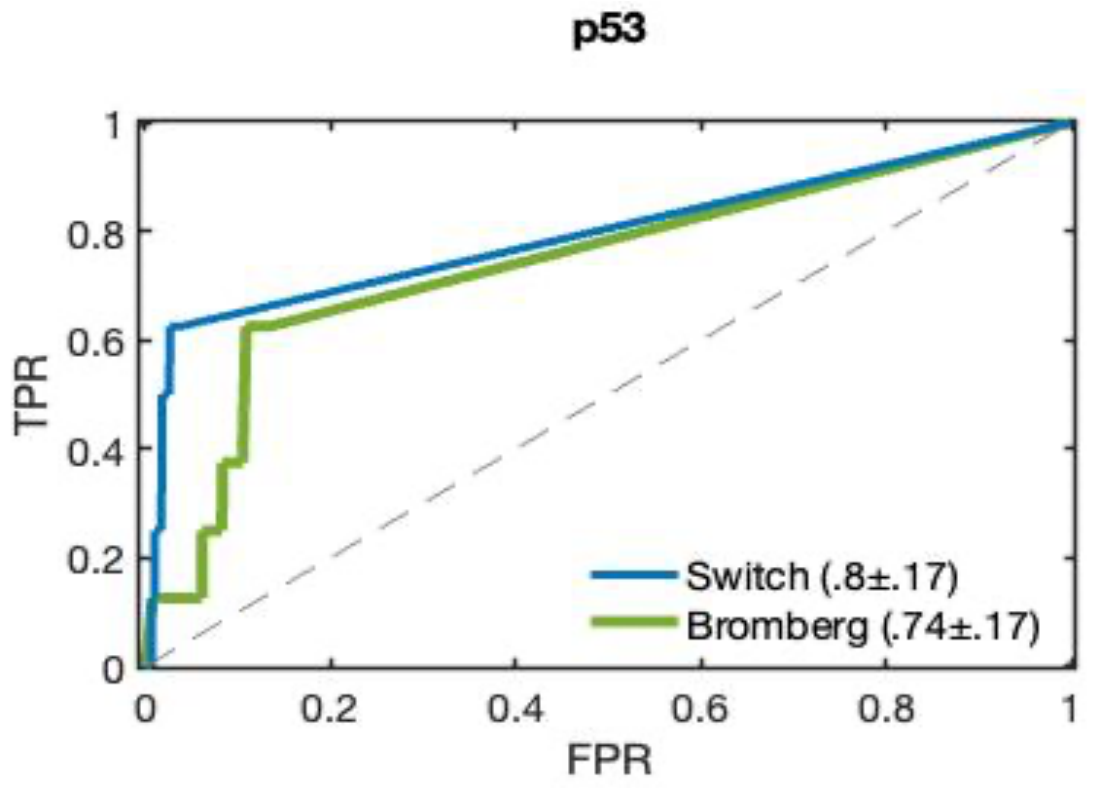

**Figure 5:** Summary of evaluation for p53 sheet in Table 2. The selected methods are picked based on AUC. ROC curves for the selected methods (primary: blue, secondary: green) are displayed. The corresponding AUC values, along with their 1.96 standard deviation, are also displayed. No reasonable baselines were available as this challenge is about the classifying variants as rescue mutations. The local $lr^+$ curve and quantities from the clinical analysis were not reported because of small number of positives, which leads to a high variance in the binning-based estimates.

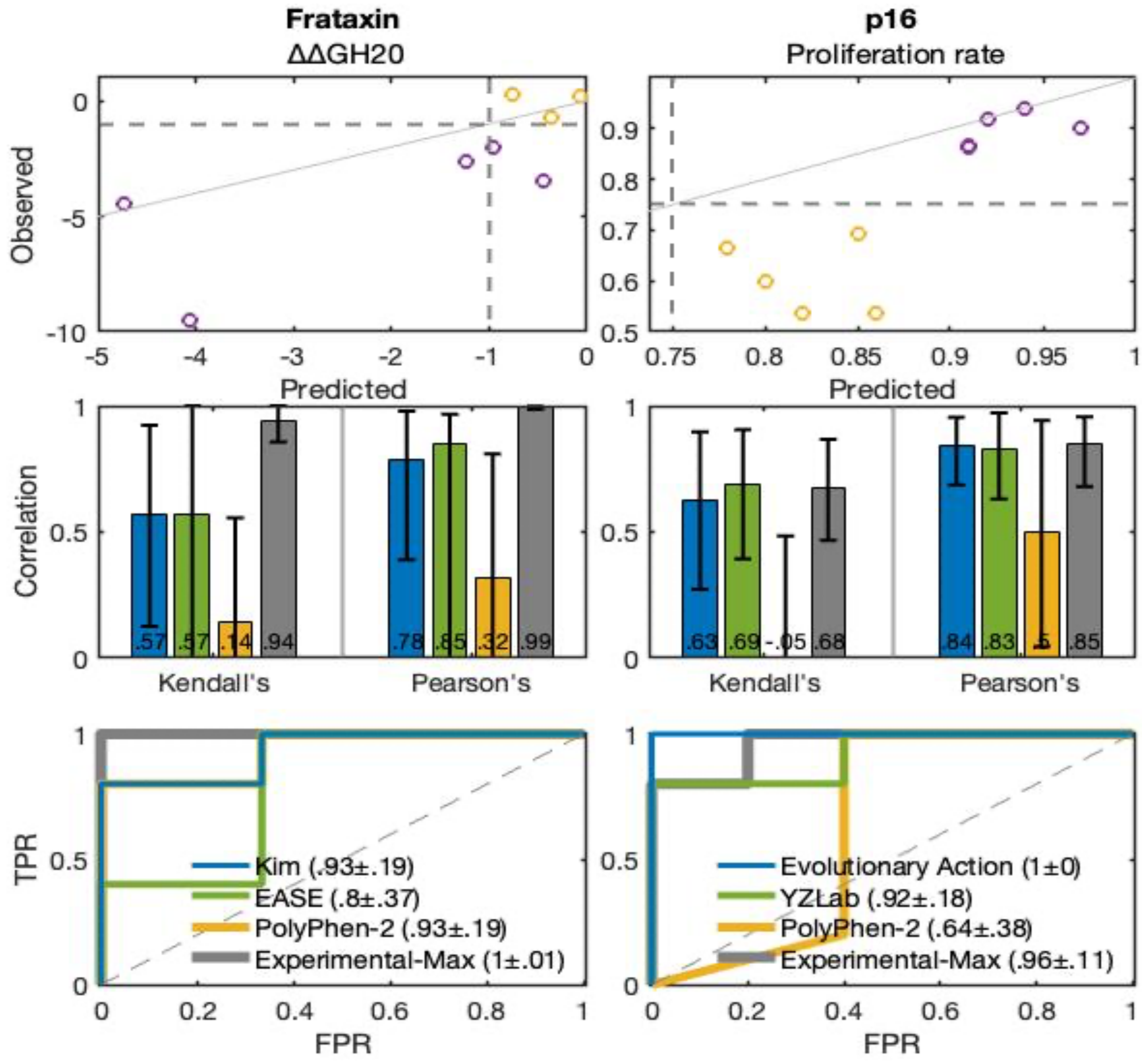

**Figure 6:** Summary of evaluation for Frataxin (left) and p16 (right) sheets in Table 2. The selected methods are picked based on Pearson's correlation, Kendall's Tau and AUC. (1) Row 1 contains the scatter plots of the experimental values versus their predictions by the primary selected method. The horizontal grey dashed line is the class boundary separating the experimental values into positives (purple) and negatives (yellow). A vertical grey dashed line is also drawn at the class boundary. A solid, light grey perfect prediction line, $x = y$, is drawn for easy comparison. (2) Row 2 displays Kendall's $\tau$ and Pearson's correlation for the selected methods (primary: blue, secondary: green), PolyPhen-2 and Experimental-Max. (3) Row 3 contains the ROC curve for the selected methods, PolyPhen-2, Experimental-Max and the random classifier (dashed, grey line). The corresponding AUC values, along with their 1.96 standard deviation, are also displayed. The local $\mathrm{lr}^+$ curve and quantities from the clinical analysis were not reported because of a small dataset size, which leads to a high variance in the binning-based estimates.

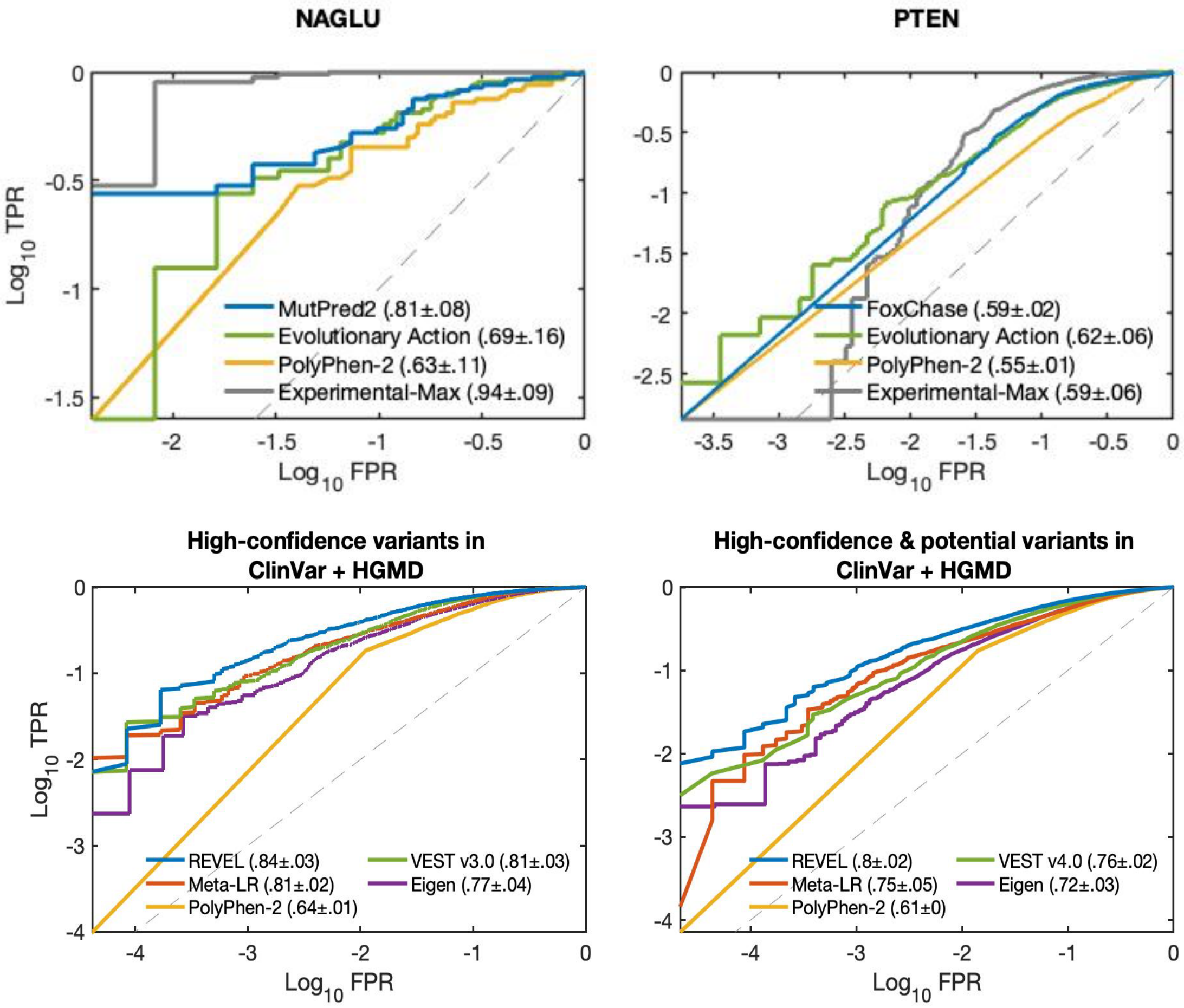

**Figure 7:** log-log ROC curves for NAGLU (top left) and PTEN (top right) sheets in Table 2 and AAM1All (bottom left) and AAM2All (bottom right) sheets for the Annotate all Missense in Table 4. For NAGLU and PTEN, the two selected methods are picked based on Pearson's correlation, Kendall's Tau and AUC and Truncated AUC. The plot displays log-log ROC for the selected methods (primary: blue, secondary: green), PolyPhen-2, Experimental-Max and the random classifier (dashed, grey line). For the Annotate all Missense sheets, the four selected methods (two meta and two non-meta predictors) are picked based on AUC and Truncated AUC. The plot displays log-log ROC for the four selected methods (primary: blue), PolyPhen-2 and the random classifier (dashed, grey line). The corresponding log-log AUC values, along with their 1.96 standard deviation, are also displayed. The remaining plots for NAGLU, PTEN, AAM1All and AAM2All are given in the Figure 2 and 3 (Main text).

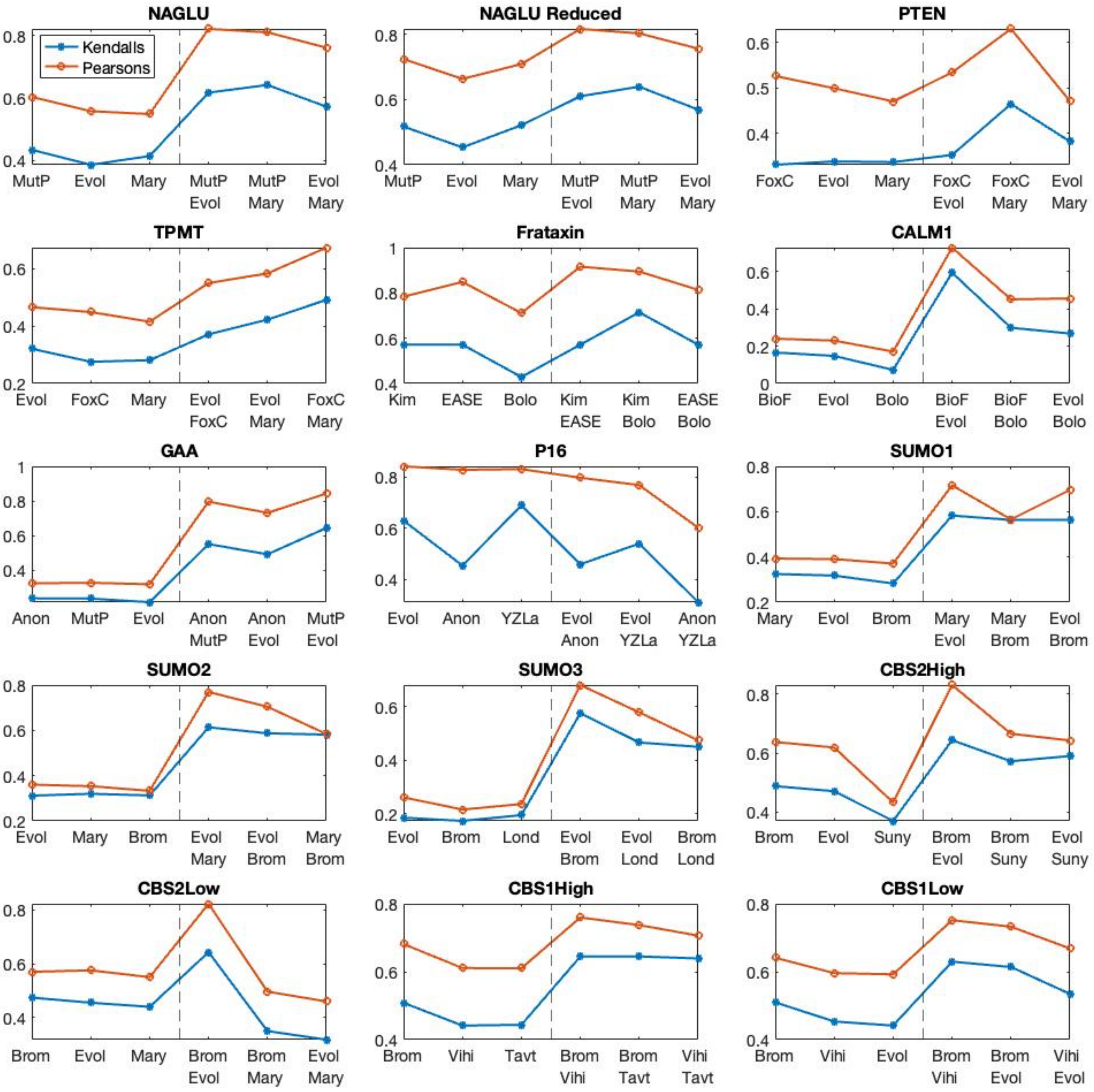

**Figure 8**: Comparison of prediction vs. experimental-value correlation and prediction vs. prediction correlation for the Biochemical challenge sheets with regression analysis in Table 2. Three selected method are picked from each sheet; see Supplementary Tables. The first four letters of the method names are used for abbreviation. Each plot displays the prediction vs. experimental-value correlations (Pearson's correlation and Kendall's Tau) for the selected methods on the left of the grey, dashed line. The pairwise correlations between pairs of predictions from the selected methods is displayed to the right of the line. As a general trend, it seems that the predictions are more correlated amongst each other as compared to how they are correlated with the experimental value. Removing ten of the hard to predict variants from the 163 NAGLU variants significantly reduces the difference between correlations on the left and the right (2nd plot). This suggests the difference between the two correlations is disproportionately due to a small number of variants.

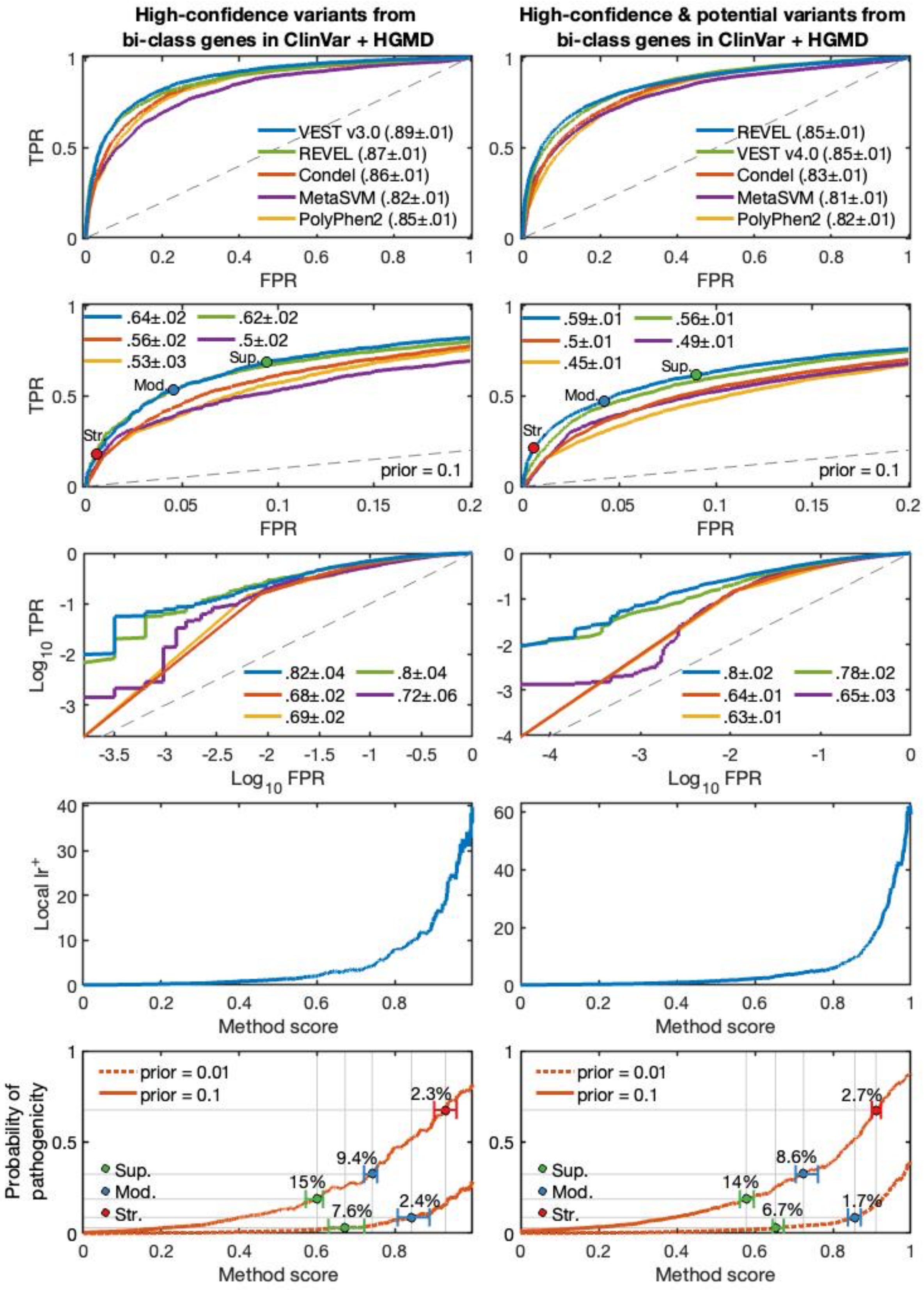

High-confidence variants from bi-class genes in ClinVar + HGMD
High-confidence & potential variants from bi-class genes in ClinVar + HGMD
VEST v3.0 (.89±.01)
REVEL (.87±.01)
Condel (.86±.01)
MetaSVM (.82±.01)
PolyPhen2 (.85±.01)
REVEL (.85±.01)
VEST v4.0 (.85±.01)
Condel (.83±.01)
MetaSVM (.81±.01)
PolyPhen2 (.82±.01)
TPR
FPR
.64±.02 .62±.02
.56±.02 .5±.02
.53±.03
.59±.01 .56±.01
.5±.01 .49±.01
.45±.01
Str. Mod. Sup.
prior = 0.1
Log10 TPR
Log10 FPR
.82±.04 .8±.04
.68±.02 .72±.06
.69±.02
.8±.02 .78±.02
.64±.01 .65±.03
.63±.01
Local lr+
Method score
Probability of pathogenicity
prior = 0.01
prior = 0.1
Sup.
Mod.
Str.
2.3% 9.4% 15% 7.6% 2.4%
2.7% 8.6% 14% 6.7% 1.7%

**Figure 9:** Summary of evaluation for AAM1BiClass (left), AAM2BiClass (right) sheets in Table 4 for the Annotate all Missense challenge. The selected methods are picked based on AUC and Truncated AUC. (1-3) Rows 1, 2 and 3 contain the ROC, Truncated ROC, and log-log ROC, respectively, for the four selected methods (two meta-predictors and two non-meta predictors), PolyPhen-2 and the random classifier (dashed, grey line). The corresponding AUC, Truncated AUC and log-log AUC values, along with their 1.96 standard deviation, are also displayed. For the primary selected method (blue), the FPR, TPR values at the reachable clinical evidence thresholds are displayed as points on its Truncated ROC curve. The corresponding class prior is also displayed. (4) Row 4 contains the local $lr^+$ curve of the primary selected method. (5) Row 5 contains the posterior probability of pathogenicity curves of the primary selected method at the screening (0.01) and diagnostic (0.1) prior. The reachable evidence thresholds and their 90% confidence intervals are displayed on the curves. The PPP value at an evidence threshold, giving the proportion of variants satisfying the threshold, is displayed as a percentage. The two priors used to compute the posterior curve, evidence thresholds and PPP are also displayed. Observe that the primary selected method on the confident set of variants is VEST3 whose ROC AUC decreased from 0.91 (Figure 3, Main text) to 0.88 as opposed to REVEL's that decreased from 0.92 (Figure 3, Main text) to 0.87. In the second column, REVEL and VEST4 remained the primary selected methods on a larger set of confident and potential variants, with an identical ROC AUC (0.85), but with REVEL achieving 3 percentage points higher performance in the low false positive rate region (Truncated AUC). In terms of clinical performance, the test data differences translated to a lower $lr^+$ at the extreme end of the prediction range and slightly more stringent thresholds, although there is almost no difference in the fraction of variants classified for Supporting, Moderate, or Strong evidential support (compare Figure 3, Main text). This suggests that the primary selected methods in this challenge are appropriately robust to be applied on a broad set of genes.

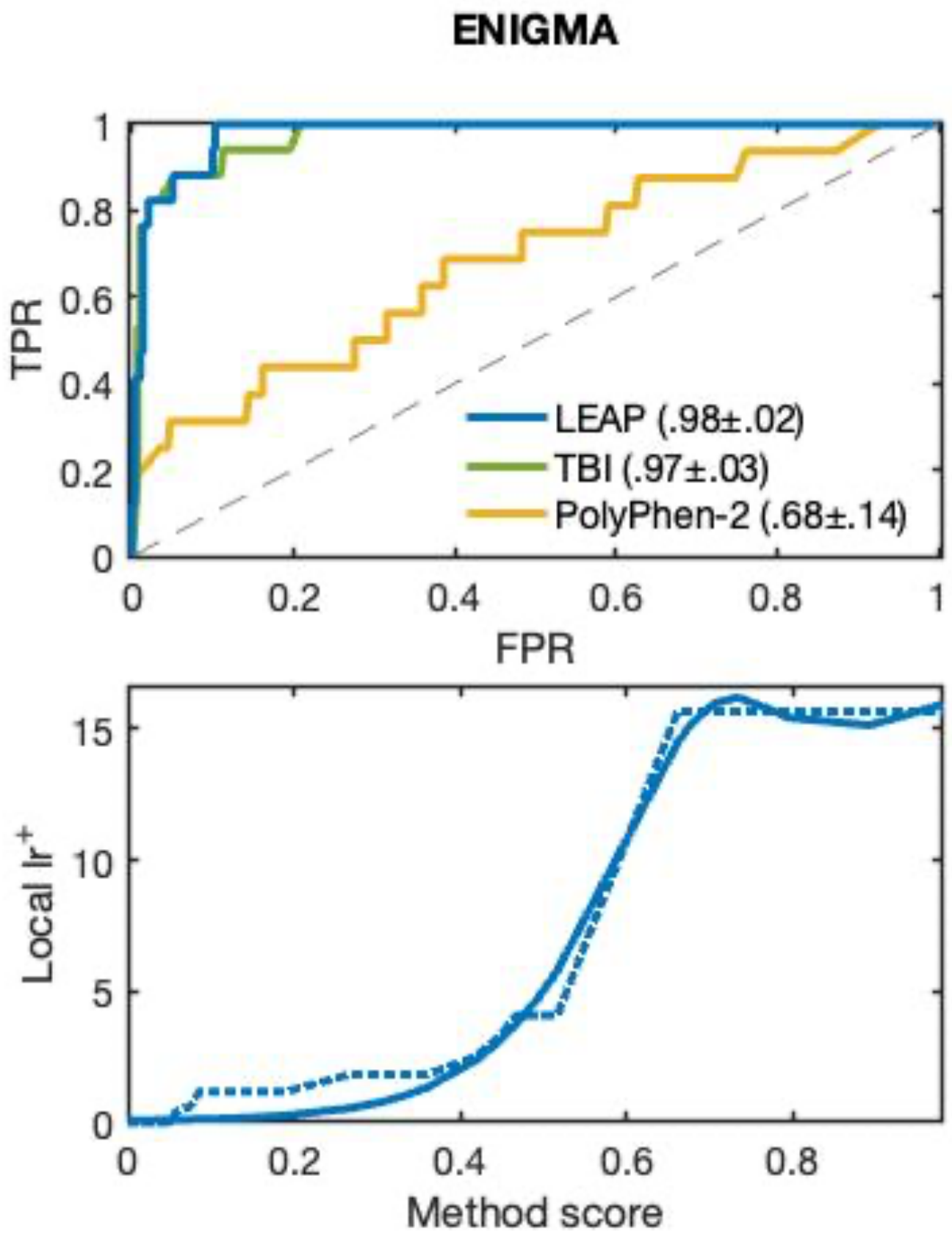

**Figure 10A:** Summary of evaluation for the ENIGMA sheet in Table 5. The selected methods are picked based on AUC. (1) The top plot contains ROC curves for the selected methods (primary: blue, secondary: green), PolyPhen-2 and the random classifier (grey dashed line). The corresponding AUC values, along with their 1.96 standard deviation, are also displayed. (2) The bottom plot contains the local $lr^+$ curve of the primary selected method, with and without smoothing.

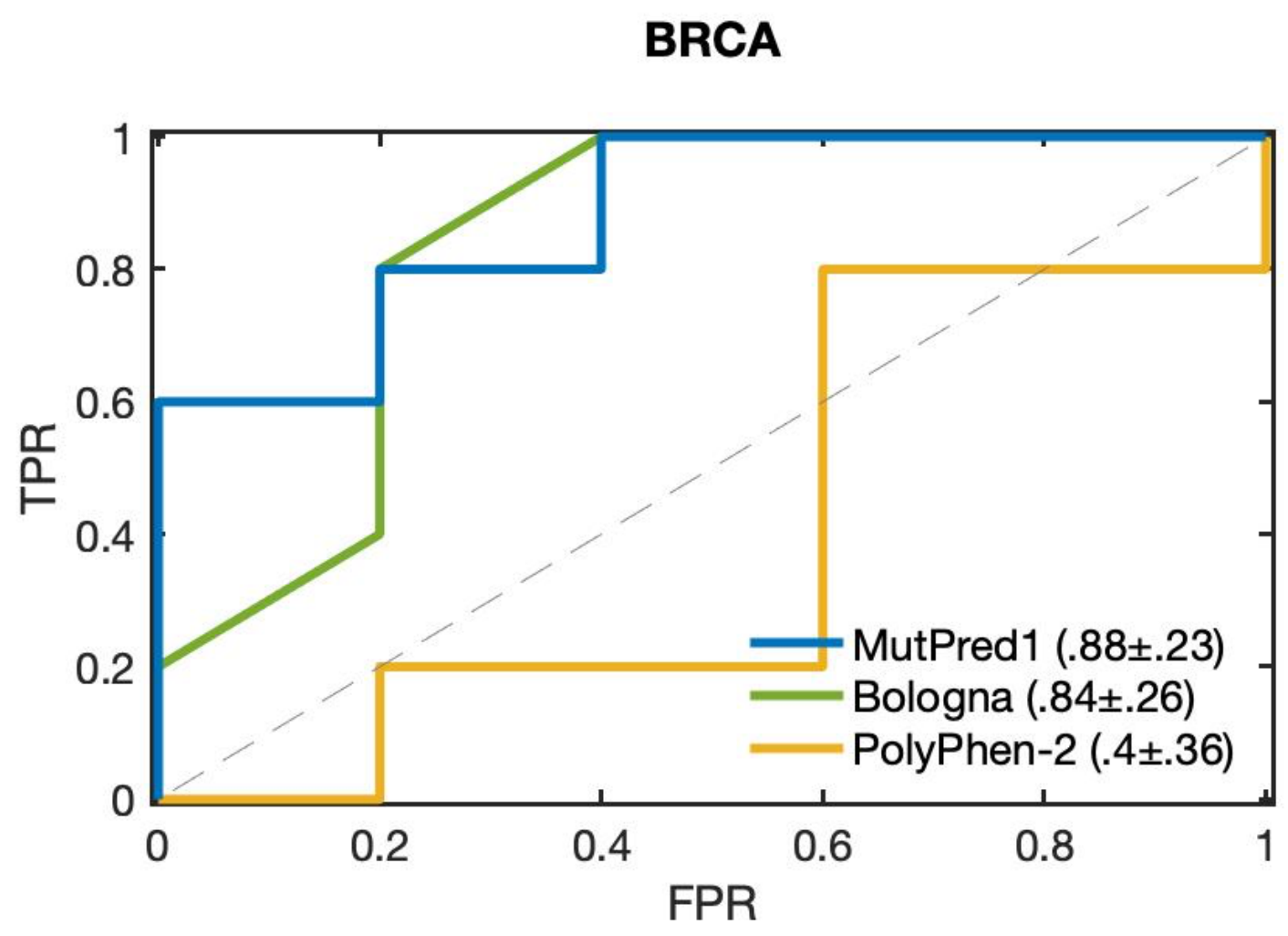

**Figure 10B:** Summary of evaluation for the BRCA sheet in Table 5. The selected methods are picked based on AUC. ROC curves for the selected methods (primary: blue, secondary: green), PolyPhen-2 and the random classifier (grey dashed line) are displayed. The corresponding AUC values, along with their 1.96 standard deviation, are also displayed. The local $lr^+$ curve was not reported because of small number of positives, which leads to a high variance in the binning based estimates.

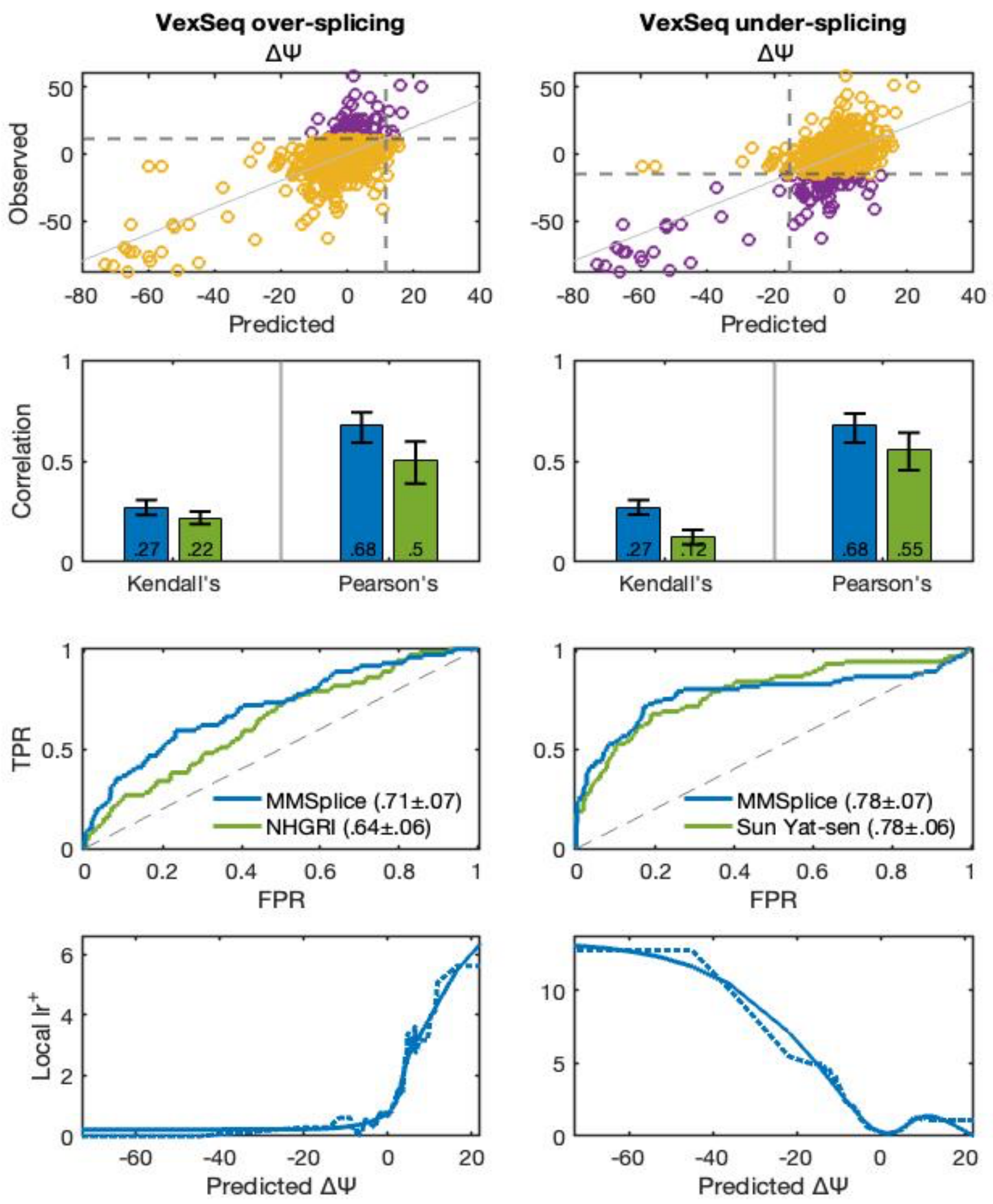

**Figure 11:** Summary of evaluation for VexSeq1 (left) and VexSeq2 (right) sheets in Table 6 for the Vex-seq challenge. The selected methods are picked based on Pearson's correlation, Kendall's Tau, and AUC. (1) Row 1 contains the scatter plots of the experimental values versus their predictions by the primary selected method. The horizontal grey dashed line is the class boundary separating the experimental values into positives (purple) and negatives (yellow). A vertical grey dashed line is also drawn at the class boundary. A solid, light grey perfect prediction line, $x = y$, is drawn for easy comparison. (2) Row 2 displays Kendall's $\tau$ and Pearson's correlation for the selected methods (primary: blue, secondary: green). (3) Rows 3 contains the ROC for the selected methods and the random classifier (dashed, grey line). The corresponding AUC values, along with their 1.96 standard deviation, are also displayed. (4) Row 6 contains the local $\text{lr}^+$ curve of the primary selected method. A smoothed version of both curves is also displayed. In spite of decent Pearson's correlation, the performance on predicting $\Delta\Psi$ is not very impressive as demonstrated by a relatively small Kendall's $\tau$. Pearson's correlation being sensitive to outliers, achieves a high value due to the few points on the lower right. The classification performance is stronger for under-splicing, compared to over-splicing, with the top ROC AUC of 0.78 and a maximum lr+ of 13.

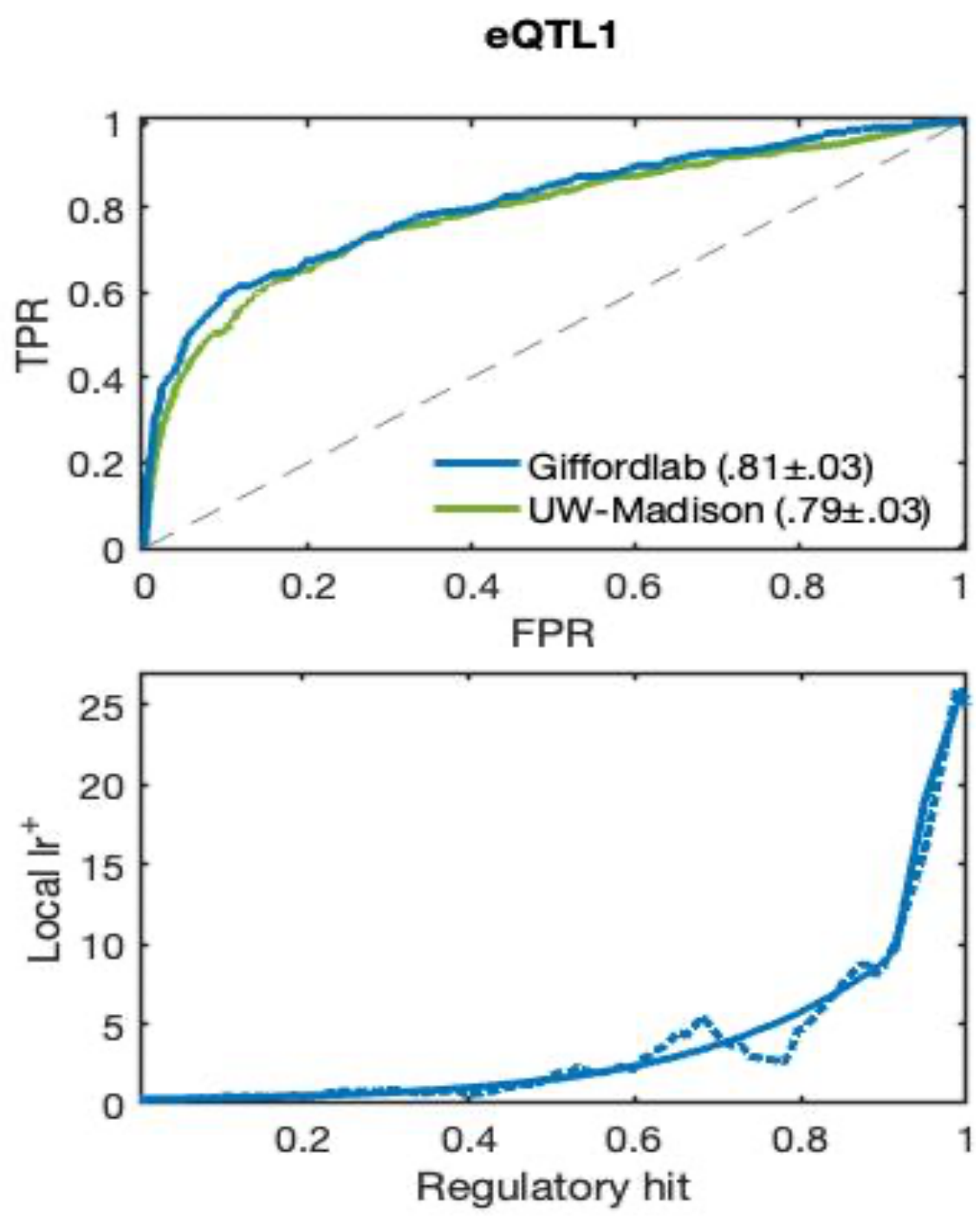

**Figure 12A:** Summary of evaluation for the eQTL1 sheet in Table 6 corresponding to the first part of the eQTL challenge for predicting regulatory hits. The selected methods are picked based on AUC. (1) The top plot contains ROC curves for the selected methods (primary: blue, secondary: green) and the random classifier (grey dashed line). The corresponding AUC values, along with their 1.96 standard deviation, are also displayed. (2) The bottom plot contains the local $lr^{+}$ curve of the primary selected method, with and without smoothing. As the ROC curve and the local $lr^{+}$curve shows, the performance by the primary selected method is good.

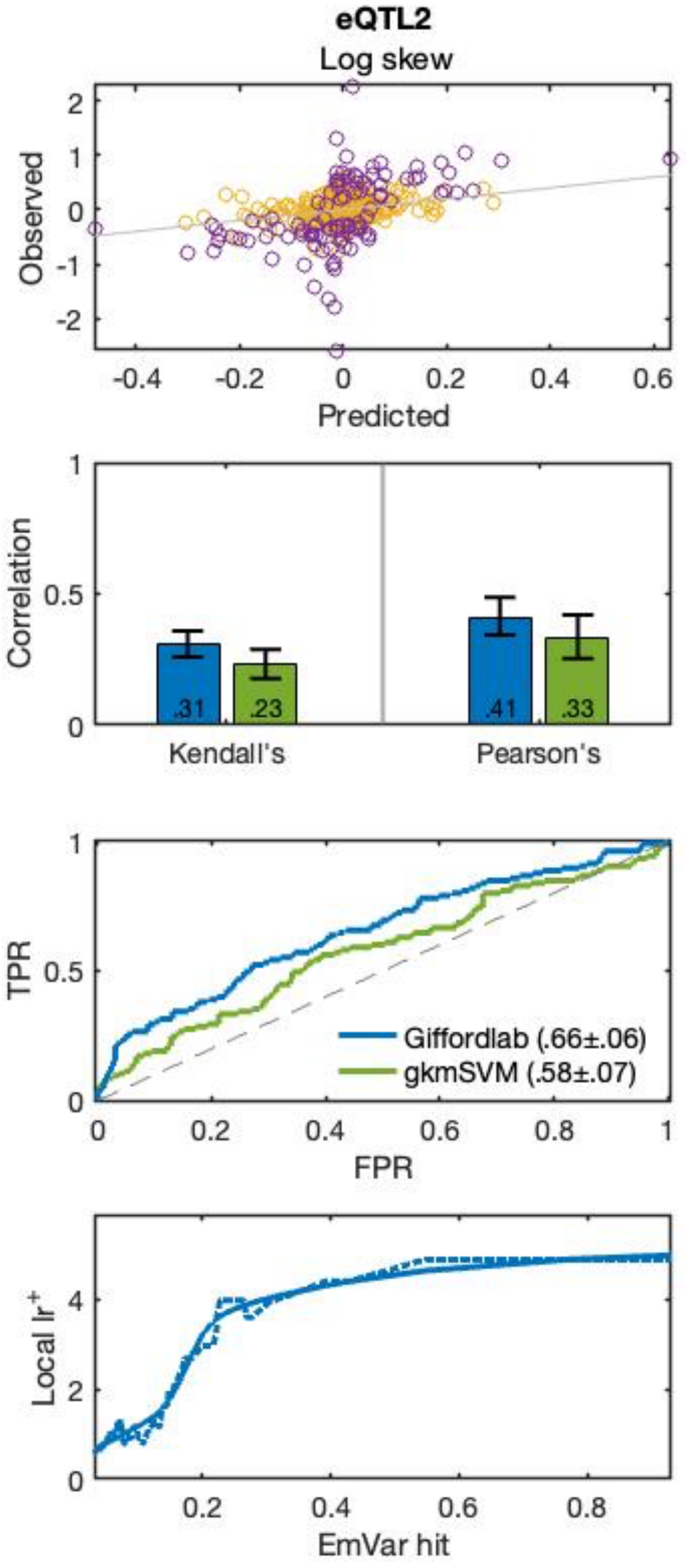

**Figure 12B:** Summary of evaluation for eQTL2 sheet in Table 6 for the second part of the eQTL challenge. The selected methods are picked based on Pearson's correlation, Kendall's Tau, and AUC. (1) Row 1 contains the scatter plots of the experimental $\log_2$ allelic skew values versus their predictions by the primary selected method. The positives ($emVar$) and negatives are shown in purple and yellow, respectively. A solid, light grey perfect prediction line, $x = y$, is drawn for easy comparison. (2) Row 2 displays Kendall's $\tau$ and Pearson's correlation for the selected methods (primary: blue, secondary: green). (3) Rows 3 contains the ROC for the selected methods and the random classifier (dashed, grey line). The corresponding AUC values, along with their 1.96 standard deviation, are also displayed. (4) Row 6 contains the local $\mathrm{lr}^+$ curve of the primary selected method. A smoothed version of both curves is also displayed. The performance in the log skew prediction task is modest (maximum Pearson's corr.: 0.41,

Kendall's $\tau$: 0.31). The performance in the *emVar* classification task is poor (maximum ROC AUC is 0.66). Few variants produce greater than a 2-fold expression changes, and experimental uncertainty is often close to that, making it difficult to draw firm conclusions.

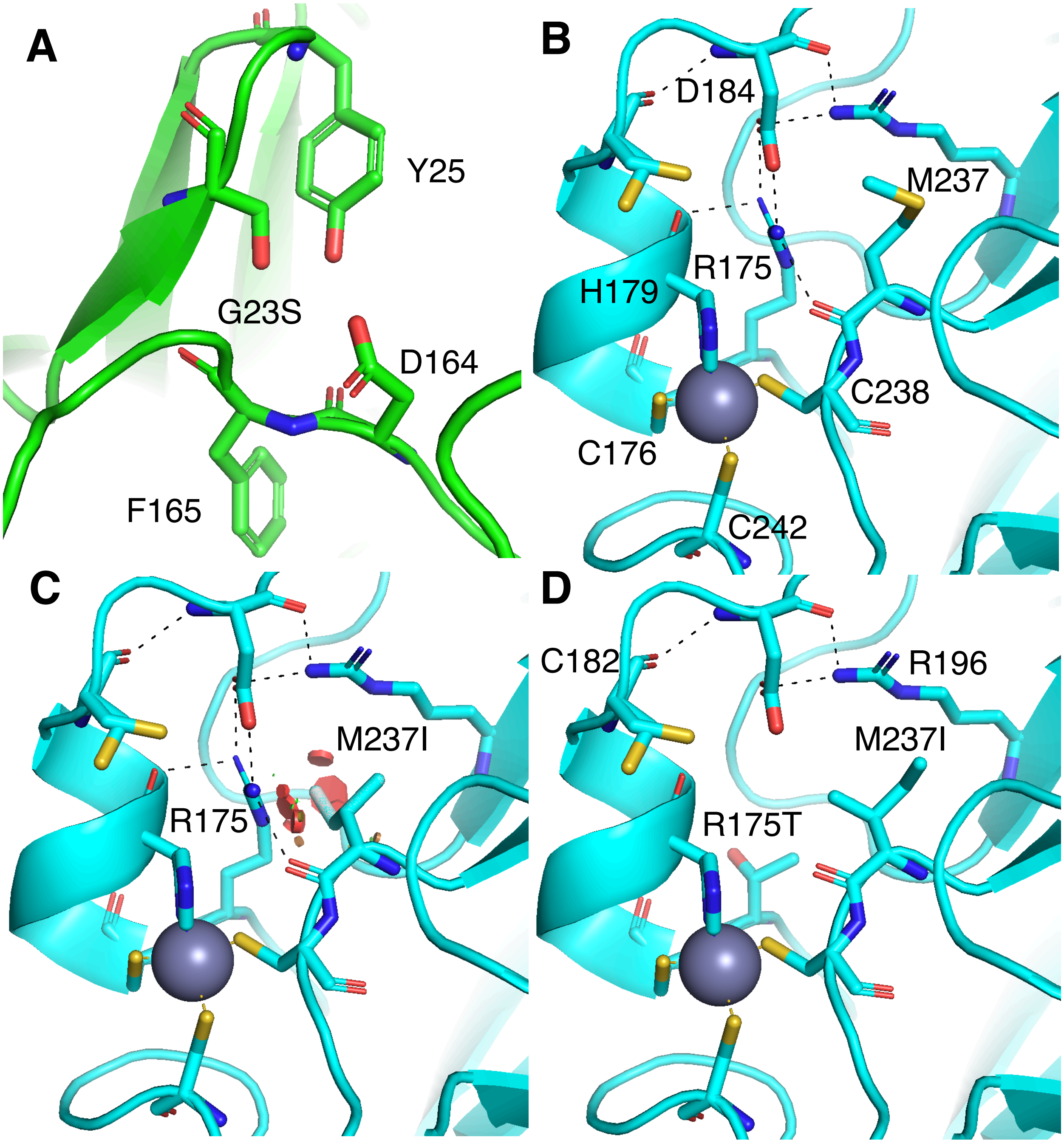

**Figure. 13**: *Examples of structure-based explanations of variant impact.* (A) In the p16 challenge, the G23S substitution results in unfavorable electrostatic interactions.
B -D: Analysis of a p53 cancer driver rescue mutation. (B) In wildtype p53, R175 affects coordination of the zinc ion (grey sphere) and is mostly buried, with its head group making electrostatic interactions with the side-chain of D184 and the main-chain carbonyl of M237. (C) The M237I cancer driver mutation sterically interferes with R175, disrupting the conformation required for zinc coordination. Steric clashes are indicated by red disks. (D) A large-scale experimental scan for rescue mutations found that replacement of R175 with a small or medium sized amino acid (R175A, R175V, R175S, R175T, R175P) restored function in the presence of M237I. As the figure shows, these R175 mutations relieve the steric clashes. The top-performing method identified the rescue mutations, using a combination of sequence conservation and stability analysis.

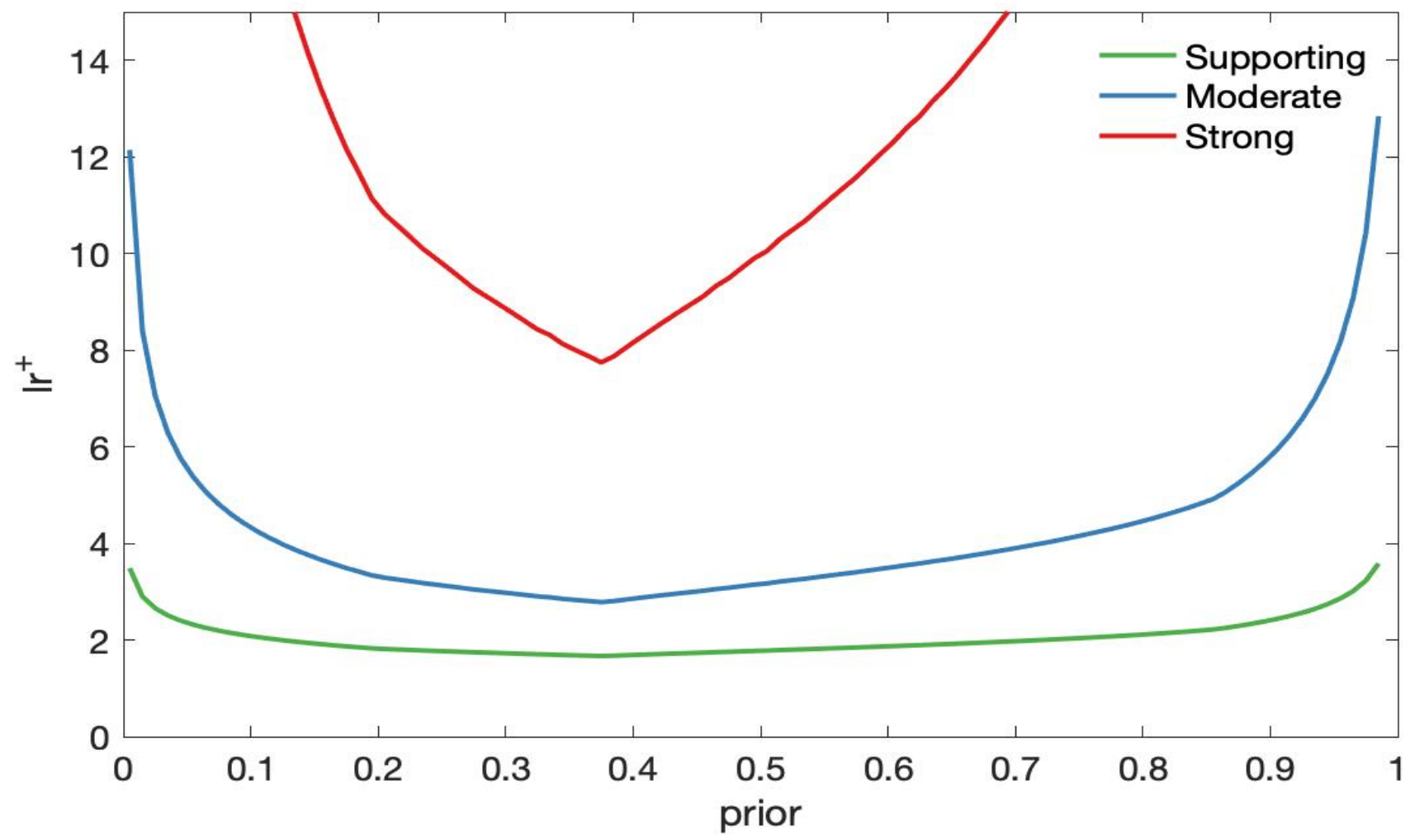

**Figure 14:** The local $lr^+$ cuttoffs for a variant to qualify as having Supporting, Moderate or Strong evidence for pathogenicity at different priors. The class prior was first selected as each of the values on the axis. The constant *c* was subsequently determined using the approach described in Methods. The values plotted as Supporting, Moderate, and Strong evidence, correspond to the eighth, fourth and square root of *c*, respectively.

## Supplementary Tables

**Table 1**. **CAGI challenges in CAGI1 through CAGI5.** Missense, SNVs and non-coding challenges involve individual genes, and often also include nonsense variants. Rare disease or Mendelian phenotypes indicate the involvement of a single gene, complex traits the involvement of multiple genes. SNVs: single nucleotide variants. Indels: insertions, deletions. WGS: whole genome sequencing.

| **Challenge** | **Edition** | **Genetic scale** | **Phenotypic characterization** | **# variants, traits or genomes** | **# sub-miss-ions** |
|---|---|---|---|---|---|
| Annotate all missense | CAGI5 | Missense | Rare disease | 81,084,849 | 5 |
| Asthma discordant monozygotic twins | CAGI2 | WGS | Complex trait, multiomics | 8 | 6 |
| Bipolar disorder[35, 63-65] | CAGI4 | Exomes | Complex trait | 1,000 | 29 |
| BRCA1 & BRCA2[65, 66] | CAGI3 | Missense, indels, non-coding (splicing) | Cancer | 100 | 14 |
| Breast cancer pharmacogenomics | CAGI2 | Other (multimodal) | Cancer | 54 | 3 |
| CALM1[11, 67, 68] | CAGI5 | Missense | Rare disease | 1,813 | 7 |
| CBS[65, 66, 69] | CAGI1 | Missense | Rare disease | 51 | 20 |
| CBS[65, 66, 69, 70] | CAGI2 | Missense | Rare disease | 84 | 20 |
| CHEK2[65, 66, 71] | CAGI1 | Missense | Cancer | 41 | 16 |
| CHEK2[37, 67, 68, 72] | CAGI5 | Missense | Cancer | 53 | 18 |
| Clotting disease[40, 67, 73] | CAGI5 | Exomes | Complex trait | 103 | 16 |
| Crohn's disease[35, 74] | CAGI2 | Exomes | Complex trait | 48 | 33 |
| Crohn's disease[35, 74] | CAGI3 | Exomes | Complex trait | 66 | 61 |
| Crohn's disease[35, 65, 74, 75] | CAGI4 | Exomes | Complex trait | 111 | 46 |
| ENIGMA BRCA1 and BRCA2[19-21, 67, 76, 77] | CAGI5 | Missense | Cancer | 430 | 10 |
| eQTL causal SNPs[27, 78, 79] | CAGI4 | Regulatory | Complex trait | 9,116 | 33 |
| FCH | CAGI3 | Exomes | Rare disease | 5 | 21 |
| Frataxin[14, 67, 68, 72, 80, 81] | CAGI5 | Missense | Cancer | 8 | 12 |
| GAA[5, 67, 72] | CAGI5 | Missense | Rare disease | 357 | 26 |
| HA | CAGI3 | Exomes | Rare disease | 4 | 18 |
| Hopkins clinical panel[44, 82] | CAGI4 | Gene panel | Rare disease | 106 | 5 |
| ID Panel[83-85] | CAGI5 | Gene panel | Rare disease | 146 | 15 |
| MaPSy[26, 86-88] | CAGI5 | Non-coding (regulatory) | Complex trait & rare disease | 4,964 | 14 |
| Mouse exomes | CAGI2 | Exomes | Rare disease | 4 | 2 |
| MRE11[66] | CAGI3 | Missense | Cancer | 42 | 23 |
| NAGLU[2, 66, 70, 71, 89] | CAGI4 | Missense | Rare disease | 163 | 17 |
| NPM-ALK[66] | CAGI4 | Missense | Cancer | 43 | 4 |
| NBS1[66] | CAGI3 | Missense | Cancer | 44 | 23 |
| p16[15, 65, 66, 70, 71] | CAGI3 | Missense | Cancer | 10 | 22 |
| p53 reactivation | CAGI2 | Missense | Cancer | 14,668 | 11 |
| PCM1[67, 72, 90, 91] | CAGI5 | Missense | Complex trait | 38 | 7 |
| PGP[54] | CAGI1 | WGS | Complex trait & Mendelian | 10 | 2 |
| PGP[54] | CAGI2 | WGS | Complex trait & Mendelian | 10 | 4 |
| PGP[54] | CAGI3 | WGS | Complex trait & Mendelian | 77 | 16 |
| PGP[54] | CAGI4 | WGS | Complex trait & Mendelian | 23 | 5 |
| PTEN[3, 67, 72] | CAGI5 | Missense | Cancer | 2,924 | 16 |
| Pyruvate kinase[66, 70, 92-94] | CAGI4 | Allostery missense | Rare disease | 113 | 5 |
| RAD50[65, 66, 71] | CAGI2 | Missense | Cancer | 69 | 14 |
| Regulation saturation[29, 95, 96] | CAGI5 | Non-coding (regulatory) | Complex trait | 17,500 | 23 |
| riskSNPs | CAGI2 | SNVs | Complex trait | 58,424 | 7 |
| riskSNPs | CAGI3 | SNVs | Complex trait | 110,477 | 13 |
| SCN5A[66] | CAGI2 | Missense | Rare disease | 3 | 7 |
| *Shewanella oneidensis* strain MR-1 | CAGI2, CAGI3 | Other (transposon insertion) | Other | 8 | 0 |
| SickKids[57, 97] | CAGI4 | WGS | Rare disease | 25 | 4 |
| SickKids[67, 97, 98] | CAGI5 | WGS | Rare disease | 24 | 9 |
| SUMO ligase[9, 66, 70, 89] | CAGI4 | Missense | Cancer, rare disease | 682 | 16 |
| TP53 splicing | CAGI3 | Non-coding (splicing) | Cancer | 3 | 5 |
| TPMT[3, 67, 72] | CAGI5 | Missense | Cancer | 3,736 | 16 |
| Vex-seq[26, 87, 99-101] | CAGI5 | Non-coding (splicing) | Complex trait & rare disease | 2,059 | 12 |
| Warfarin exomes[35] | CAGI4 | Exomes | Complex trait | 103 | 9 |

**Table 3 (excel file):**
Summary of results on fully analyzed biochemical effect challenges. The sheet "Full" shows the performance of primary selected method, the baseline model (PolyPhen-2) and Experimental-Max (if available) on seven performance measures for all sheets in Table 2. The primary selected method is obtained as the first predictor from the ranking procedure described in the next section. The measures include AUC, Truncated AUC, Person's correlation, Spearman's correlation, Kendall's tau, R-squared and Coverage (defined in the next section). The sheet "Clinical" includes a binarized information on whether the primary selected method, for each sheet (Table 2) with clinical analysis, reaches supporting, moderate, or strong evidential support in the clinic for three different class priors: data prior, 0.1 and 0.01 (see next section). The sheet "Reduced" shows the summary for 10 challenges (copied from "Full") used in the paper to report average performance over all challenges. Only one sheet was selected for any challenge with multiple sheets (CBS1, CBS2 and SUMO) in Table 2. Any sheet without a regression analysis (PCM1, L-PYK1, L-PYK2 and P53) were further excluded from the reduced set. Truncated AUC was not evaluated for two of challenges (Frataxin and P16) in the reduced set, since a clinical analysis was not performed due to small dataset sizes.

**Tables 2, 4, 5, 6 and 7 (excel files):**
The table below describes the content of Tables 2, 4, 5, 6 and 7.

| | Data/challenges | Excel sheet names with analysis type | |
|---|---|---|---|
| Table 2 | Biochemical Effect | Regression, Classification, Clinical | NAGLU, PTEN*, TPMT*, CBS1High, CBS1Low, CBS2High, CBS2Low, SUMO1*, SUMO2*, SUMO3*, CALM1, GAA |
| | | Regression, Classification | Frataxin*, p16* |
| | | Classification, Clinical | PCM1, LPYK1, LPYK2 |
| | | Classification | p53* |
| Table 4 | Annotate all Missense | Classification, Clinical | AAM1All, AAM2All, AAM1BiClass, AAM2BiClass, AAM1CV, AAM2CV, AAM1HGMD, AAM2HGMD |
| Table 5 | Cancer | Classification | ENIGMA, BRCA |
| Table 6 | Expression and Splicing | Regression, Classification | RegSatEnh1, RegSatEnh2, RegSatProm1, RegSatProm2, VexSeq1, VexSeq2, eQTL2 |
| | | Regression | MaPSy1, MaPSy2 |
| | | Classification | eQTL1, MaPSy3 |
| Table 7 | Complex disease | Classification | Crohns |

* These Biochemical effect sheets involve genes implicated in cancer.
A short description of the sheet without self-explanatory names is given below. For details see Analyzed Challenges (supplementary text).

- AAM1All: Annotate all missense data with pathogenic and benign (P, B) variants from ClinVar and disease mutations (DM) from HGMD.
- AAM2All: Annotate all missense data with pathogenic, likely pathogenic, benign and likely benign (P, LP, B, LB) variants from ClinVar and disease mutations and questionable disease mutations (DM, DM?) from HGMD.
- AAM1BiClass: Annotate all missense data with pathogenic and benign (P, B) variants from ClinVar and disease mutations (DM) from HGMD restricted to the bi-class genes
- AAM2BiClass: Annotate all missense data with pathogenic, likely pathogenic, benign and likely benign (P, LP, B, LB) variants from ClinVar and disease mutations and questionable disease mutations (DM, DM?) from HGMD, restricted to the bi-class genes.
- AAM1CV: Annotate all missense data with pathogenic and benign (P, B) variants from ClinVar.
- AAM2CV: Annotate all missense data with pathogenic, likely pathogenic benign and likely benign (P, LP, B, LB) variants from ClinVar.
- AAM1HGMD: Annotate all missense data with benign (B) variants from ClinVar and disease mutations (DM) from HGMD.
- AAM2HGMD: Annotate all missense data with benign and likely benign (B, LB) variants from ClinVar and disease mutations and questionable disease mutations (DM, DM?) from HGMD.
- CBS1High, CBS1Low: the data from CAGI 1 CBS challenge with relative yeast growth measured in high and low pyridoxine concentration, respectively.
- CBS2High, CBS2Low: the data from CAGI 2 CBS challenge with relative yeast growth measured in high and low pyridoxine concentration, respectively.
- SUMO1, SUMO2, SUMO3: The data from the three subsets of variants created for the SUMO challenge.
- LPYK1, LPYK2: the data from the two subsets of variants created for the L-PYK challenge for predicting the absence of enzymatic activity.
- RegSatEnh1, RegSatEnh2: contains data from the Regulation-Saturation challenge restricted to the enhancers. In RegSatEnh1 the positive label corresponds to increased expression, whereas in RegSatEnh2 it corresponds to decreased expression.
- RegSatProm1, RegSatProm2: contains data from the Regulation-Saturation challenge restricted to the promoters. In RegSatProm1 the positive label corresponds to increased expression, whereas in RegSatProm2 it corresponds to decreased expression.
- MaPSy1, MaPSy2: the data from the MaPSy challenge for separate analysis of allelic ratios measured in vivo and in vitro, respectively.
- MaPSy3: the data from the MaPSy challenge for ESM prediction.
- eQTL1: the data for the first subset of variants from the eQTL challenge corresponding to the prediction of regulatory hit.
- eQTL2: the data for the second subset of variants from the eQTL challenge corresponding to the prediction of emVar hit and log skew allelic ratio.
- VexSeq1, VexSeq2: contains data from the VexSeq challenge. In VexSeq1 the positive label corresponds to over-splicing, whereas in VexSeq2 it corresponds to under-splicing.

The table below gives the measures corresponding to the three analysis types that are reported in the tables. Only the measures corresponding to the applicable analysis type is reported in a given sheet. A subset of the Reported measures are used as Selection measures to determine the order in which methods are displayed in the tables as well as for picking the methods displayed in the figures. Coverage (COV), the proportion of variants for which a method makes a prediction, is additionally reported in all sheets.

| Analysis type | Reported measures | Selection measures |
|---|---|---|
| Regression | R-squared ($R^2$), RMSE, Pearson's correlation ($r$), Spearman's rank correlation, Kendall's Tau ($\tau$) | Pearson's correlation, Kendall's Tau ($\tau$) |
| Classification | AUC | AUC |
| Clinical | Truncated AUC, log-log AUC, TPR, FPR, $LR^+$, $LR^-$, DOR, MCC, PPV, PPP | Truncated AUC |

Each clinical measure, except Truncated AUC and Log-log AUC, is computed w.r.t. a threshold for the method score, corresponding to one of the three evidence levels: supporting, moderate and strong. The value of an evidence threshold depends on the prior (proportion of pathogenic variants) as discussed in the Methods section. We consider the following three priors for all biochemical effect sheets with clinical analysis.

1) **Data prior**: The proportion of pathogenic variants present in the dataset. The value of the data prior is determined from the challenge specific class boundary used to create binary classes (ground truth for classification) from the experimental values. For details on how the class boundaries were determined for each challenge refer to Analyzed Challenges (supplementary text).
2) **0.1**: An approximation of the proportion of pathogenic variants encountered in a clinic in a diagnostic setting; see Methods.
3) **0.01**: An approximation of the proportion of pathogenic variants encountered in a clinic in a screening setting; see Methods.

For each of three class priors, three evidence thresholds are determined, giving a total of nine thresholds. Clinical measures are computed for all nine thresholds. In case of Annotate all missense sheets, only 0.1 and 0.01 are considered as the class priors. This gives a total of six thresholds for which the clinical measures are reported. The class priors used to compute the evidence thresholds, are also listed in the sheets containing the clinical measures. $LR^+$ and $LR^-$ reported in the sheets are global. Clinically relevant local $lr^+$ values are not reported in the sheets, since the local $lr^+$ values at the clinical thresholds are completely determined by the assumed prior as given in Figure 14 (Supplementary text).

**Coverage (Cov):**

We define coverage of a method as the percent of variants for which the method makes a prediction. If both regression and classification analyses are applicable to a sheet and the size of the classification data is different from that of the regression data, then the coverage is computed relative to the size of the classification set; i.e., the proportion of variants in the classification set for which the method makes a prediction. The variants for which an experimental value could not be recorded are excluded from the regression and classification set and consequently do not affect the coverage calculation.

**Methods reported:**

A sheet contains evaluation of one representative method from each group that made a submission for the corresponding CAGI challenge. Additionally, performance of baseline methods and Experimental-Max (if applicable) were also reported. Each method takes two columns in the sheet: one containing the measures evaluated for the method on the entire data set and the other containing confidence intervals (CI) of those measures from 1000 bootstrap samples. In case of Experimental-Max, the first column contains the mean of each measure computed with 1000 Experimental-Max predictions, which are also used to compute the confidence interval. The coverage (COV) of the method is reported without a confidence interval.

The order in which the methods are reported in the sheet is determined as follows. All available methods, including multiple submissions from the same group, baseline methods and Experimental-Max, are first ranked separately on the four selection measures: Pearson's correlation, Kendall's $\tau$, AUC and Truncated AUC. In this manner, each method gets four rank values. An average of the four ranks is taken to determine the method's final rank. Additionally, each method is assigned a nominal value by taking an average over the four measures. The nominal values are used to resolve ties in the average ranks. Any ties still present are resolved randomly. If a group has multiple methods, the representative method is picked as the method with the highest rank in that group. The representative methods, except any baselines and Experimental-Max, are listed first in the order of their ranks left to right, followed by the baselines and then the Experimental-Max.

If any method has a coverage below 0.9, it is initially excluded from the main ranking. All such methods are ranked separately in the same manner as described above in a secondary ranking. Then the main ranking is extended to include these methods at its end. This process ensures that any low coverage method, in spite of better performance on the four measures, appears after all the high coverage methods. Any baseline methods and the Experimental-Max still appear at the end.

If regression is not applicable to a sheet, the two correlation coefficients are excluded from the ranking. Similarly, if classification is not applicable, AUC and Truncated AUC are excluded. If classification is applicable, but the clinical analysis is not, then Truncated AUC is excluded.

PolyPhen-2 is used as a baseline for all biochemical effect challenges, except P53, the two cancer challenges and Annotate all missense. SIFT was used as an additional baseline for Annotate all missense. No reasonable baselines were available for expression, splicing and complex disease challenges. In addition to the methods submitted for the Annotate all missense, all available predictors from dbNSFP v3.5 were also analyzed.

**Dataset properties reported**

In each sheet there is a small table, below the performance evaluation table, giving the size and the "data prior" (proportion of positives in the data). If the sheet also had a regression type analysis, the size of the data used for regression evaluation might be different from the size of the data for classification evaluation. If the sizes are indeed different, then regression data size is reported separately from the classification data size.